\documentclass{ws-ijbc}
\usepackage{ws-rotating}     
\usepackage{graphicx}
\usepackage{epstopdf}
\usepackage{wrapfig}
\begin{document}

\catchline{}{}{}{}{} 

\markboth{Author's Name}{Paper Title}

\title{SIMPLE SCENARIOS OF ONSET OF CHAOS IN THREE-DIMENSIONAL MAPS}

\author{ALEXANDER GONCHENKO}

\address{Research Institute of Applied Mathematics and Cybernetics, \\
Lobachevsky State University of Nizhni Novgorod, \\
ul. Ul'yanova 10, Nizhny Novgorod, 603005 Russia\\
agonchenko@mail.ru}

\author{SERGEY GONCHENKO}

\address{Research Institute of Applied Mathematics and Cybernetics, \\
Lobachevsky State University of Nizhni Novgorod, \\
ul. Ul'yanova 10, Nizhny Novgorod, 603005 Russia\\
gonchenko@pochta.ru}

\author{ALEXEY KAZAKOV}

\address{Research Institute of Applied Mathematics and Cybernetics, \\
Lobachevsky State University of Nizhni Novgorod, \\
ul. Ul'yanova 10, Nizhny Novgorod, 603005 Russia\\
kazakovdz@yandex.ru}

\author{DMITRY TURAEV}

\address{Department of Mathematics, \\
Imperial College London, \\
London SW7 2AZ, United Kingdom\\
dturaev@imperial.ac.uk}

\maketitle

\begin{history}
\received{(to be inserted by publisher)}
\end{history}

\begin{abstract}
We give a qualitative description of two main routes to chaos in three-dimensional maps. We discuss
Shilnikov scenario of transition to spiral chaos and a scenario of transition to discrete Lorenz-like and figure-eight
strange attractors. The theory is illustrated by numerical analysis of three-dimensional Henon-like maps
and Poincare maps in models of nonholonomic mechanics.
\end{abstract}


\newtheorem{lm}{Lemma}
\newtheorem{rem}{Remark}

\section{Shilnikov chaos in flows.}\label{Shf}

In 1965, Shilnikov discovered that a homoclinic loop to a saddle-focus can imply chaos. The notion itself did not exist then;
the ``chaos theory'' emerged and became popular only 10-20 years later. Chaos was found in many nonlinear models of hydrodynamics,
optics, chemical kinetics, biology, etc. It also occurred that strange attractors in models of various origins often have a spiral
structure, i.e. the chaotic orbits seem to move near a saddle-focus homoclinic loop. That the homoclinic loop to
a saddle-focus with a positive saddle value implies chaos - this is Shilnikov theorem \cite{Sh65,Sh70},
but why
the converse is also so often true, how can chaos imply a homoclinic loop to a saddle-focus? This question quite
preoccupied Shilnikov in the middle 80s. He found
\cite{Sh86}
that if a system depends on a parameter and evolves,
as the parameter changes, from a stable (``laminar'') regime to a chaotic (``turbulent'') motion, then
this process is naturally accompanied by a creation of a saddle-focus equilibrium in the phase space and,
no matter what particular way to chaos the system chooses, it is also natural for the stable and unstable manifolds
of this saddle-focus to get sufficiently close to each other, so a creation of a homoclinic loop becomes easy.

This idea is not mathematically formalizable, it is an empirical statement, which makes it even more important:
as it is not mathematics, it cannot be derived from any abstract notion. It relates the beginning
of the route to chaos (Andronov-Hopf bifurcation) with the end (formation of a spiral attractor) in a simple
and model-independent way. In this paper we further develop this idea (see also \citet{GGS13}) and discuss
new basic scenarios of chaos formation which should be typical for three-dimensional maps and four-dimensional
flows (higher dimensions will, surely, bring more diversity). The first of these scenarios (Section \ref{ShD})
has already been mentioned in \citet{Sh86}. The other scenario (see Section \ref{L8}) is not related to saddle-foci
and is more ``Lorenz-like'', however it does not require the symmetry the classical Lorenz system possesses.


First, we recall in more detail the scenario from \citet{Sh86} of the creation of spiral chaos.
Shilnikov considered a smooth three-dimensional system
\begin{equation} \dot{x} = X (x, R) \label{S65:eq1}
\end{equation}
that depends on a certain parameter $R$ (the choice of notation for the parameter had a hydrodynamic motivation;
one may think of $R$ as being somehow related to Reynolds number). Let the increase of $R$ lead the system
from a stable regime to a chaotic one. That is, at some $R<R_1$ the system has a stable equilibrium state $O$,
at $R=R_1$ it loses stability, the new stable regime also loses stability with the increase of $R$,
and so on. Without additional symmetries or degeneracies, or other equilibria coming into play, it is natural
to assume that the loss of stability at $R=R_1$ corresponds to the Andronov-Hopf bifurcation, so that a single
periodic orbit $L$ is born from $O$ at $R>R_1$, and this periodic orbit inherits the stability of $O$.
The point $O$ is a saddle-focus at $R>R_1$, and at small positive values of $R-R_1$ the two-dimensional unstable
\begin{wrapfigure}{o}{0.3\textwidth}
\centerline{\epsfig{file=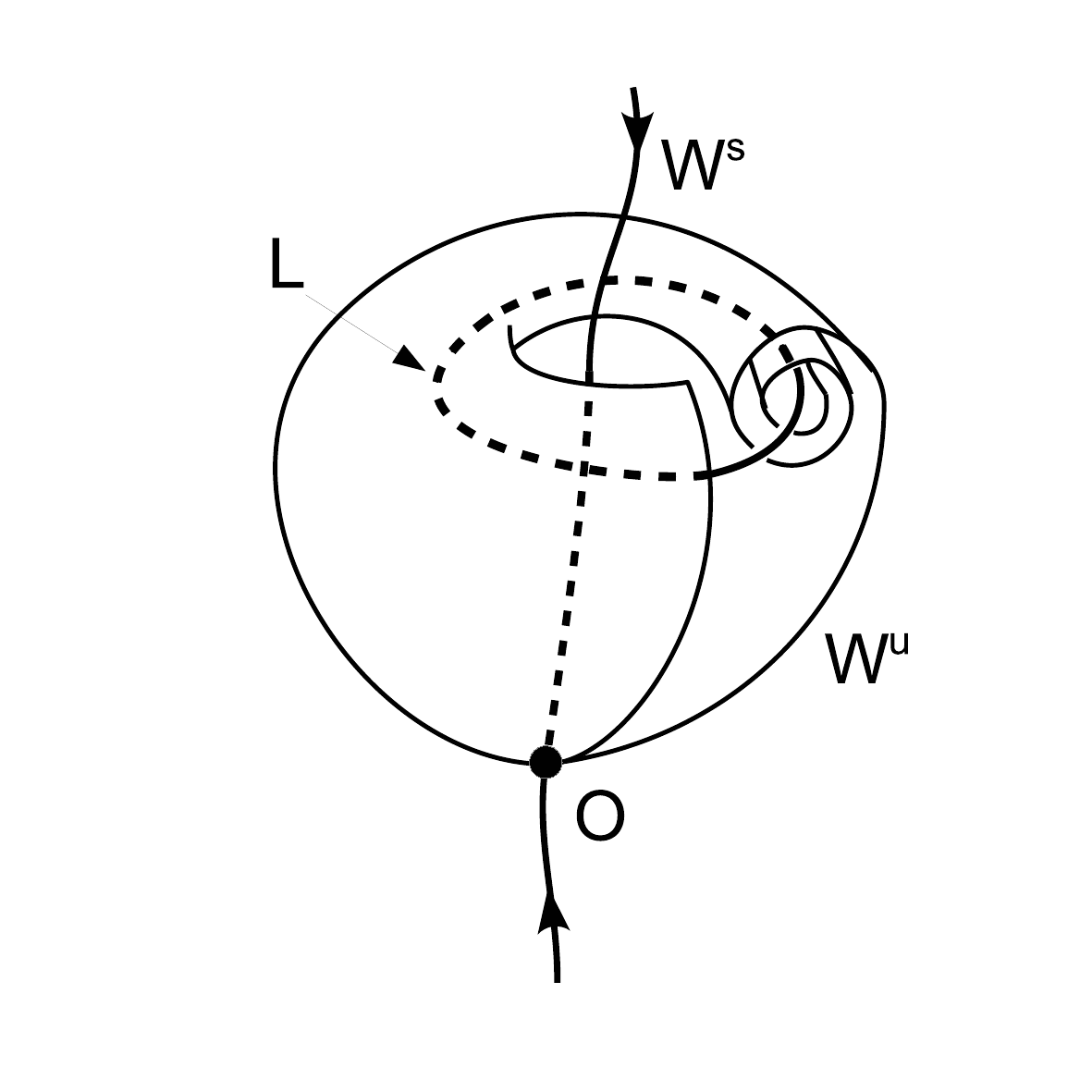, width=0.3\textwidth, height=45mm }}
\vspace{-0.5cm}
\caption{{\footnotesize A funnel-type configuration of $W^u(O)$.}} \label{sh86-1}
\end{wrapfigure}
manifold $W_O^u$ is a disc with boundary $L$. As the system evolves towards chaos with $R$ increasing,
the stable periodic orbit $L$ may also lose the stability, via a period-doubling or a secondary Andronov-Hopf
bifurcation that corresponds to a birth of an invariant torus from $L$. In any case, before the periodic orbit loses
stability its multipliers must become complex at $R>R_2$ for some $R_2>R_1$ (the multipliers of $L$ are real positive
at $R$ close to $R_1$, so they must become complex before one of them becomes equal to $-1$). At $R>R_2$
the manifold $W_O^u$ will wind onto $L$ and form a funnel-type configuration (Fig.~\ref{sh86-1}).
This funnel will attract all orbits from some open region $D$. After the funnel is formed, the creation of
a homoclinic loop to $O$ as $R$ grows further becomes very natural: the throat of the funnel may become smaller
or it may change its position, so that $W_O^u$ and $W_O^s$ may start getting closer to each other until
a homoclinic loop is formed at some $R=R_3$. If the complex characteristic exponents of the saddle-focus $O$
are nearer to the imaginary axis than the real negative one (this condition is automatically fulfilled at
$R=R_1$, so we may assume it continues to hold at $R=R_3$ too), then the existence of the homoclinic loop to $O$
implies complex orbit behavior (infinitely many suspended Smale horseshoes) in a neighborhood of the loop \cite{Sh65,Sh70}.
In case the throat of the funnel can be cut by a cross-section such that all the orbits that intersect the cross-section
come inside, the unstable manifold $W_O^u$ (more precisely, its part from $O$ till the cross-section, plus
the cross-section itself) will bound a forward-invariant region; at $R=R_3$ the attractor which lies in
this region will contain the homoclinic loop and the chaotic set around the loop. The orbits in this set spiral
around the saddle-focus, so the characteristic shape of the ``spiral attractor'' can be observed. When $R$ changes
the loop splits, but a large portion of the chaotic set will survive; also new, multi-round homoclinic loops
may appear, etc. One can have this scenario of the transition to chaos for $n$-dimensional systems
with any $n\geq 3$, e.g. just by adding $(n-3)$ contracting directions.

The main point of this observation is that the Andronov-Hopf bifurcation of the stable equilibrium $O$
not only creates a stable periodic orbit $L$, it also transforms $O$ to a saddle-focus, and instead of following
details of the further evolution of the stable regimes (the periodic orbit $L$, the periodic orbit born from $L$
after, for example, the first period-doubling, etc.) it may be more useful for the understanding of the transition
to chaos to continue to watch what happens to the primary equilibrium $O$ and how the shape of
its unstable manifold evolves. Studying typical dynamical features of attractors that can
exist in a Shilnikov funnel could be an interesting research direction. A model for the Poincare map in the funnel
was proposed in \citet{Sh86}. Based on the analysis of this map a birth of an invariant torus in the funnel was
studied in \citet{AfrVoz1,AfrVoz2}.  In \citet{Belykh} there was shown that certain type of a funnel is consistent
with the existence of a hyperbolic Plykin attractor. The wild attractor built in \citet{Bamon} can also be
inscribed in a Shilnikov funnel (in dimension $n\geq 5$).

The above described scenario appears to give the simplest (hence, the most general) route to chaos.
It involves a very small number of objects responsible for chaos formation: the equilibrium $O$,
its unstable manifold, and the periodic orbit $L$. However, there can be more complicated schemes.
For example, Shilnikov also noticed that the Andronov-Hopf bifurcation at $R=R_1$ can be
different from what is described above. Namely, we assumed that this bifurcation is soft, i.e. the equilibrium $O$
is stable at the bifurcation moment and its stability is transferred to the stable periodic orbit $L$ at $R>R_1$.
However, there can also be a subcritical Andronov-Hopf bifurcation at $R=R_1$: the periodic orbit $L$ can be saddle,
it exists then at $R<R_1$, and at the moment $L$ merges with $O$ (i.e. at $R=R_1$ already) the equilibrium $O$
becomes a (weak) saddle-focus. The unstable manifold $W_O^u$ at $R=R_1$ is the limit of the unstable manifold
of $L$. Thus, already at $R=R_1$, the manifold $W_O^u$ may have a non-trivial shape, e.g. it may form a funnel,
so a large forward-invariant region associated to this funnel is created at $R=R_1$. If a chaotic set $\Lambda$
(not necessarily an attractor) had already been formed at $R<R_1$ inside this region, then we can observe
a sudden transition from the stable regime $O$ to a well-developed spiral chaos at $R=R_1$. A similar way
of a sudden transition from a stable equilibrium to a large invariant torus was considered in \citet{AfrVoz1,AfrVoz2}.
The chaotic set $\Lambda$ can be created in several ways. For example, at some $R$ smaller than $R_1$
a saddle-node periodic orbit can emerge and, as $R$ grows, decompose into a saddle periodic orbit $L$ and
a stable periodic orbit $L_+$. In the three-dimensional case the stable manifold of $L$ is two-dimensional,
and $L$ divides it into two halves. Let one of the halves tend to $O$ and the other half, $W_L^{u+}$ to $L_+$.
As $R$ grows, the orbit $L_+$ may lose stability in some way and, eventually, homoclinic intersections of $W_L^{u+}$
with $W_L^s$ may form. The homoclinic to a saddle periodic orbit is accompanied by a nontrivial hyperbolic set
$\Lambda^\prime$ \cite{Sh67}. If $L$ keeps the homoclinics as it merges with $O$, then the weak saddle-focus
$O$ will have a homoclinic loop at $R=R_1$. Chaotic dynamics associated with this so-called Shilnikov-Hopf bifurcation was
studied in \citet{Bel70,whoelse}. If $L$ loses its homoclinics near $R=R_1$, a portion $\Lambda$ of the hyperbolic
set $\Lambda^\prime$ may still survive until $R=R_1$.

In systems with symmetry, instead of the Andronov-Hopf bifurcation a pitchfork bifurcation may happen to
a symmetric stable equilibrium $O$. Then, instead of a stable periodic orbit $L$, a pair of stable,
symmetric to each other equilibria $O_1$ and $O_2$ will be born; the equilibrium $O$ will become a
saddle with one-dimensional unstable manifold that tends to $O_{1,2}$. After the equilibria $O_{1,2}$
acquire complex characteristic exponents, the unstable separatrices of $O$ will start winding around $O_{1,2}$;
the further increase of a parameter may lead then to formation of a symmetric pair of homoclinic
loops and chaos like in the Lorenz model or in systems with ``double-scroll'' attractors \cite{Arn1,Arn2,Chua}.
In dimension $n\geq 4$ a symmetric wild Lorenz-like attractor may emerge in this way \cite{TS98}.
Without a symmetry, similar scenarios are also possible (see e.g. \citet{AshSh}):
in a system with a stable equilibrium $O_1$ a saddle-node equilibrium may emerge which decomposes
into a saddle equilibrium $O$ and a stable equilibrium $O_2$, so that one separatrix of $O$ tends to
$O_2$ and the other tends to $O_1$. After that, as parameters change, chaos may form around these
three equilibria and their unstable manifolds.

Returning to the simplest scenario, note that the spiral attractor formed in the
funnel does not need to be the ``true'' strange attractor. Bifurcations of a homoclinic loop to a saddle-focus can lead
to the birth of stable periodic orbits along with the hyperbolic sets \cite{OSh}. Therefore, stable periodic orbits
can coexist with hyperbolic sets in the funnel. If the period of these orbits is large, or
their domains of attraction are narrow, then they will be practically invisible
and the attractor will appear chaotic. Such attractors were called quasiattractors in \citet{ASh83a}.
We discuss this notion in more detail in Section \ref{QP}. We also give conditions (following \citet{OSh,TS98})
for the absence of stable periodic orbits and the true chaoticity of the attractor.

\section{Shilnikov scenario for maps} \label{ShD}

The second basic scenario which was described in \citet{Sh86} requires the dimension $n$ of the system to be at least $4$.
We assume that system (\ref{S65:eq1}) has a stable periodic orbit $L$, which undergoes a soft Andronov-Hopf bifurcation
at $R = R_1$ (i.e. its multipliers cross the unit circle and a stable two-dimensional invariant torus is born from $L$).
One may consider a cross-section $S$ to $L$, then the point $O=S\cap L$ is a saddle-focus fixed point of the Poincare
map on $S$.  The intersection of the invariant torus with the cross-section is an invariant curve $C$; it bounds the
unstable manifold $W_O^u$. At small $R-R_1$ a neighbourhood $D$ of $W_O^u\cup C$ is an absorbing
domain (the image by the Poincare map of the closure of $D$ lies strictly inside $D$), and $W_O^u\cup C$
is the attractor in $D$. We assume that for the entire range of $R$ values under consideration there exists a
continuously dependent on $R$ absorbing domain $D$ which contains $O$ along with $W_O^u$. As $R$ grows, the manifold $W_O^u$ may
start winding onto $C$, and a funnel will form. Then $W_O^u$ may come closer to the stable manifold $W_O^s$, so at a certain
interval of values of $R>R_1$ the saddle-focus fixed point $O$ will have homoclinic orbits in $D$.
The corresponding attractor was called in \citet{Sh86} {\em Poincare attractor}. The idea was that when we do not
consider this attractor on a cross-section and look at it in the phase space of the continuous-time dynamical
system (\ref{S65:eq1}), it will appear different from the spiral attractor described in the previous Section.
The main element of the spiral attractor is an equilibrium state and its unstable manifold, the main element
of the Poincare attractor is the saddle-focus periodic orbit $L$ and homoclinics to it (transverse homoclinics
to periodic orbits were discovered by Poincare, so the name).

We, however, will focus more on the attractors of discrete-time dynamical systems, i.e. we will not assume that
the map under consideration is the Poincare map for some smooth flow. Then the chaotic attractor in the funnel
formed by the unstable manifold of a saddle-focus fixed point $O$ of our map can have a shape very similar
to that of the spiral attractor for systems with continuous time. Therefore, we will also call it spiral
or Shilnikov attractor, or discrete Shilnikov attractor. One of the differences of this attractor from
the spiral attractor for flows is that, in the case of maps, the homoclinics to $O$ exist for intervals
of parameter values (not for a discrete set of parameter values as it is typical for flows).
The boundaries of such interval correspond to homoclinic tangencies.

\begin{figure}[h]
\begin{minipage}[h]{0.3\linewidth}
\center{\includegraphics[width=1\linewidth, height=0.8\linewidth]{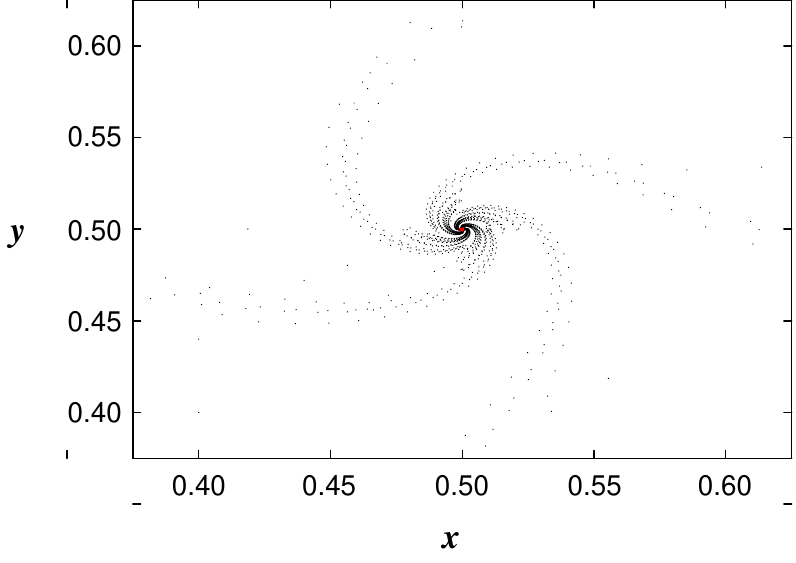} \\ (a) $M_2=0.8$}
\end{minipage}
\hfill
\begin{minipage}[h]{0.3\linewidth}
\center{\includegraphics[width=1\linewidth, height=0.8\linewidth]{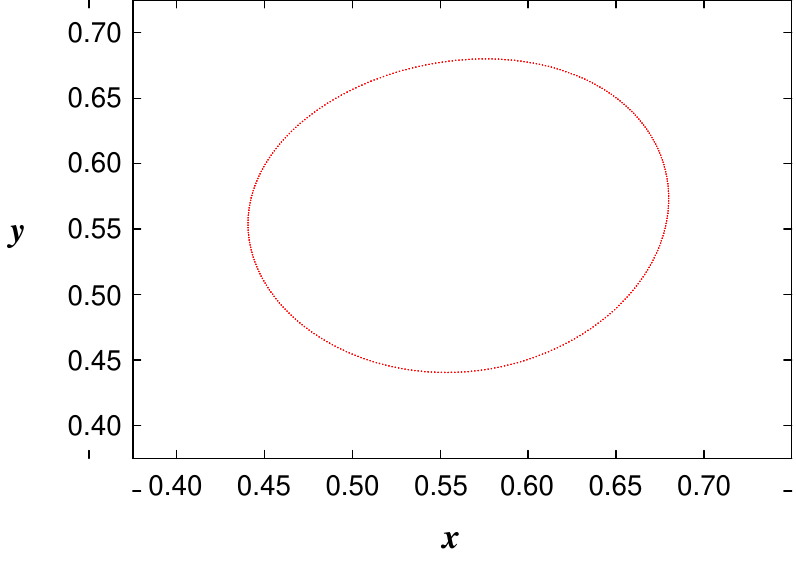} \\ (b) $M_2=0.875$}
\end{minipage}
\hfill
\begin{minipage}[h]{0.3\linewidth}
\center{\includegraphics[width=1\linewidth, height=0.8\linewidth]{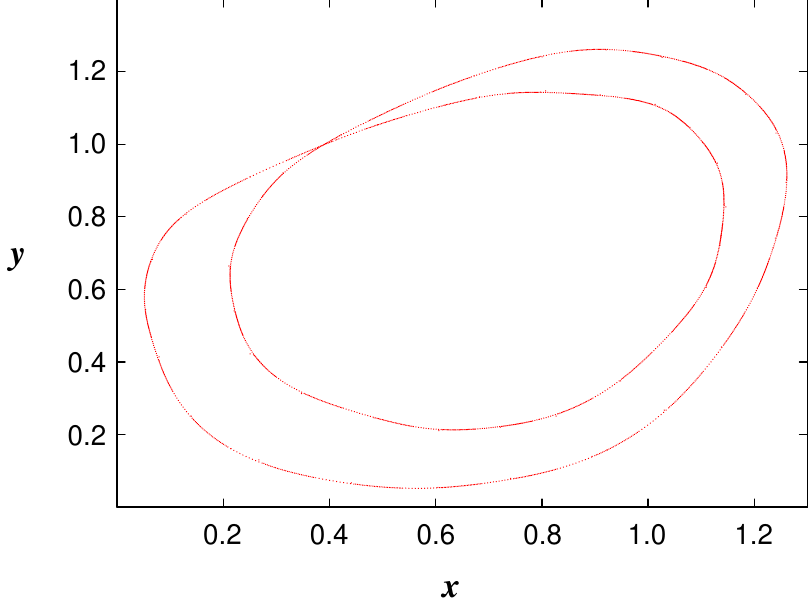} \\ (c)  $M_2=1.21$}
\end{minipage}
\vfill
\begin{minipage}[h]{0.3\linewidth}
\center{\includegraphics[width=1\linewidth, height=0.8\linewidth]{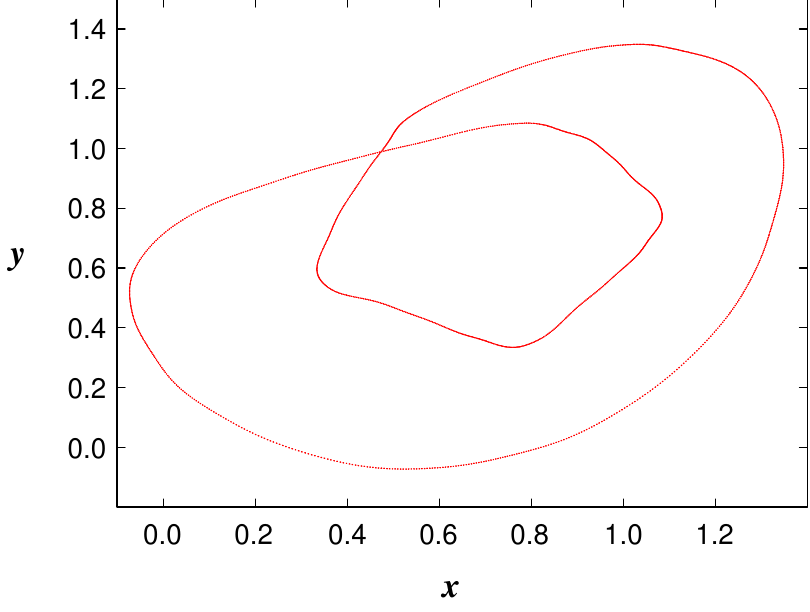} \\ (d) $M_2=1.24$}
\end{minipage}
\hfill
\begin{minipage}[h]{0.3\linewidth}
\center{\includegraphics[width=1\linewidth, height=0.8\linewidth]{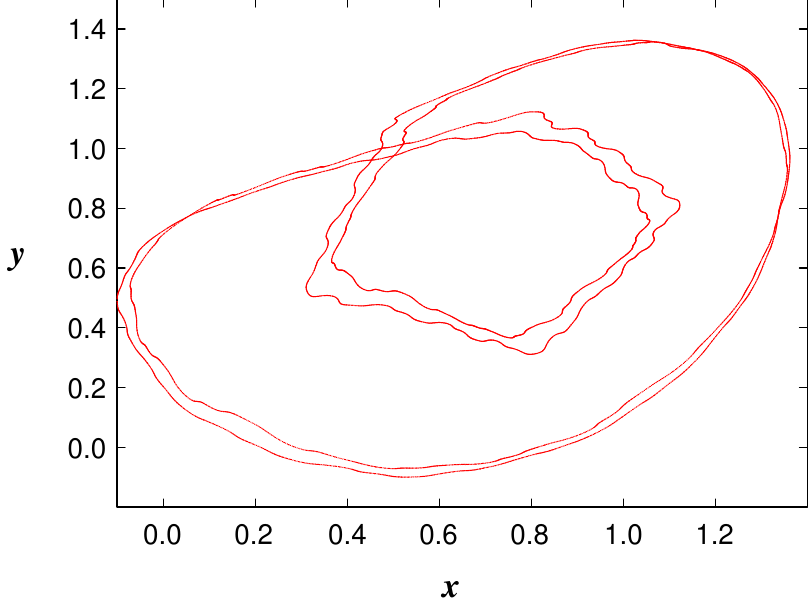} \\ (e) $M_2=1.245$}
\end{minipage}
\hfill
\begin{minipage}[h]{0.3\linewidth}
\center{\includegraphics[width=1\linewidth, height=0.8\linewidth]{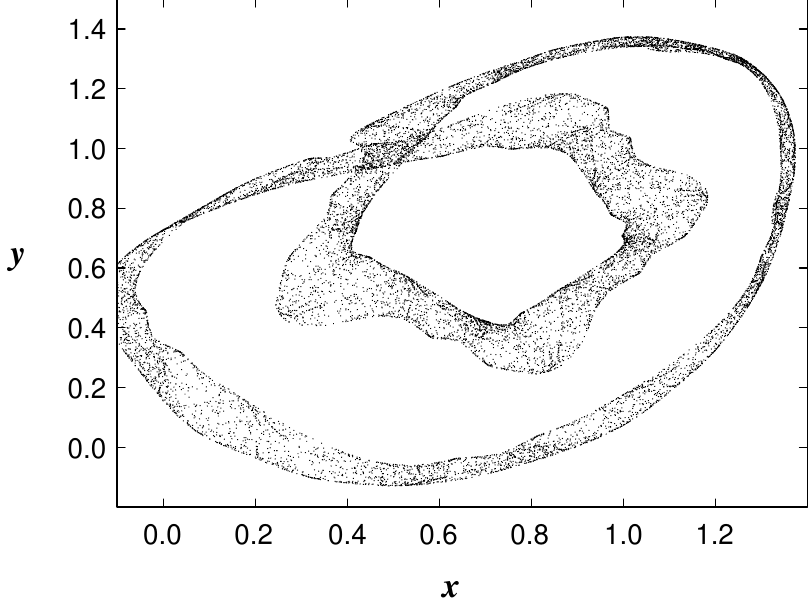} \\ (f)  $M_2=1.25$}
\end{minipage}
\vfill
\begin{minipage}[h]{0.3\linewidth}
\center{\includegraphics[width=1\linewidth, height=0.8\linewidth]{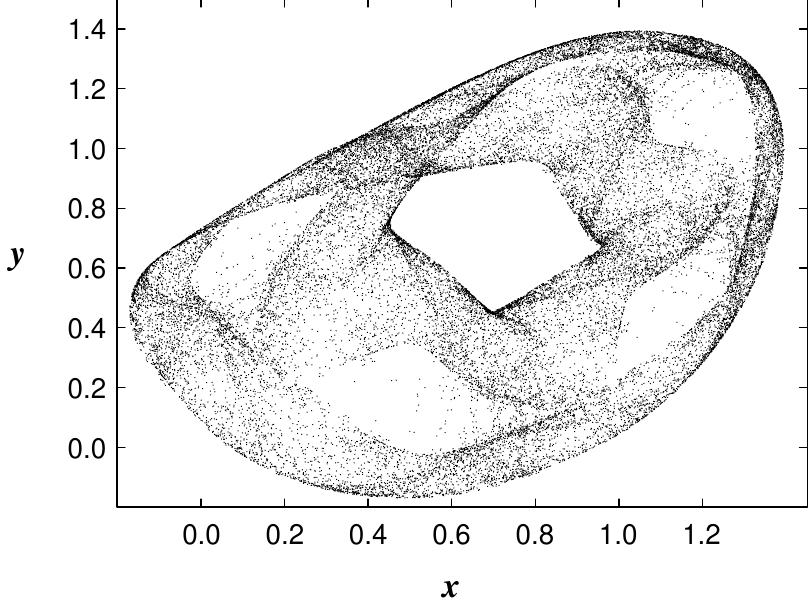} \\ (g) $M_2=1.258$}
\end{minipage}
\hfill
\begin{minipage}[h]{0.3\linewidth}
\center{\includegraphics[width=1\linewidth, height=0.8\linewidth]{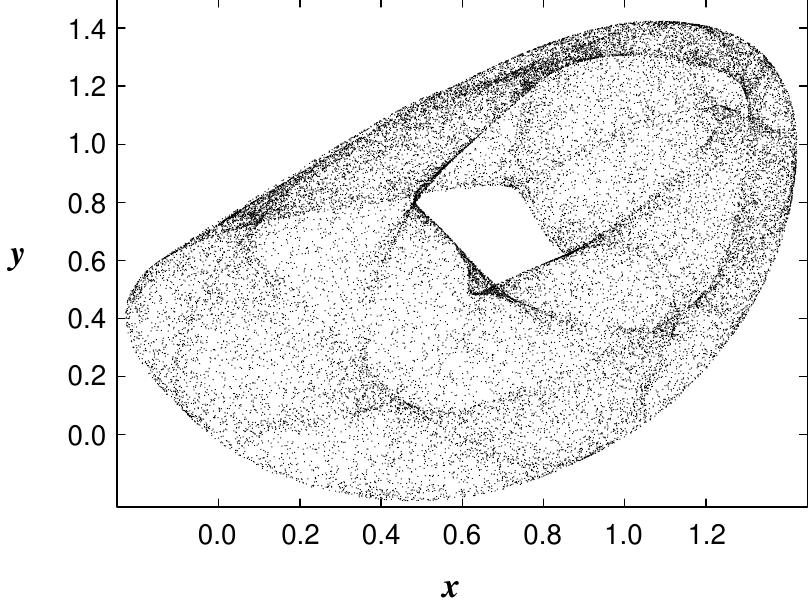} \\ (h)  $M_2=1.27$}
\end{minipage}
\hfill
\begin{minipage}[h]{0.3\linewidth}
\center{\includegraphics[width=1\linewidth, height=0.8\linewidth]{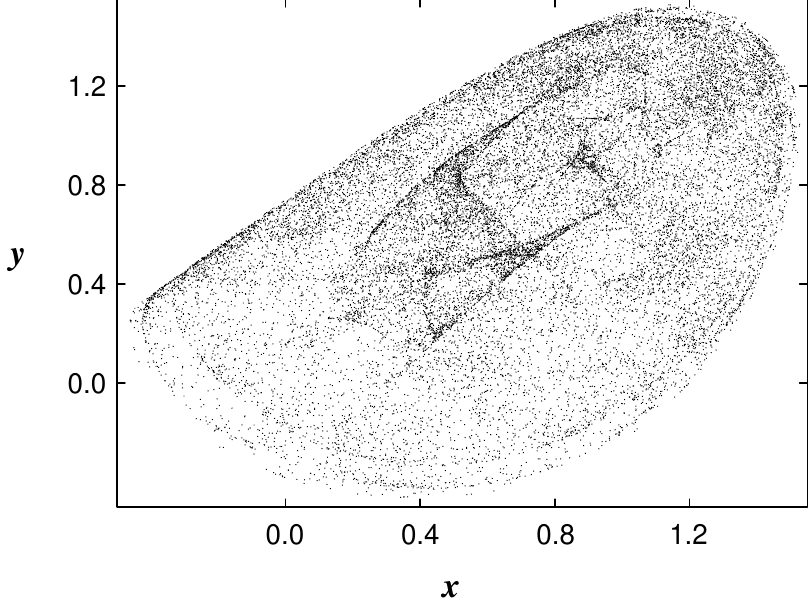} \\ (i)  $M_2=1.29$}
\end{minipage}
\caption{{\footnotesize Evolution of attractors of Henon-like map (\ref{HM2})
at fixed $B=0.7$ and $M_1=0$ (a) a stable fixed point; (b)  a stable
closed invariant curve; (c)  the invariant curve has doubled;
(d),(e) the second doubling and loss of smoothness;
(f)--(h)  breakdown of the invariant
curve and onset of chaos; (i)  Shilnikov attractor.}}
\label{fig:h2b07m10-1}
\end{figure}

In the figures below we show the discrete Shilnikov attractor in the three-dimensional Henon-like map
\begin{equation}
\bar x = y, \;\; \bar y = z, \;\; \bar z = M_1 + B x + M_2 z - y^2 \label{HM2}
\end{equation}
(this map emerges in the study of homoclinic tangencies in multidimensional systems \cite{GST93c}).
In Fig.~\ref{fig:h2b07m10-1}(i) one can notice that the attractor (here - the numerically obtained limit
of iterations of a randomly chosen initial point) is strikingly similar to the spiral attractor for flows.
The beginning of the route to the spiral chaos is quite flow-like here:
the closed invariant curve $C$ (Fig.~\ref{fig:h2b07m10-1}b) bifurcates as a single entity
to a double-round invariant curve (Fig.~\ref{fig:h2b07m10-1}c); the double-round curve
loses stability and bifurcates to the 4-round curve (Fig.~\ref{fig:h2b07m10-1}e). Next,
the bifurcation scenario changes: the invariant curve does not double anymore,
it loses smoothness, gets destroyed, and chaos is created (Fig.~\ref{fig:h2b07m10-1}f).

\begin{figure}[h]
\begin{minipage}[h]{0.24\linewidth}
\center{\includegraphics[width=1\linewidth]{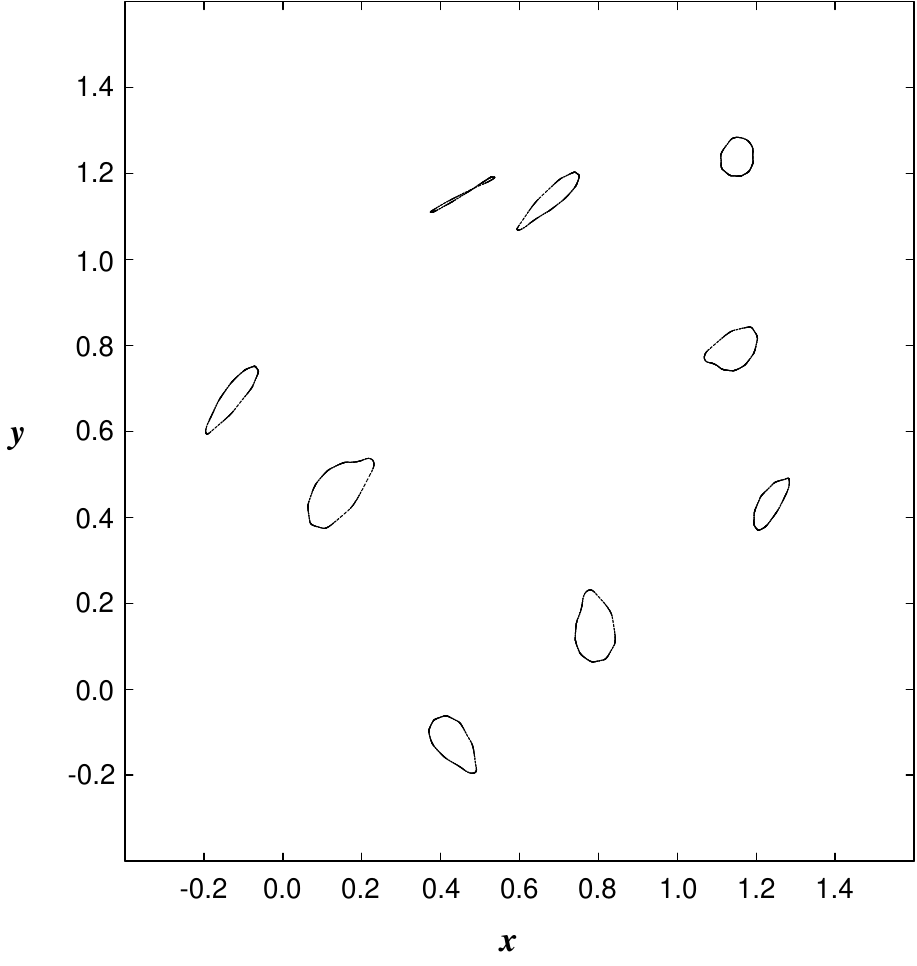} \\ (a) $M_1=0.138$}
\end{minipage}
\hfill
\begin{minipage}[h]{0.24\linewidth}
\center{\includegraphics[width=1\linewidth]{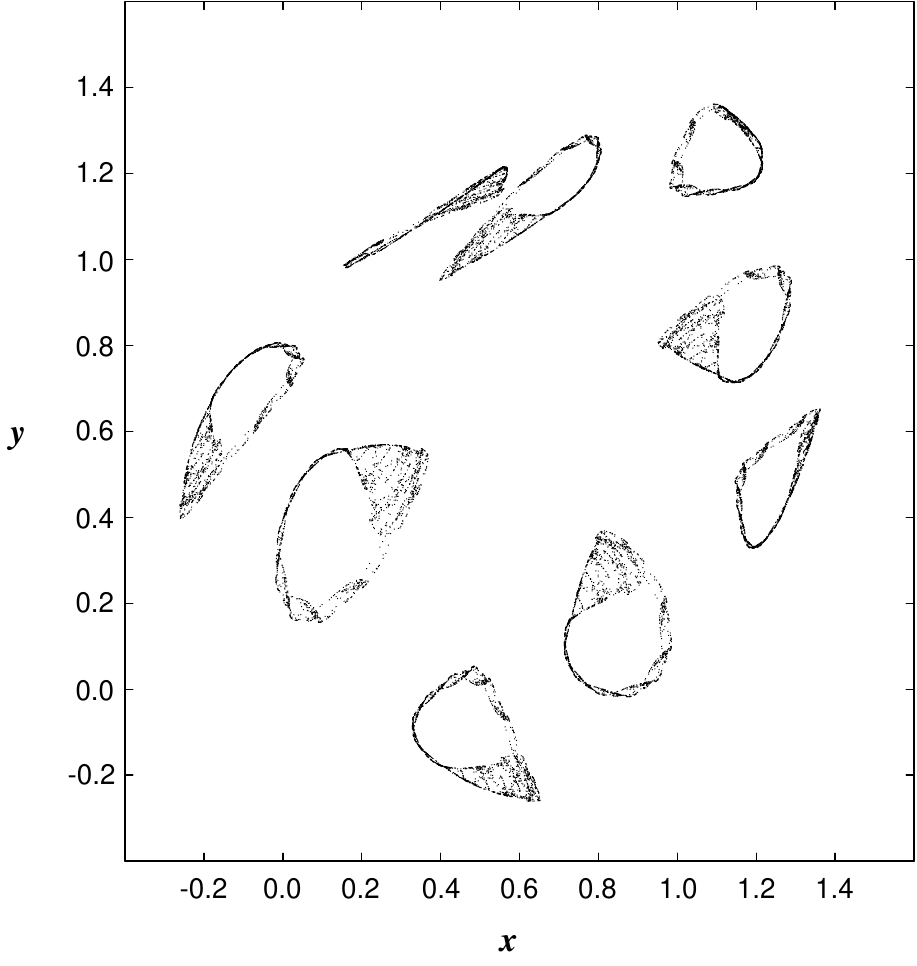} \\ (b) $M_1=0.143$}
\end{minipage}
\hfill
\begin{minipage}[h]{0.24\linewidth}
\center{\includegraphics[width=1\linewidth]{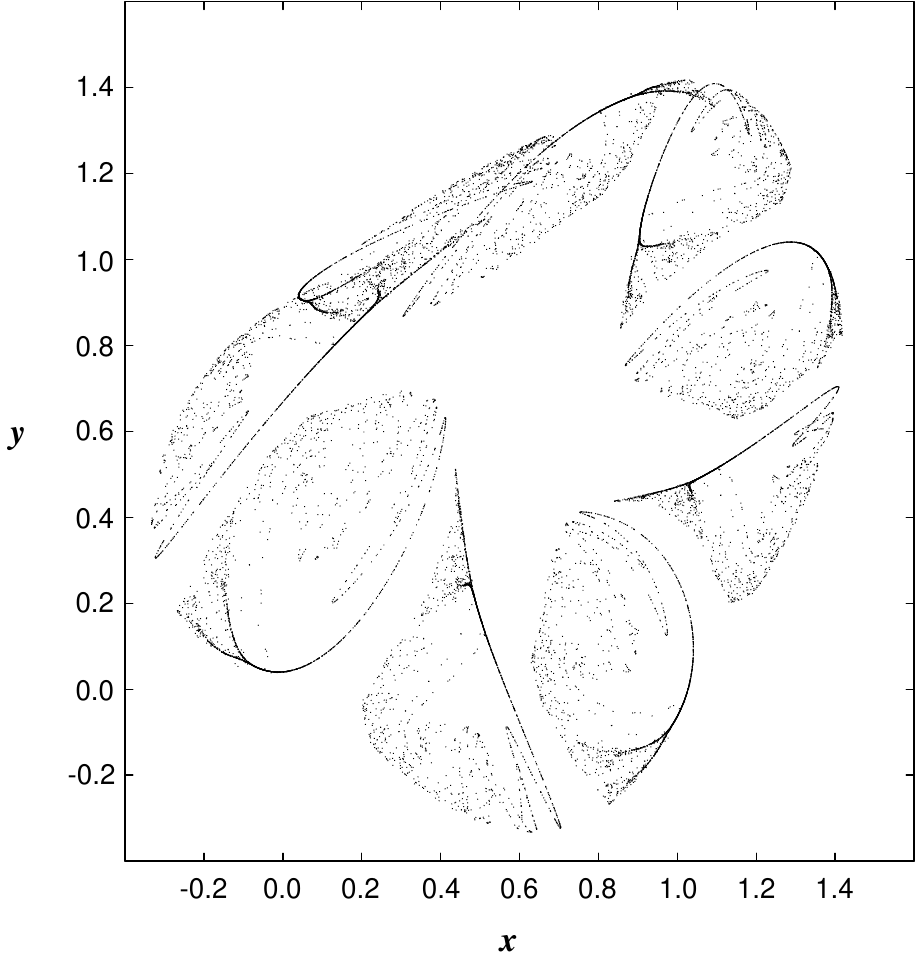} \\ (c) $M_1=0.157$}
\end{minipage}
\hfill
\begin{minipage}[h]{0.24\linewidth}
\center{\includegraphics[width=1\linewidth]{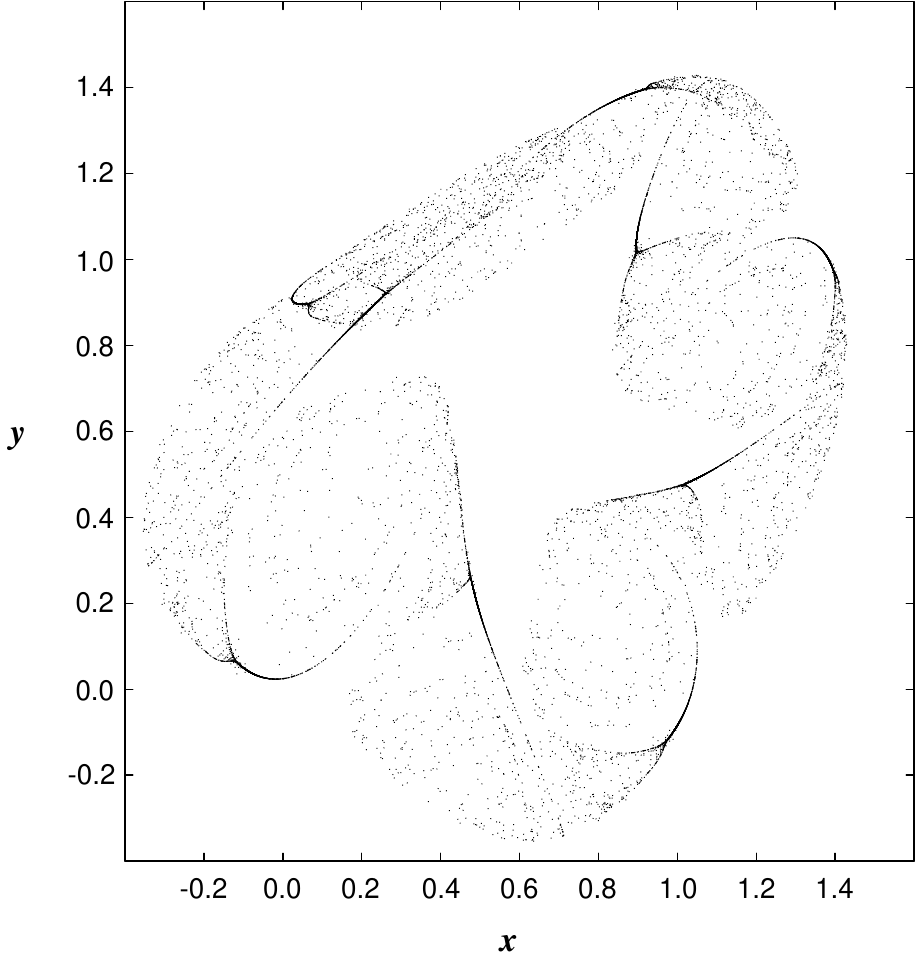} \\ (d) $M_1=0.161$}
\end{minipage}
\caption{{\footnotesize  Attractors of map (\ref{HM2}) for $B=0.7$, $M_2=1.055$.
(a) 9-component closed periodic curve; (b) spiral attractor
``sitting'' on the resonant orbit of period 9, (c) the components of period-9 spiral attractor start to collide;
(d) ``super-spiral'' attractor.} } \label{fig:supersf9}
\end{figure}

In general, it is difficult for a closed invariant curve to bifurcate as a single object.
When we change parameters, pair of resonant periodic orbits, saddle $L$ and stable $L_+$, emerge
and become visible on the invariant curve. The transition to, say, double-round closed invariant
curve would require a simultaneous period-doubling of the two resonant periodic orbits, which is
a codimension-2 phenomenon, i.e. it would require a special parameter tuning. Therefore, when a resonance materializes
on the invariant curve, it is more
natural to expect a breakdown of the invariant curve following one of Afraimovich-Shilnikov scenarios \cite{AfrSh}.
In particular, the stable resonant periodic orbit $L_+$ can itself undergo the Andronov-Hopf bifurcation
and became a saddle-focus $L_{sf}$; the unstable manifold of the $L_{sf}$ will be bounded by a multi-component
closed invariant curve (the number of components equals to the period of $L_{sf}$).
As parameters change, the unstable manifold of this periodic saddle-focus can form  a periodic funnel
and a periodic spiral attractor can form inside it.

\begin{figure}[htb]
\begin{minipage}[h]{0.3\linewidth}
\center{\includegraphics[width=\linewidth,height=0.7\linewidth]{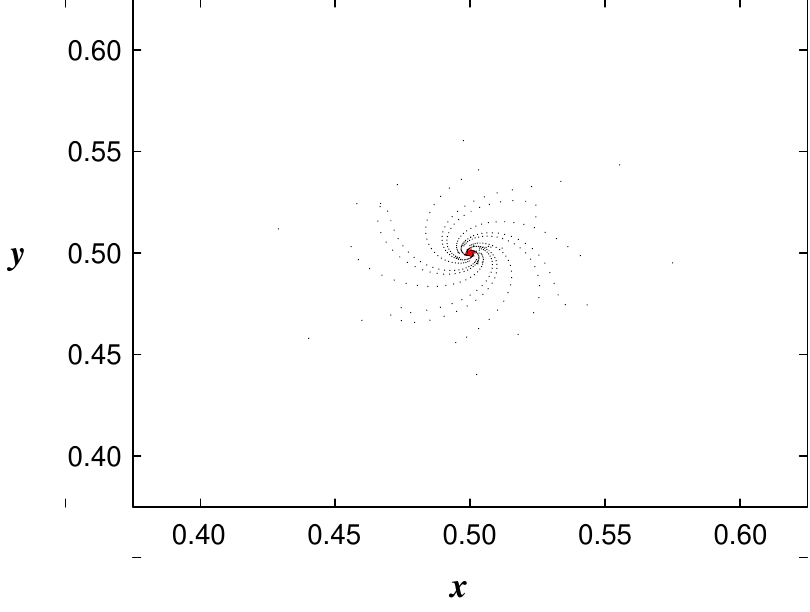} \\ (a) $M_1=0$}
\end{minipage}
\hfill
\begin{minipage}[h]{0.3\linewidth}
\center{\includegraphics[width=\linewidth,height=0.7\linewidth]{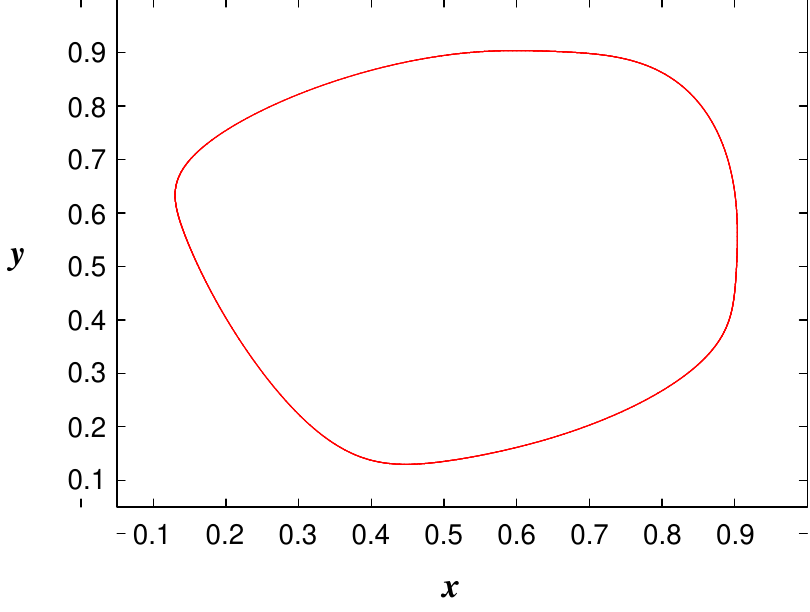} \\ (b) $M_1=0.1$}
\end{minipage}
\hfill
\begin{minipage}[h]{0.3\linewidth}
\center{\includegraphics[width=\linewidth,height=0.7\linewidth]{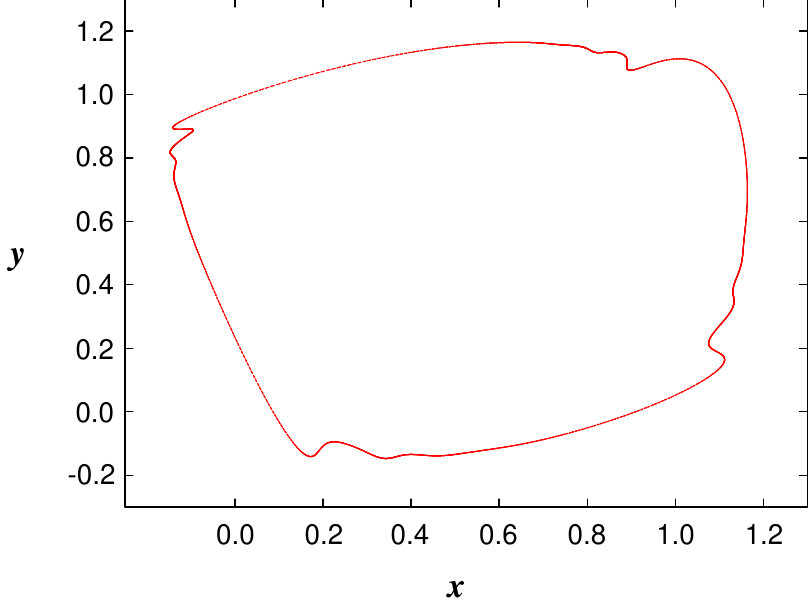} \\ (c) $M_1=0.26$}
\end{minipage}
\vfill
\begin{minipage}[h]{0.3\linewidth}
\center{\includegraphics[width=\linewidth,height=0.7\linewidth]{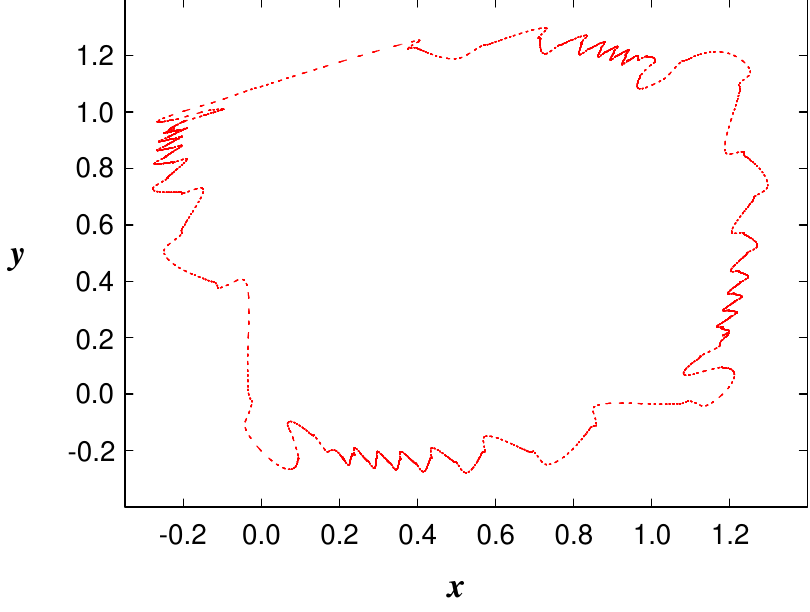} \\ (d) $M_1=0.335$}
\end{minipage}
\hfill
\begin{minipage}[h]{0.3\linewidth}
\center{\includegraphics[width=\linewidth,height=0.7\linewidth]{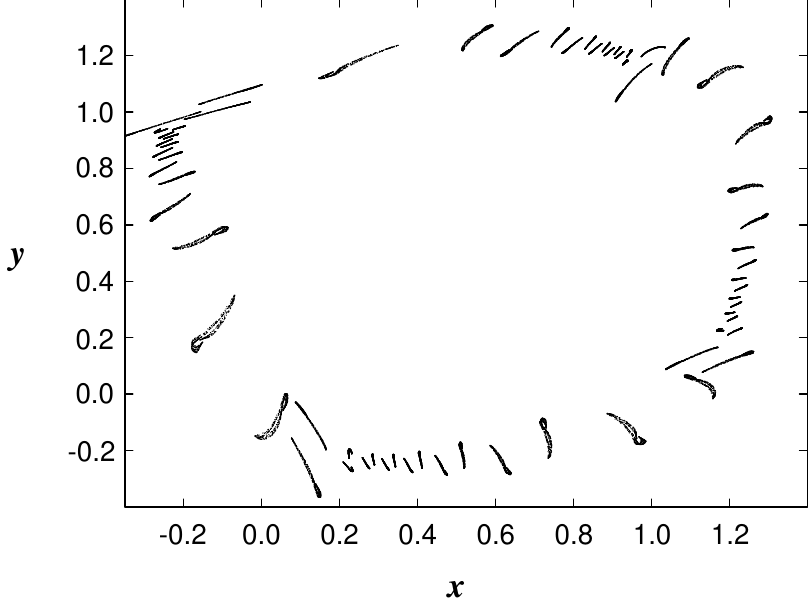} \\ (e) $M_1=0.34$}
\end{minipage}
\hfill
\begin{minipage}[h]{0.3\linewidth}
\center{\includegraphics[width=\linewidth,height=0.7\linewidth]{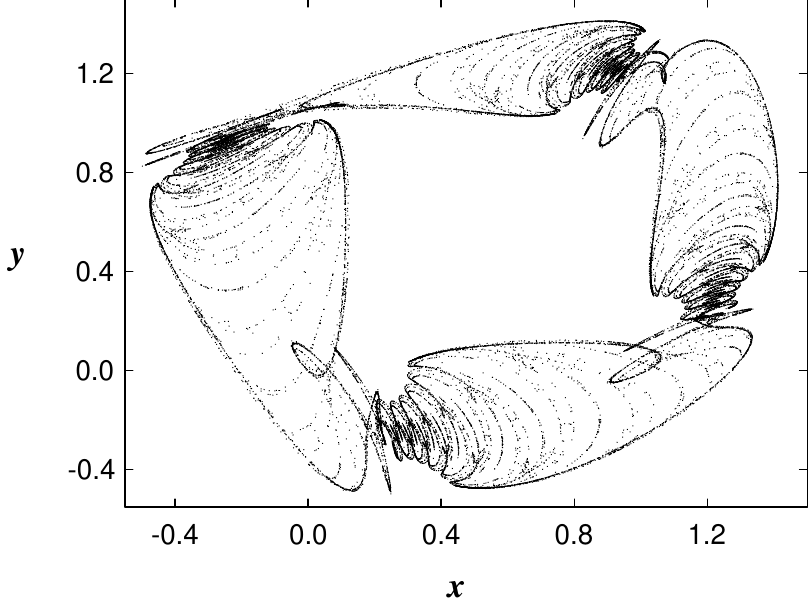} \\ (f) $M_1=0.35$}
\end{minipage}
\caption{{\footnotesize  Attractors of map  (\ref{HM2}) at $B=0.7$, $M_2=0.8$
(a) stable fixed point; (b),(c) stable closed invariant curve;
(d),(e) loss of smoothness and breakdown of the invariant curve;
(f) super-spiral attractor ``sitting'' on resonant points of period 4.} } \label{fig:supersf}
\end{figure}

An example of such behavior is shown in Fig.~\ref{fig:supersf9}. The periodic spiral attractor consists of several disjoint
components; their number equals to the period of the resonant saddle-focus (see Fig.~\ref{fig:supersf9}b). Note that the components
may collide to each other (Figs.~\ref{fig:supersf9}c,d) as parameters change. This means that the two-dimensional unstable manifold
of the resonant saddle-focus $L_{sf}$ that used to bound the components of the periodic spiral attractor starts to intersect the codimension-1
stable manifold of the other resonant periodic orbit $L$. The attractor now includes both of the resonant periodic orbits; we call
such attractor ``super-spiral''. Another example of such type of attractor is given in Fig.~\ref{fig:supersf}.

\begin{figure}[h]
\begin{minipage}[h]{0.3\linewidth}
\center{\includegraphics[height=0.7\linewidth]{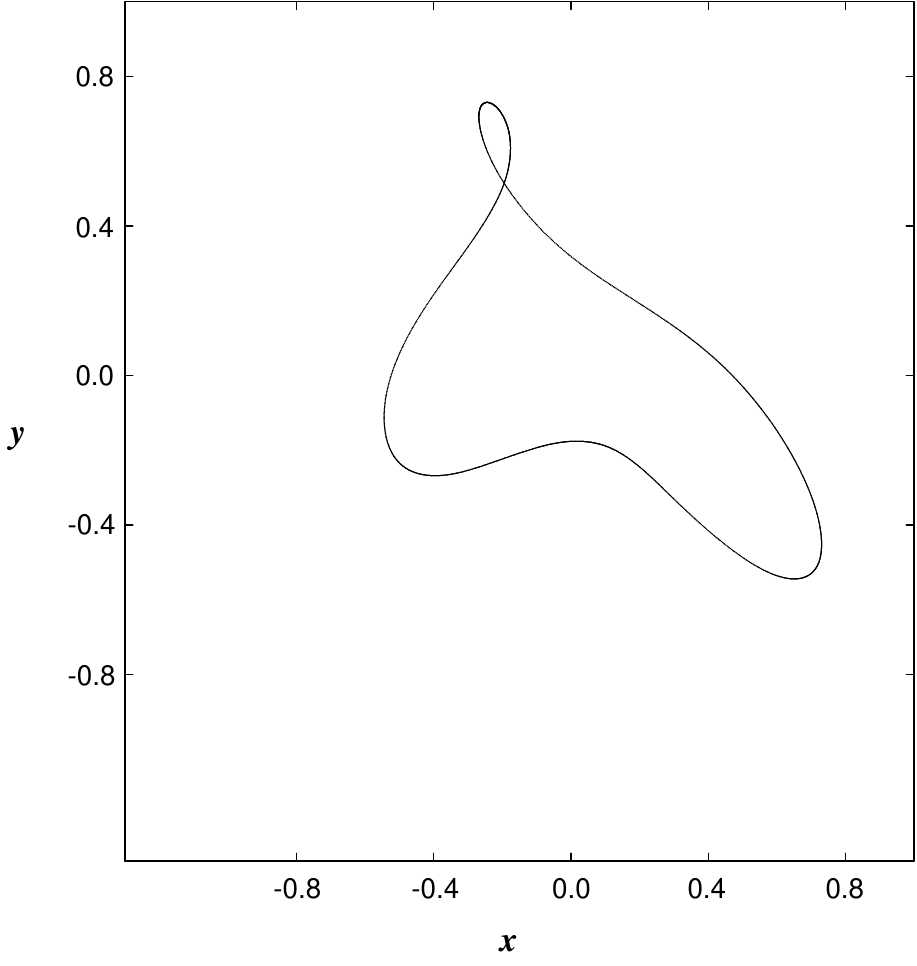} \\ (a) $M_2=-0.55$}
\end{minipage}
\hfill
\begin{minipage}[h]{0.3\linewidth}
\center{\includegraphics[height=0.7\linewidth]{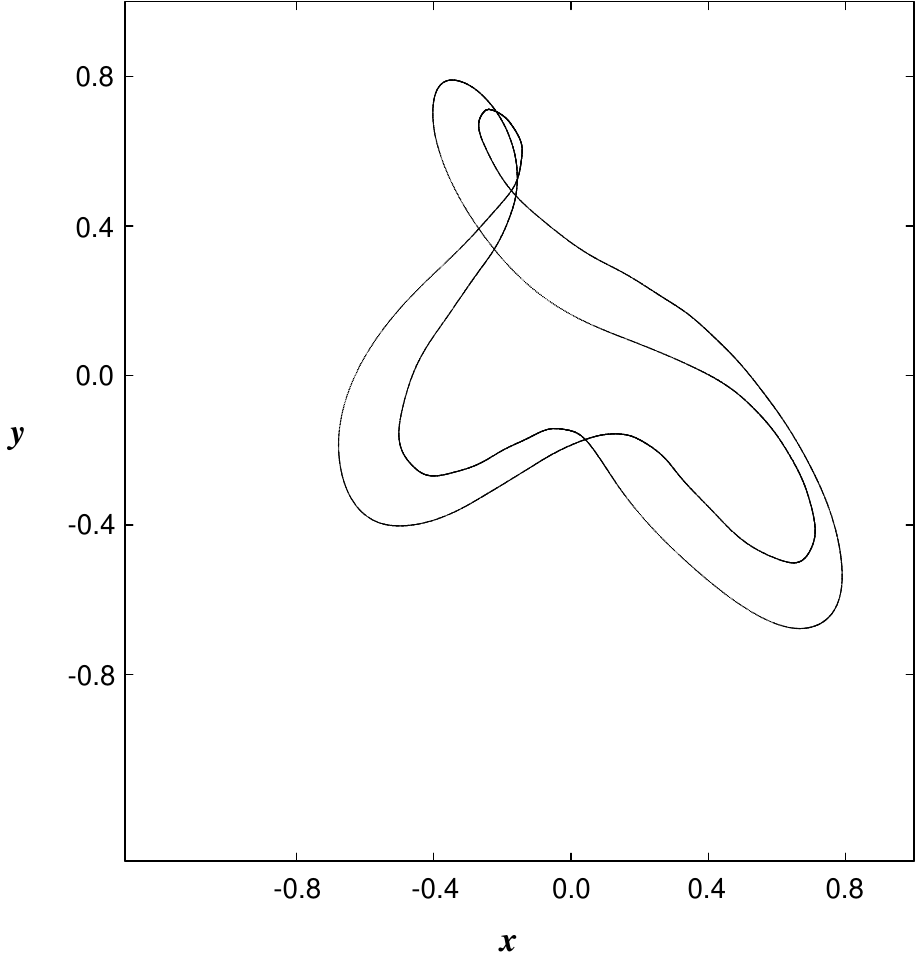} \\ (b) $M_2=-0.51$}
\end{minipage}
\hfill
\begin{minipage}[h]{0.3\linewidth}
\center{\includegraphics[height=0.7\linewidth]{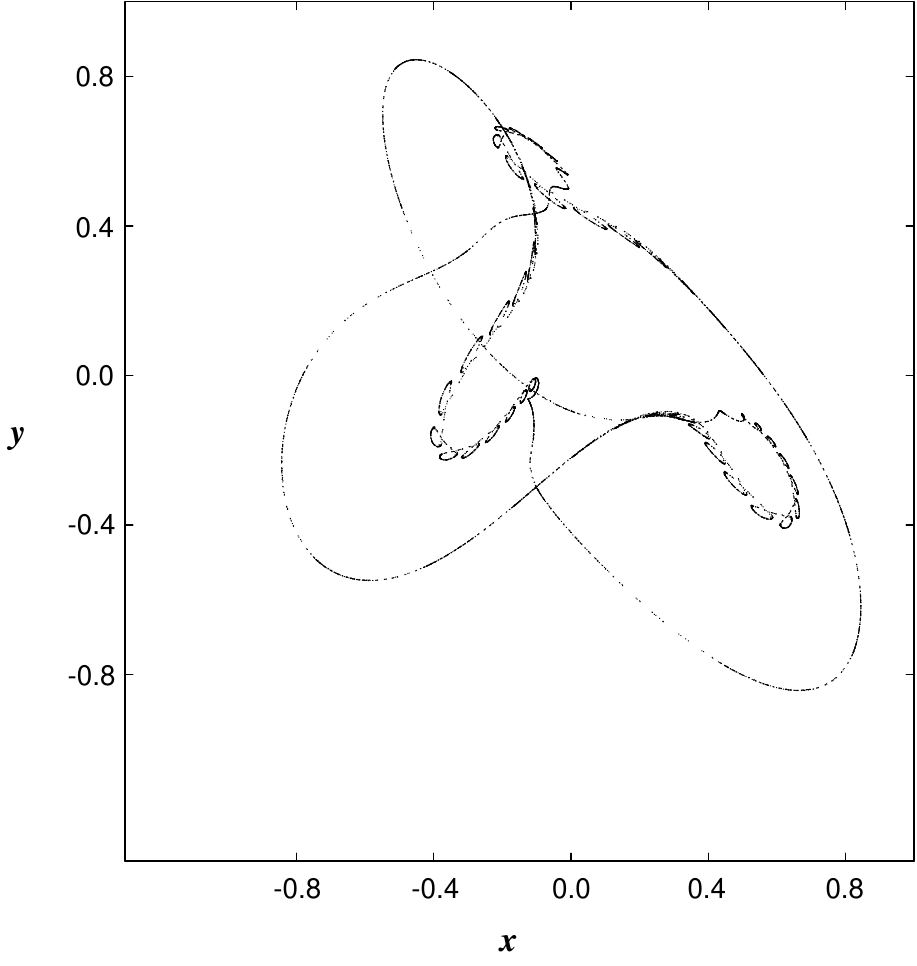} \\ (c) $M_2= -0.465$}
\end{minipage}
\vfill
\begin{minipage}[h]{0.3\linewidth}
\center{\includegraphics[height=0.7\linewidth]{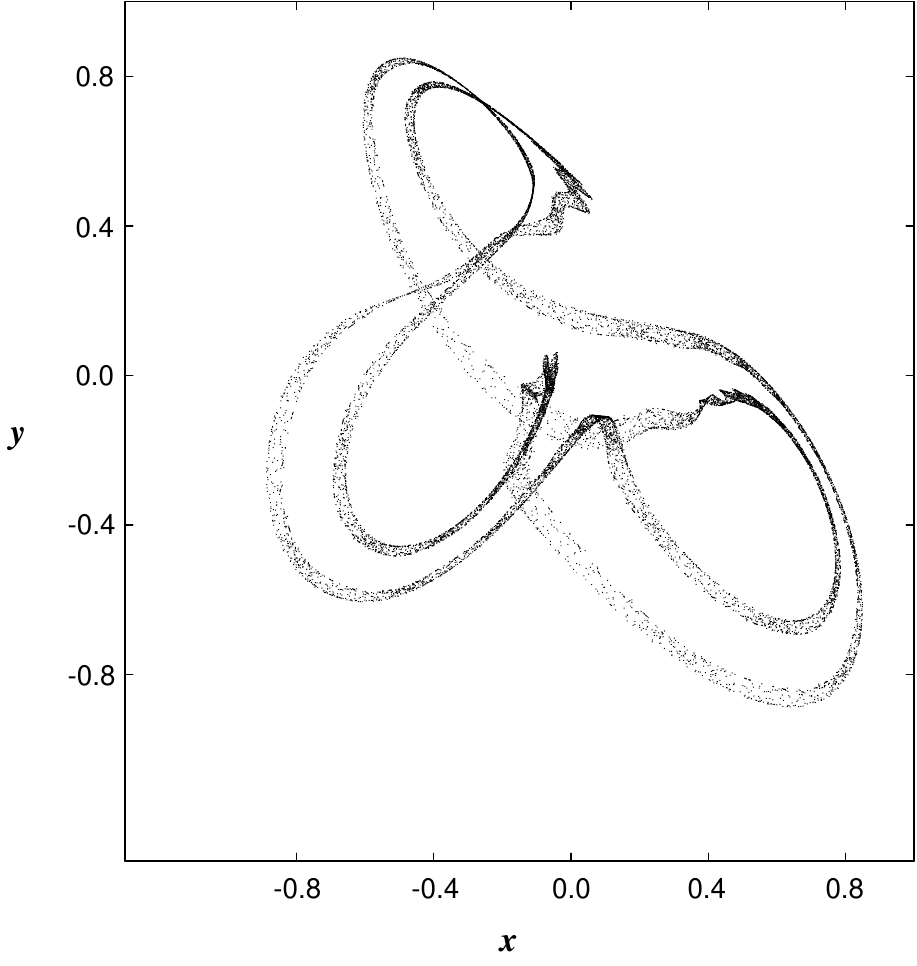} \\ (d) $M_2=-0.425$}
\end{minipage}
\hfill
\begin{minipage}[h]{0.3\linewidth}
\center{\includegraphics[height=0.7\linewidth]{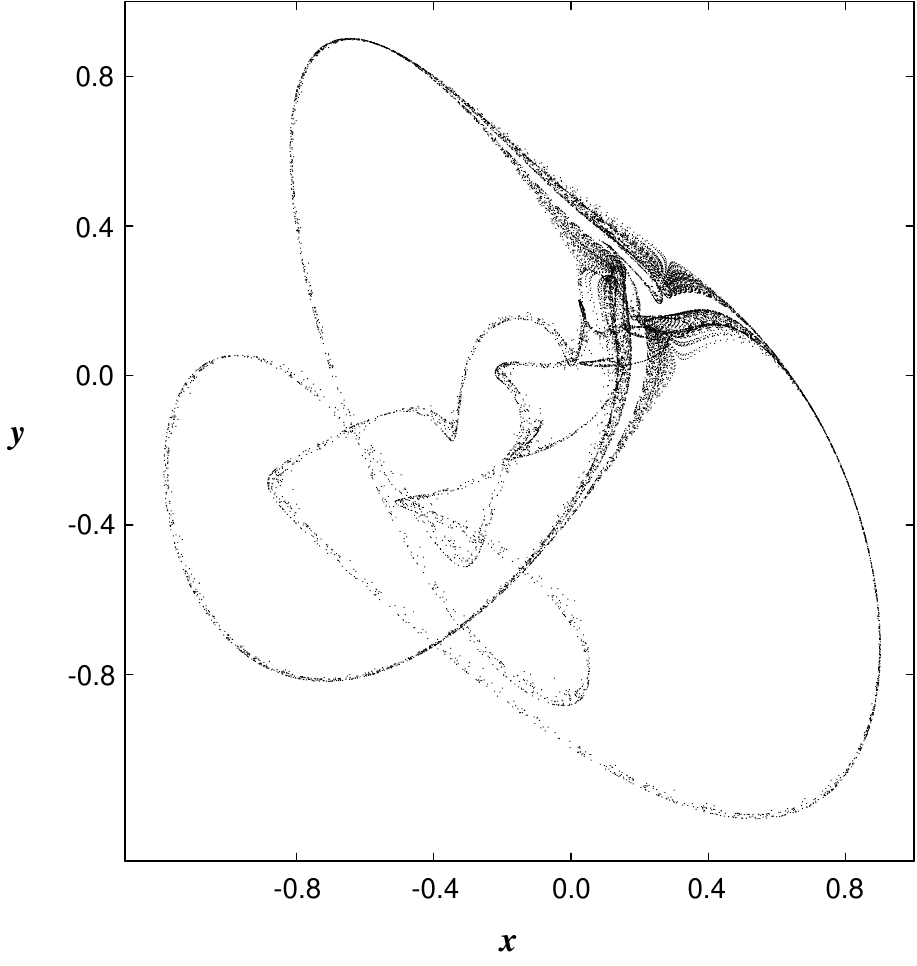} \\ (e) $M_2=-0.345$}
\end{minipage}
\hfill
\begin{minipage}[h]{0.3\linewidth}
\center{\includegraphics[height=0.7\linewidth]{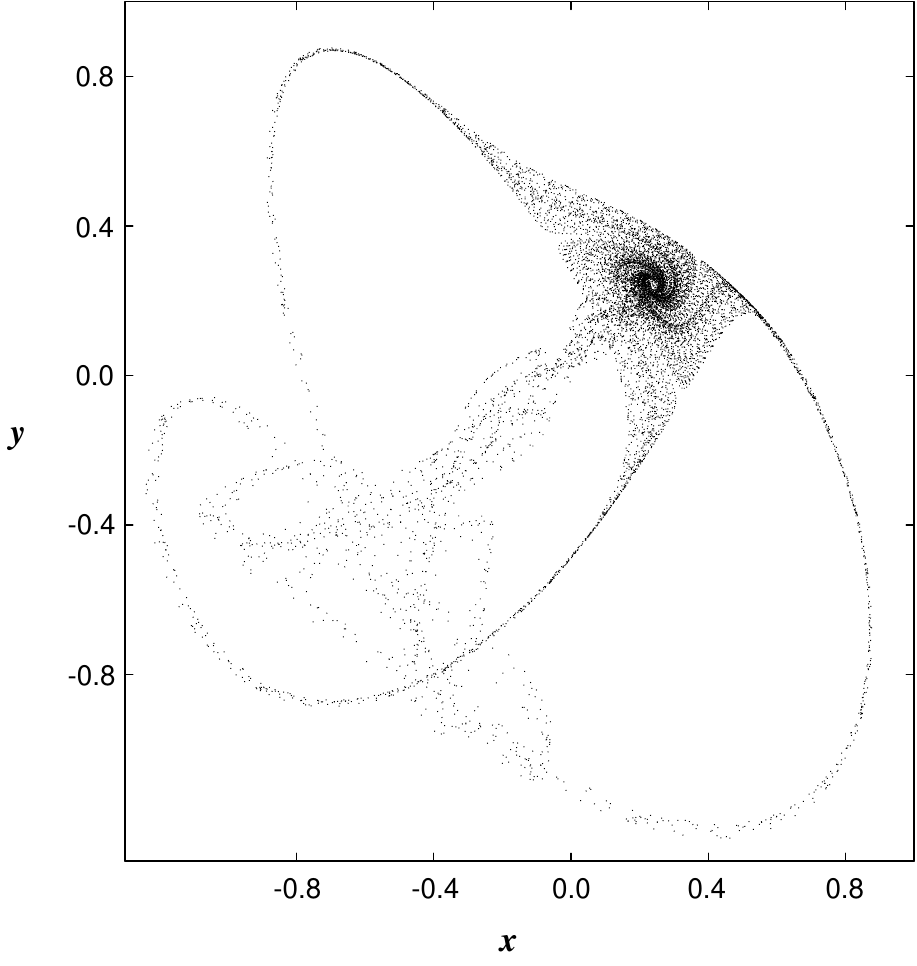} \\ (f) $M_2=-0.26$}
\end{minipage}
\caption{{\footnotesize Attractors of map (\ref{HM2}) at $B=0.7$, $M_1=-0.195$.
(a) closed invariant curve; (b)  the curve has doubled; (c),(d) period-3 orbit gets involved, and chaotic attractor is created;
(e),(f) the saddle-focus fixed point gets included into the attractor.}} \label{fig:1:31}
\end{figure}

Recall that these resonant spiral/superspiral attractors exist within the
funnel formed by the unstable manifold of the original saddle-focus
fixed point. As parameters change, the stable and unstable manifolds of the resonant periodic orbits may intersect the unstable and,
respectively, stable manifold of this fixed point, so the periodic structure of the attractor may be lost and it may start to look
more flow-like. The interplay between the original fixed point and the resonant periodic orbits can proceed in many different ways.
An example is shown in Fig.~\ref{fig:1:31} where the shape of attractor
is \begin{wrapfigure}{o}{0.3\textwidth}
\centerline{\epsfig{file=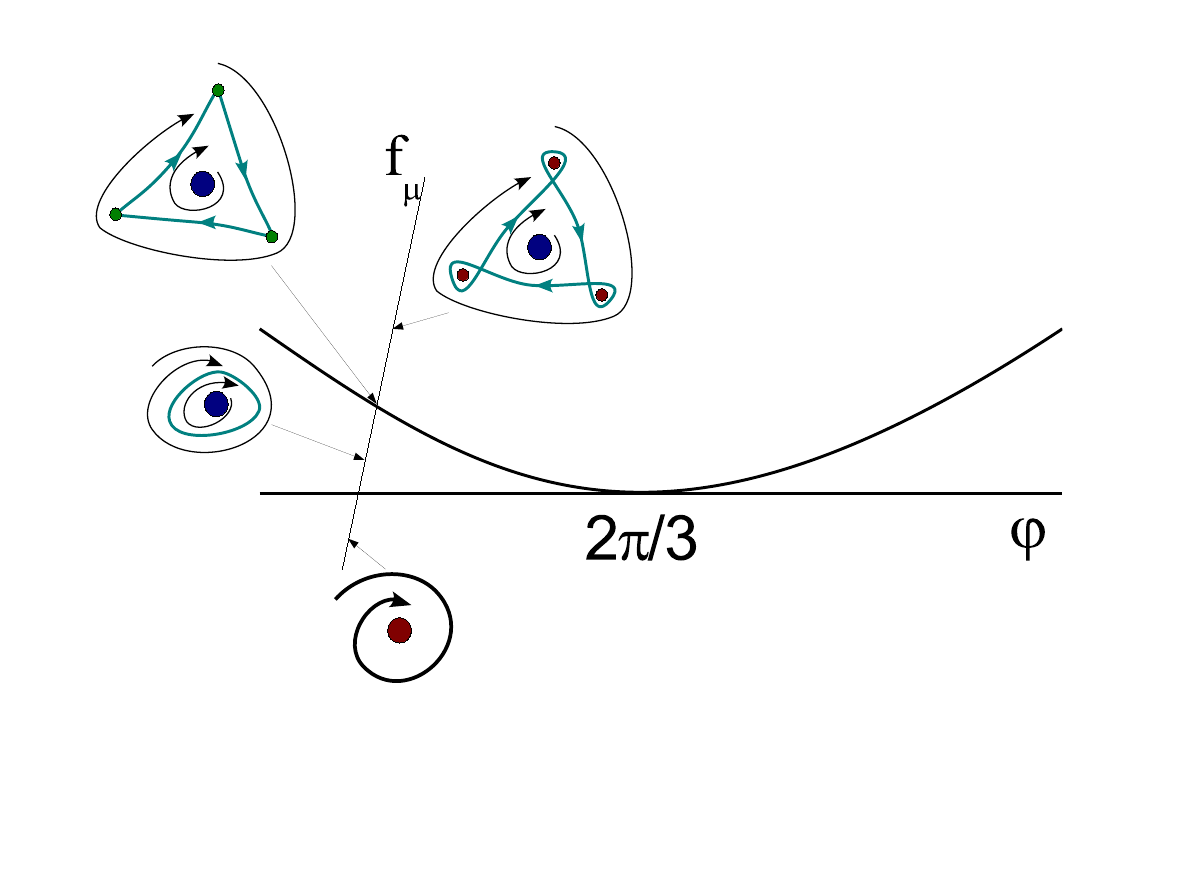, width=0.3\textwidth, height=45mm }}
\vspace{-0.5cm}
\caption{{\footnotesize Resonance 1:3.}} \label{fig:2pi3}
\end{wrapfigure}
determined by the two-dimensional unstable manifold of the fixed point and the one-dimensional unstable manifold of an orbit of period 3.


Resonance 1:3 is strong in the sense that if a periodic point undergoes the
Andronov-Hopf bifurcation with a pair of multipliers $e^{\pm i\varphi}$ where $\varphi$ is close to $\frac{2\pi}{3}$,
then the closed invariant curve that is born at this bifurcation may
fast get destroyed by colliding with a homoclinic structure of a nearby orbit of triple period;
the corresponding bifurcation diagram is in Fig.~\ref{fig:2pi3} (see \citet{Gav77,Arn77,Kuznetsov}).
This determines the
triangle shape of the funnel. Since the zone in the parameter space that is
associated with the resonance 1:3 is quite wide, this characteristic shape should be observed quite often.

\section{Lorenz-like scenario for maps.}

\subsection{Quasiattractors and true strange attractors. Pseudo\-hy\-per\-bo\-li\-ci\-ty}\label{QP}

It is well-known that hyperbolic attractors and Lorenz attractors
are two types of ``true'' chaotic attractors. Namely, every orbit in such attractor has positive maximal
Lyapunov exponent and this property is robust (it persists at small changes of the the system). Hyperbolic
attractors are structurally stable; Lorenz attractors are not, but their chaoticity is persistent \cite{ABS77, ABS83, G76, W77, GW79}.

We note that this property (of keeping ``strangeness'' at small smooth perturbations) does not seem to hold
for many ``physical'' attractors observed in numerical experiments, where the apparent chaotic behavior can
easily correspond to a stable periodic orbit with a very large period (plus inevitable noise); see more discussion in \citet{N74,ASh83a}.
In particular, H\'enon-like strange attractors \cite{BC,MV93} that are often in two-dimensional maps may transform into
stable long-period orbits by arbitrarily small changes of parameters \cite{Ures95}. The same is true for spiral attractors
of various types and, in particular, for Poincare-Shilnikov attractors presented in the previous section.
The point is that homoclinic tangencies to the saddle-focus periodic orbit can emerge within the spiral attractor.
When three-dimensional volumes are contracted bifurcations of such tangency lead to the birth of periodic sinks \cite{GST93c,GST96,GOT12}.

In general, non-transverse homoclinics and heteroclinics are ubiquitous in non-hyperbolic attractors. Without special
restrictions \cite{T96} such bifurcations lead to the birth of stable long-period orbits, so ``windows of stability'' emerge
in chaos, and the better the accuracy of observations the more of these stability windows can be seen. This makes the whole concept
of strange attractor questionable (in respect to its applicability to reality). In order to resolve this problem
Afraimovich and Shilnikov introduced the term {\em quasiattractor}, or $\varepsilon$-quasiattractor \cite{ASh83a}, that means
an attractive closed invariant set which contains a saddle periodic orbit with a transverse homoclinic (i.e. a chaotic component) and
may contain stable periodic orbits too, but the period of every stable orbit must be larger than $\varepsilon^{-1}$. So, for $\varepsilon$
small enough, even if there are stable periodic orbits within the attractor, they will not be recognized.

The spiral attractor discussed in the previous Section is, thus, a quasiattractor which contains a transverse homoclinic to
a saddle-focus fixed point or a periodic orbit. The discrete Lorenz-like and figure-eight attractors which we discuss below are
examples of a strange attractor (maybe, a quasiattractor) which contains a transverse homoclinic to a fixed or periodic point
which is a saddle, i.e. its leading (nearest to the unit circle) multipliers are real. An important feature of these attractors
is that they can, under certain conditions, be true strange attractors, i.e. one can guarantee the robust absence of stable
periodic orbits.

A universal structure which prevents the birth of stable periodic orbits was proposed in \citet{TS98}.
Namely, if an attractor has the so-called {\em pseudohyperbolicity} property, then neither the system itself nor any close system
can have stable periodic orbits in a certain neighbourhood of the attractor. This property (the term volume-hyperbolicity can also be used
in the same context \cite{Bonatti}) is formulated as follows. Let a map $F$ (the case of a flow is treated analogously) have an
absorbing domain ${\cal D}$ (a strictly forward-invariant neighborhood of an attractor $A$) and let the tangent space at each point
$x\in {\cal D}$ admit a decomposition into the direct sum of two subspaces $E^{ss}_x$ and $E^{uc}_x$ which are invariant with respect to
the differential $DF$ and which depend continuously on $x$. Moreover, let $DF$ be strongly contracting along $E^{ss}$ and let it expand
volume in $E^{uc}$. Then the map is pseudohyperbolic in $\cal D$, and every orbit in the attractor $A\subset {\cal D}$ has positive
maximal Lyapunov exponent; moreover this property persists at small smooth perturbations of the system (see \citet{TS98, TS08} for more detail).
Note that one can derive easily verifiable sufficient conditions for the pseudohyperbolicity, as given by the following result
(a reformulation of Lemma 1 of \citet{TS98}).
\begin{lm}\label{lmph}
Let a map $T$ be defined on a closure of an open region $\cal D$, and $T(cl({\cal D}))\subset {\cal D}$. Suppose that in some coordinates $(x,z)$
on $\cal D$ the map $T:(x,z)\mapsto(\bar x,\bar z)$ can be written as $\bar z=f(x,z), \qquad \bar x=g(x,z)$
where $f,g$ are at least $C^2$-smooth, and $\displaystyle det(\frac{\partial g}{\partial x})\neq 0$. Denote
$\displaystyle A=\frac{\partial f}{\partial z}-\frac{\partial f}{\partial x}(\frac{\partial g}{\partial x})^{-1}\frac{\partial g}{\partial z}$,
$\displaystyle B=\frac{\partial f}{\partial x}(\frac{\partial g}{\partial x})^{-1}$,
$\displaystyle C=(\frac{\partial g}{\partial x})^{-1}\frac{\partial g}{\partial z}$,
$\displaystyle D=(\frac{\partial g}{\partial x})^{-1}$. If
$$\max\left\{\sup_{(x,z)\in{\cal D}}\sqrt{\|A\|\;\|D\|}, \sup_{(x,z)\in{\cal D}}\|A\|, \sup_{(x,z)\in{\cal D}}\sqrt{|det D|}\right\}
+\sqrt{\sup_{(x,z)\in{\cal D}}\|B\|\;\sup_{(x,z)\in{\cal D}}\|C\|}<1,$$
then the attractor of the map $T$ in the absorbing domain $\cal D$ is pseudohyperbolic.
\end{lm}
This lemma is based on the Afraimovich-Shilnikov ``annulus principle'' \cite{AfrShRing1,AfrShRing2,AfrShRing3,book}
which gives sufficient conditions for the existence of what is now called a dominated splitting. It also generalizes the hyperbolicity conditions proposed in \citet{ABS77,ABS83} for the Poincare map of the Lorenz attractor.

Hyperbolic and Lorenz attrtactors satisfy the pseudohyperbolicity property, however there are other pseudiohyperbolic attractors.
For example, in \citet{TS98} an example of a {\em wild-hyperbolic} strange attractor was constructed for a four-dimensional flow.
Unlike hyperbolic and Lorenz attractors, wild hyperbolic ones may contain homoclinic tangencies.
However, these tangencies are such that their bifurcations do not lead to stable periodic
orbits (as the conditions from \citet{GST93c,GST96,GST08} for the birth of periodic sinks from
homoclinic tangencies are automatically violated by the pseudohyperbolicity).

Another example of a wild-hyperbolic attractor with the pseudohyperbolicity property can be obtained by a
small time-periodic perturbation of a flow
that possesses a Lorenz attractor \cite{TS08}. By taking a discrete forward orbit of the corresponding Poincar\'e map
(the map for the period of the perturbation), we obtain a strange attractor which looks quite similar to the canonical
(continuous time) Lorenz attractor. We call such attractors {\em discrete Lorenz attractors}
(see exact definitions in \citet{GGOT13}). Importantly, a normal form for the bifurcations of
periodic points with the triplet of multipliers $(-1,-1,+1)$ is an (exponentially small) periodic perturbation of
the Shimizu-Morioka system \cite{SST93}; this system is known to have a Lorenz attractor \cite{ShA86,ShA93}. Therefore,
discrete Lorenz-like attractors can appear at the bifurcations of an arbitrary map which has a periodic orbit that
undergoes the $(-1,-1,+1)$-bifurcation \cite{SST93,GOST05,GGOT13}.

In particular, a class of Henon-like maps was considered in \citet{GGOT13}:
\begin{equation}
\bar x=y,\quad \bar y=z, \quad \bar z=Bx + f(y,z), \label{3dHenon}
\end{equation}
where $f$ is a smooth function. The Jacobian of such map is constant and equals $B$. The fixed points are given by
$x=y=z=x_0,\quad x_0(1-B)=f(x_0,x_0)$. The characteristic equation at the fixed point is
$\lambda^3-A\lambda^2-C\lambda-B=0$, where $A=f'_z(x_0,x_0)$, $C=f'_y(x_0,x_0)$. At $(A=-1, C=1, B=1)$, the fixed point has multipliers
$(-1,-1,+1)$. Take a smooth three-parameter family of maps (\ref{3dHenon}) which at zero parameter values has
a fixed point with multipliers $(-1,-1,+1)$, and let the fixed point exist for a region of
parameter values adjoining to zero. Move the origin to the fixed point. The map takes the form
\begin{equation}
\bar x=y,\quad \bar y=z, \quad\bar z=(1-\varepsilon_1)x + (1-\varepsilon_2)y -(1+\varepsilon_3) z +
\alpha y^2 + \beta yz + \gamma z^2 + \dots, \label{Henon}
\end{equation}
where $\varepsilon_{_{1,2,3}}$ are small,
$\alpha=\frac{1}{2}f_{yy}''(x_0,x_0),\qquad \beta=f_{yz}''(x_0,x_0),\qquad\gamma=\frac{1}{2}f_{zz}''(x_0,x_0)$,
and the dots stand for cubic and higher order terms.
\begin{lm} \cite{GGOT13}
Assume
\begin{equation}\label{gabeta}
(\gamma-\alpha)(\alpha-\beta+\gamma)>0.
\end{equation}
Then map (\ref{Henon}) has a pseudohyperbolic Lorenz-like attractor for all $\varepsilon$ from an open,
adjoining to $\varepsilon=0$, subregion of $\{\varepsilon_1>0, \;\varepsilon_1+\varepsilon_3>0,
|\varepsilon_2-\varepsilon_1-\varepsilon_3|\leq L(\varepsilon_1^2+\varepsilon_3^2)\}$ with some $L>0$.
\label{lm11}
\end{lm}

Example for which the hypothesis of the lemma holds is given by the map
\begin{equation}
\bar x = y, \;\; \bar y = z, \;\; \bar z = M + B x + C y - z^2,
\label{HM1}
\end{equation}
for which a discrete Lorenz attractor was found in \citet{GOST05} for an open domain of the parameters $(M,B,C)$
adjoining to the point $(M= - 1/4, B=1, C = 1)$. At these values of the parameters, the map has a fixed point
$x=y=z=\frac{1}{2}$. After shifting the coordinate origin to this point we have the map in the form
$$\bar x=y,\qquad \bar y=z,\qquad \bar z = x+y -z - z^2,$$
i.e. the fixed point has multipliers $(-1,-1,+1)$, and $\alpha=0,\quad \beta=0,\quad\gamma=-1$.
As we see, condition (\ref{gabeta}) of the lemma holds.
Numerically obtained portraits of Lorenz-like attractors in
this map see in Fig.~\ref{fig1lor}.

Another example is given by
\begin{equation}\label{mapc}
\bar x=y,\qquad \bar y=z+\gamma y^2,\qquad \bar z = M_0+Bx+M_1y +Az + \delta y^3+\beta yz.
\end{equation}
Introduce $z_{new}= z +\gamma y^2$. Then, map (\ref{mapc}) takes the standard H\'enon form
\begin{equation}\label{mapc1}
\bar x=y,\qquad \bar y=z,\qquad \bar z = M_0+Bx+M_1y +Az -A\gamma y^2 + \gamma z^2 + \beta yz + (\delta-\beta\gamma) y^3.
\end{equation}
Let $x=y=z=x_0$ be a fixed point of map (\ref{mapc}), i.e.
$M_0=x_0(1-B-M_1-A)-(\delta-\beta\gamma)x_0^3-(1+\beta\gamma-A\gamma)x_0^2$.
By shifting the coordinate origin to this point, we write the map in form (\ref{Henon}):
$$\bar x=y,\quad \bar y=z,\quad \bar z = Bx+(M_1+(\beta-2A\gamma) x_0+3(\delta-\beta\gamma)x_0^2)y +(A+(\beta+2\gamma) x_0)z +
\alpha y^2+\beta yz+\gamma z^2+\dots.$$
where $\alpha=3(\delta-\beta\gamma)x_0-A\gamma$.
The fixed point has the multipliers $(-1,-1,+1)$ at
$B=1,\qquad A+x_0(\beta+2\gamma)=-1,\qquad M_1=1+(2A\gamma-\beta) x_0-3(\delta-\beta\gamma)x_0^2$.
Condition (\ref{gabeta}) reads as
$x_0(3\delta-2\beta\gamma+2\gamma^2)\left[(3\delta-2\beta\gamma+2\gamma^2)x_0-\beta+2\gamma\right]<0$.
For every given $\beta$, $\gamma$ and $\delta$ one can always find $x_0$ for which this is fulfilled, provided
$\delta\neq \frac{2}{3}\gamma(\gamma-\beta)$ and $\beta\neq 2\gamma$ is fulfilled. Therefore,
by Lemma \ref{lm11}, for every fixed $\beta$, $\gamma$, $\delta$ which satisfy these inequalities
there is an open region in the space of parameters $(M_0,M_1,B,A)$ which corresponds to the existence of the Lorenz-like attractor.
Numerically obtained portraits of Lorenz-like attractors in this map see in Figs.~\ref{fig2Lor},\ref{fig3Lor}.

Further, discrete Lorenz attractors were found numerically in other models,
including systems of nonholonomic mechanics \cite{GG13,GGK13} (see Section \ref{nhm}). Below we describe the simplest
scenarios leading to discrete Lorenz attractors and their ``figure-eight''
analogues.

\subsection{Discrete attractors of ``Lorenz-like'' and ``figure-eight'' shapes}\label{L8}

In this Section we describe a basic scenario of transition to chaos in three-dimensional maps, which is different from
the Shilnikov scenario of Section \ref{ShD}. Here, the first bifurcation that determines the future
shape of the strange attractor is the period-doubling bifurcation (for the spiral attractor the similar role
is played by the Andronov-Hopf bifurcation).

Consider a one parameter family $f_\mu$ of three-dimensional orientable diffeomorphisms and assume that
for the values of $\mu$ from some interval $I$ the diffeomorphism $f_\mu$
has an absorbing domain ${\cal D}_\mu$. Let $\mu_1$ and $\mu_2$ be certain values from $I$ such that $\mu_1<\mu_2$.
Assume that at $\mu\leq \mu_1$ the forward orbit of every point in ${\cal D_\mu}$ tend to a stable fixed point $O_\mu$.
Assume that at $\mu=\mu_1$ the point $O_\mu$ undergoes a soft (supercritical) {\em period doubling bifurcation}.
As a result, a stable period-2 orbit $P_{\mu}=(p_1,p_2)$, where $f_\mu(p_1)=p_2$ and $f_\mu(p_2)=p_1$, is born from $O_\mu$ at  $\mu>\mu_1$,
and the point $O_\mu$ becomes a saddle. We denote its multipliers as $\lambda_1,\lambda_2,\lambda_3$, where $\lambda_1<-1$,
and $|\lambda_3|<|\lambda_2|<1$. Note that we have here two cases: $\lambda_2<0, \lambda_3>0$ and $\lambda_2>0, \lambda_3<0$.

Let the saddle point $O_\mu$ have a transverse homoclinic orbit at $\mu>\mu_2$. Then, the maximal attractor
\begin{wrapfigure}{o}{0.43\textwidth}
\centerline{\epsfig{file=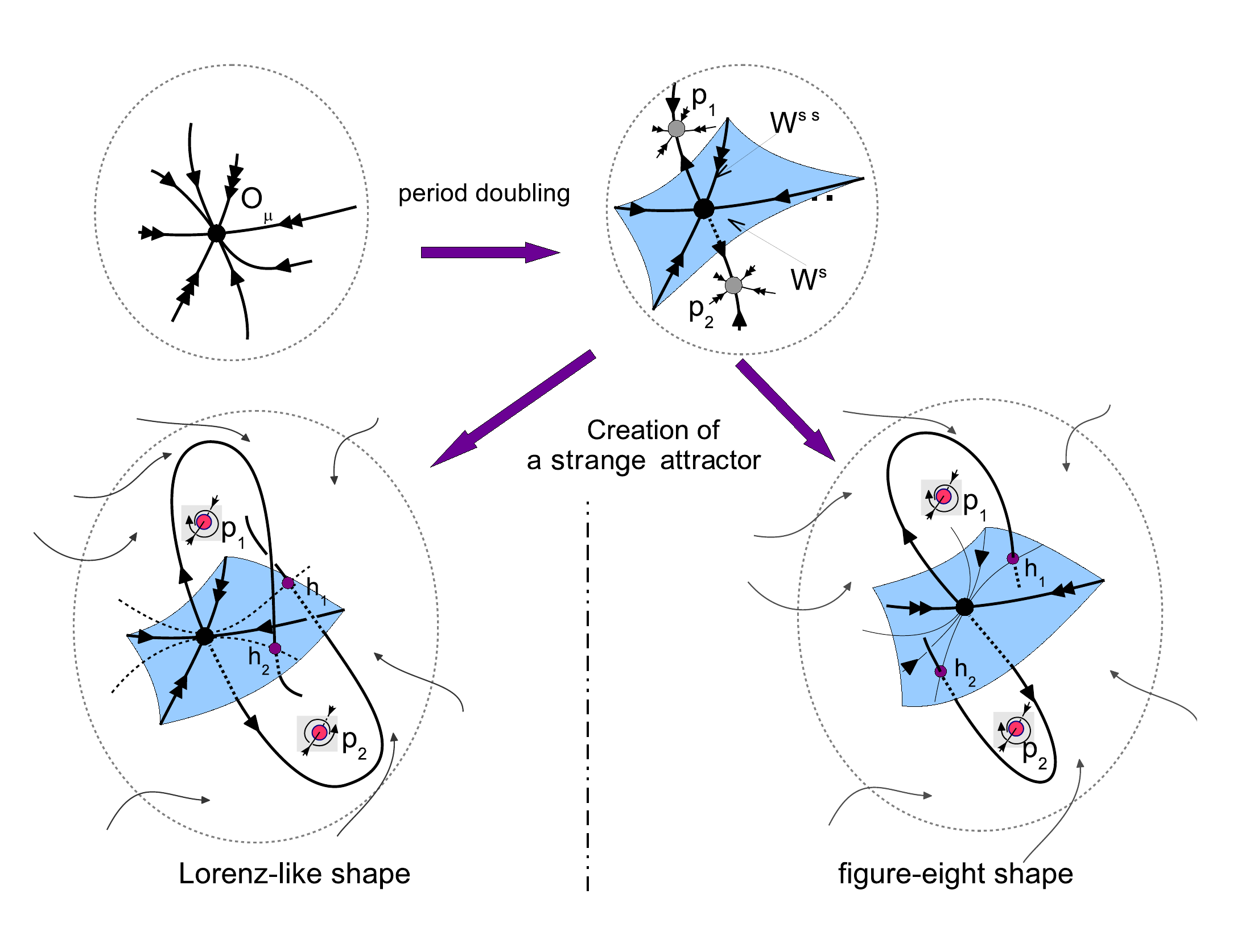, width=0.43\textwidth, height=68mm }}
\vspace*{-0.3cm}
\caption{{\footnotesize Two shapes of the attractor}} \label{scen2}
\end{wrapfigure}
$A_\mu$ in the absorbing domain ${\cal D}_\mu$ contains the non-trivial hyperbolic set associated with this homoclinic, i.e.
we may speak about a quasiattractor (if there are no obvious stable periodic orbits in it). There are two distinct possibilities for
the shape of this attractor, which mainly depends on the signs of the stable multipliers (see Fig.~\ref{scen2}). Recall that the unstable
multiplier $\lambda_1$ of $O_\mu$ is negative, therefore its unstable separatrices
(the two components into which $O_\mu$ divides
its unstable manifold) are mapped to each other by $f_\mu$. Thus, the homoclinic orbit belongs to both of these
separatrices, i.e.
they both intersect the stable manifold $W^s(O_\mu)$. Typically, the homoclinic intersection does not belong to the strong stable manifold
$W^{ss}(O_\mu)$ which is tangent to the eigenspace that coresponds to the non-leading multiplier $\lambda_3$.
The manifold $W^{ss}(O_\mu)$ divides $W^s(O_\mu)$ into two parts. These parts are invariant with respect to $f_\mu$ if the leading stable
multiplier $\lambda_2$ is positive, and they are taken to each other by $f_\mu$ if $\lambda_2<0$. Thus, we have two cases:\\~\\
if $\lambda_2>0$, then both unstable separatrices of $O_\mu$ can intersect $W^s(O_\mu)$ on one side from $W^{ss}(O_\mu)$ --
we say that the attractor $A_\mu$ has a {\em Lorenz-like} shape in this case;\\~\\
if $|\lambda_2<0$, then the unstable separatrices must intersect $W^s(O_\mu)$ on both sides from $W^{ss}(O_\mu)$ --
we say that  $A_\mu$ has a {\em figure-eight} shape.\\

\begin{wrapfigure}{o}{0.46\textwidth}
\centerline{\epsfig{file=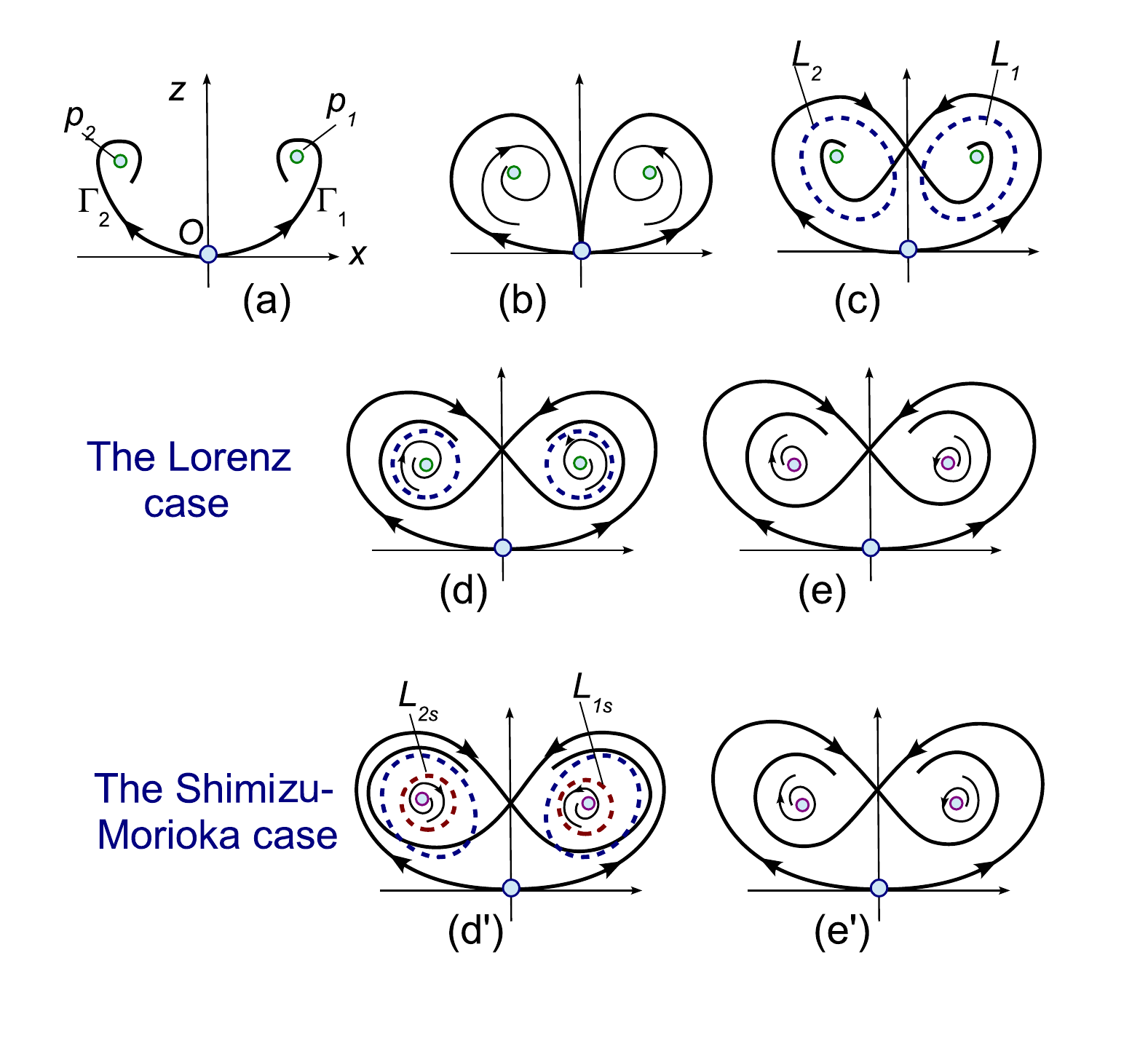, width=0.48\textwidth, height=8.4cm}}
\vspace*{-0.5cm}
\caption{{\footnotesize A sketch of the Poincare map for a
small time-periodic perturbation of the Lorenz or Shimizu-Morioka models. (a)
After a period-doubling, the fixed point becomes a saddle, and a stable period-2 orbit $(p_1,p_2)$ is created; (b)
creation of a thin homoclinic butterfly structure; (c) a saddle period-2 closed curve $(L_1,L_2)$ detaches from the butterfly;
(d) the Lorenz-like attractor gets separated from the stable orbit $(p_1,p_2)$; (e) the orbit $(p_1,p_2)$ becomes unstable
(subcritical Andronov-Hopf bifurcation);
(d') the Lorenz-like attractor gets separated from the stable period-2 closed curve born from
$(p_1,p_2)$ at a supercritical Andronov-Hopf bifurcation; (e') the stable and saddle curves of period 2 get destroyed.}}
\label{Fig-LSMcase}
\end{wrapfigure}

Note that the loss of stability of the period-2 orbit $P_{\mu}$ does not need to be correlated
with the emergence of the homoclinics to the fixed point $O_\mu$. In fact, many different variants are
possible. For example, a cascade of period-doublings may continue and the transition to chaos may precede the
creation of homoclinics to $O_\mu$ (this is typical if the Jacobian is small and
the map is close to one-dimensional). For non-small Jacobians, the period-2 orbit may undergo an Andronov-Hopf bifurcation.
A supercritical bifurcation leads to the birth of a stable invariant curve with two closed connected components, see
Fig.~\ref{Fig-LSMcase}. At the further growth of $\mu$ this curve can get destroyed and transformed to a ``homoclinic structure''
involving $O_\mu$. If the Andronov-Hopf bifurcation is subcritical, then a saddle closed period-2 curve merges with $P_{\mu}$.
The homoclinics to $O_\mu$ can already exist in this case, so the period-2 curve is formed at the fringes of the homoclinic structure.
Depending on the situation, the two-dimensional stable manifold of this curve may serve as a barrier that separates the attraction domains
of the period-2 orbit $P_{\mu}$ and the Lorenz-like attractor that contains $O_\mu$, or there may be no such attractor
separate from $P_{\mu}$ (this happens when the closure of the unstable manifold of $O_\mu$ contains $P_mu$; then we should
speak about the Lorenz-like attractor only after $P_\mu$ loses stability). Similar scenarios
(where the period-2 orbit is replaced by a pair of symmetric equilibria and the period-2 closed curve
is replaced by a pair of symmetric limit cycles) are known to lead to the onset of the Lorenz attractor in the Lorenz model
(with subcritical Andronov-Hopf) \cite{ABS77,Sh80,BSS} and the Shimizu-Morioka model (with supercritical Andronov-Hopf) \cite{ShA86,ShA93}; see Fig.~\ref{Fig-LSMcase}. Therefore, the above described transition\\~\\
Fixed~Point $\Rightarrow$ Period-2~Orbit $\Rightarrow$ Stable/Saddle~Period-2~Curve $\Rightarrow$ discrete Lorenz-like~Attractor\\~\\
should be typical for Poincare maps for small periodic perturbations of these systems and, hence, for arbitrary maps near
the moment of bifurcations of periodic orbits with multipliers $(-1,-1,+1)$ (see e.g. Lemma \ref{lm11} in the previous Section).

As we mentioned in the previous Section, the Lorenz-like and figure-eight attractors of three-dimensional maps
can be true strange attractors, provided they satisfy the pseudohyperbolicity property. The pseudohyperbolicity
should be verified at each point of the absorbing domain. In particular, at the fixed point $O_\mu$ this property requires
that the saddle value $\sigma = |\lambda_1\lambda_2|$ is greater than 1 (as at least some two-dimensional areas must be expanded
by the linearization of the map at the fixed point). This is a necessary condition that is easiest to check; if it is not satisfied,
then stable periodic orbits will be born from homoclinic tangencies to $O_\mu$, i.e. the attractor $A_\mu$ will be a quasiattractor.
However, this condition does not need to be sufficient. One may also numerically estimate Lyapunov exponents at some randomly chosen
orbit in $A_\mu$. The pseudohyperbolicity requires the positivity of the sum of the two largest Lyapunov exponents. Again, this is not
yet a sufficient condition for the true chaoticity of the attractor, even if the
orbit appears to be dense in $A_\mu$: one also needs to verify that the angle between the invariant subspace corresponding
to the two largest Lyapunov exponents and the subspace that corresponds to the rest of Lyapunov exponents stays bounded away from zero.
The most robust approach to the prove of hyperbolicity is, of course, based on Lemma~\ref{lmph}. In the examples below we do not go into
such depths in verification of the pseudohyperbolicity. However, we do the simple checks of the saddle value and Lyapunov exponents. Also,
the similarity of the shape of our discrete Lorenz-like attractors with the classical Lorenz attractor is often very high, therefore we are
quite certain these attractors are pseudohyperbolic, hence truly chaotic.

\section{Numerical experiments with H\'enon-like maps}
\label{numerical}

\begin{figure}[h]
\begin{minipage}[h]{0.3\linewidth}
\center{\includegraphics[width=1\linewidth, height=0.8\linewidth]{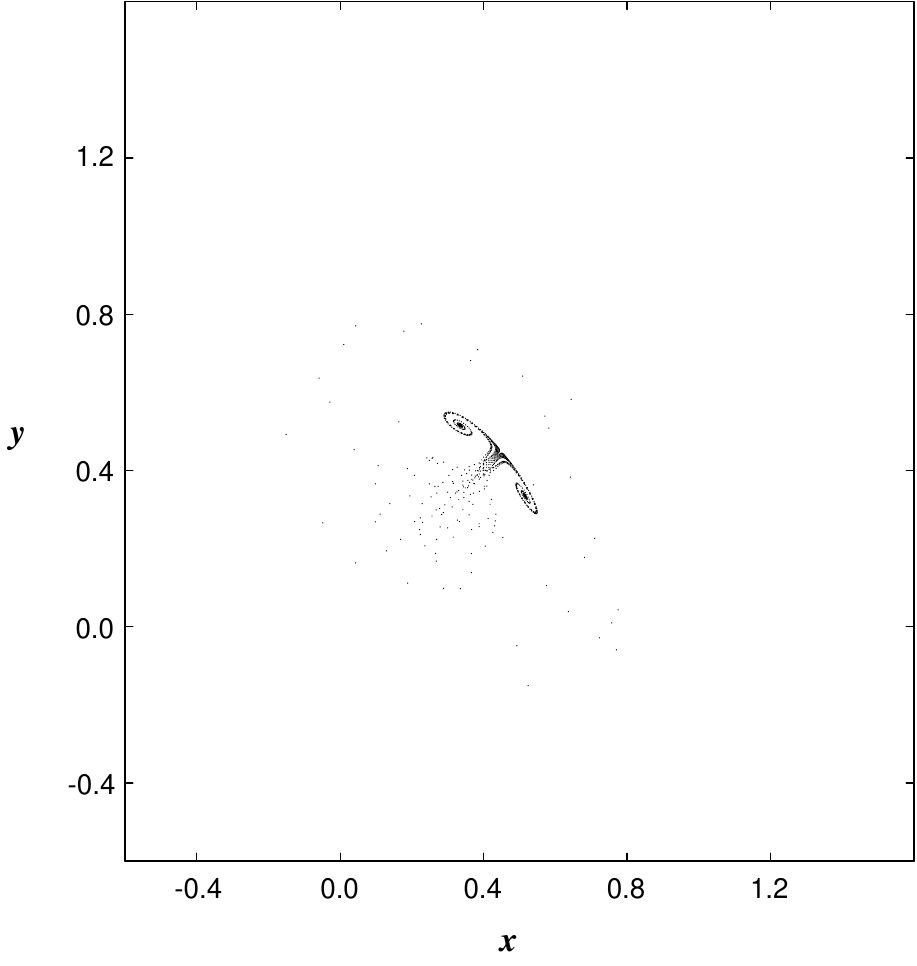} \\ (a) $M_1=-0.045$}
\end{minipage}
\hfill
\begin{minipage}[h]{0.3\linewidth}
\center{\includegraphics[width=1\linewidth, height=0.8\linewidth]{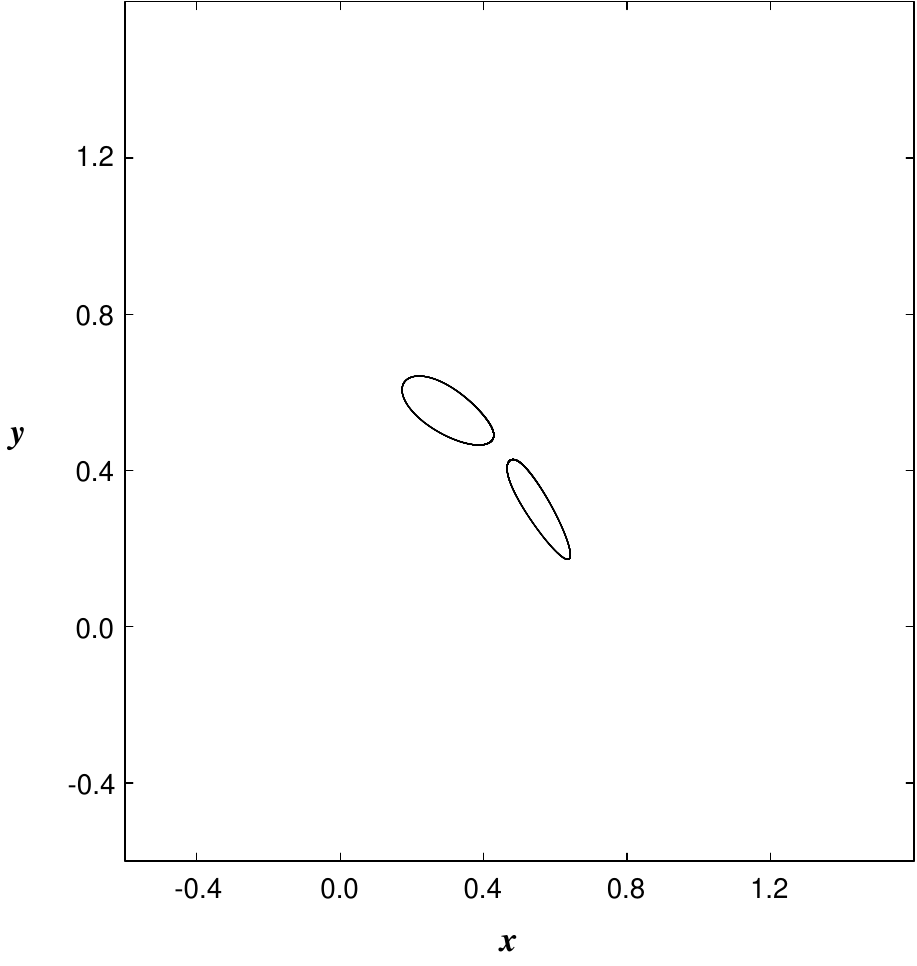} \\ (b) $M_1=-0.0342$}
\end{minipage}
\hfill
\begin{minipage}[h]{0.3\linewidth}
\center{\includegraphics[width=1\linewidth, height=0.8\linewidth]{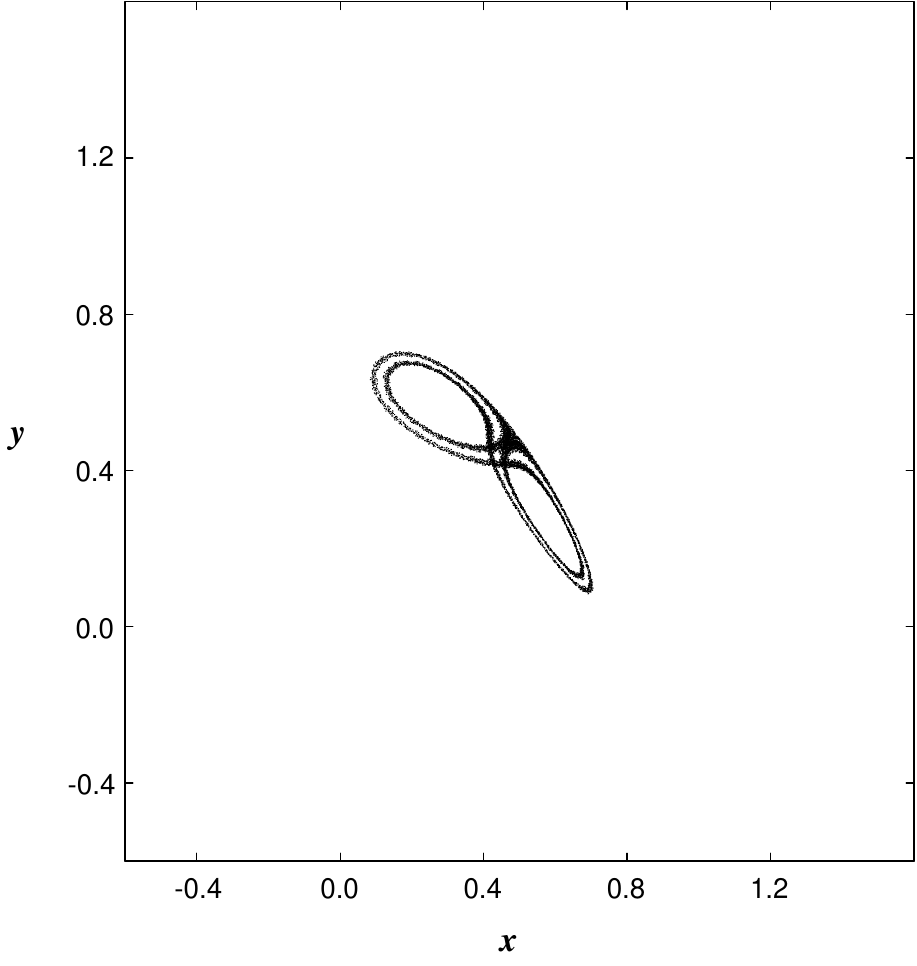} \\ (c) $M_1=-0.028$}
\end{minipage}
\vfill
\begin{minipage}[h]{0.3\linewidth}
\center{\includegraphics[width=1\linewidth, height=0.8\linewidth]{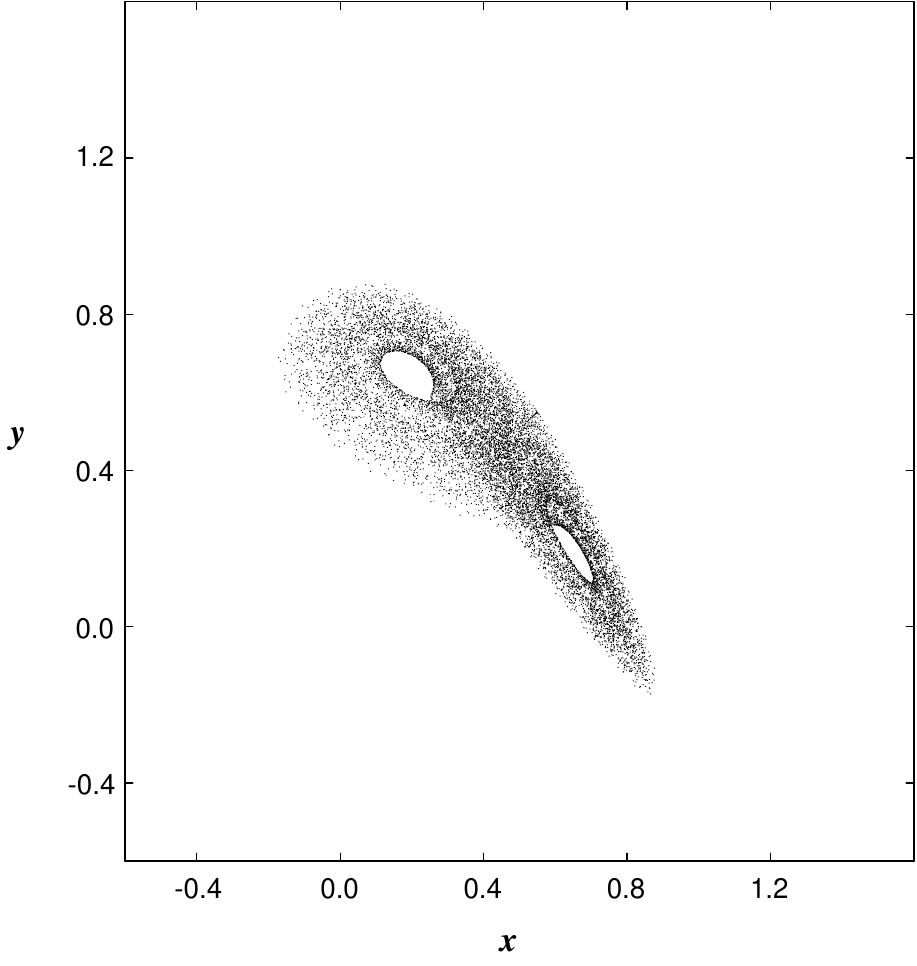} \\ (d) $M_1=0$}
\end{minipage}
\hfill
\begin{minipage}[h]{0.3\linewidth}
\center{\includegraphics[width=1\linewidth, height=0.8\linewidth]{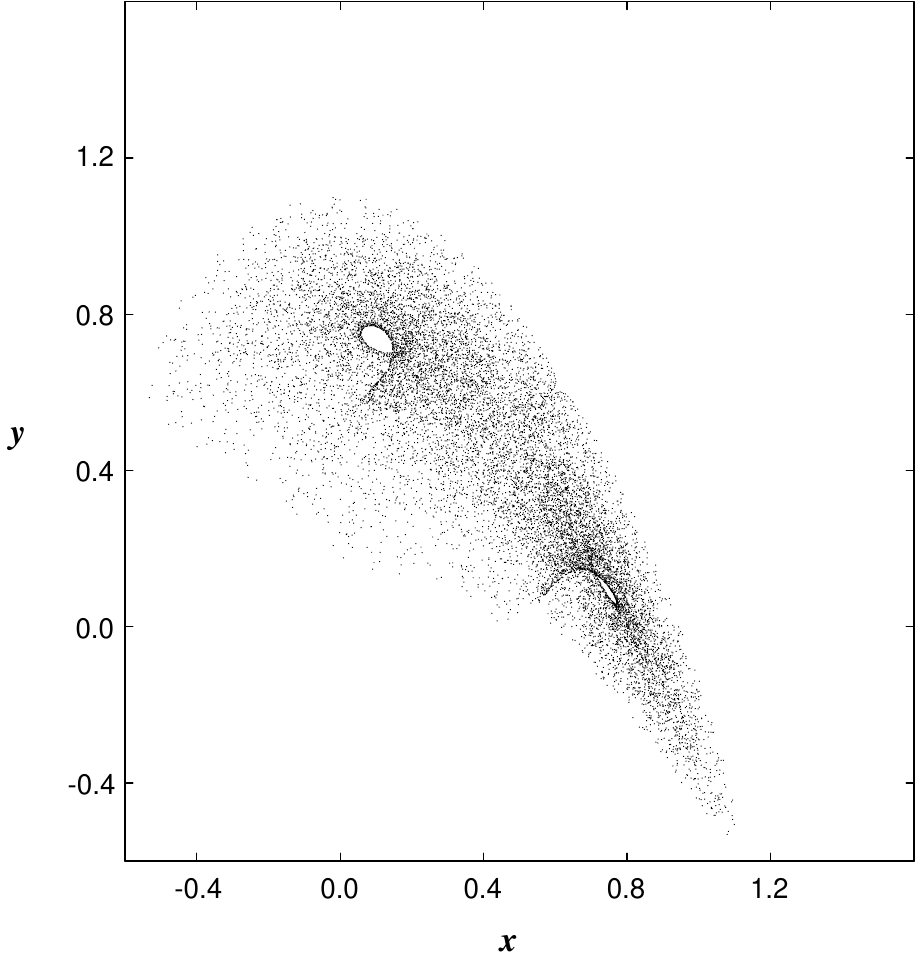} \\ (e) $M_1=0.05$}
\end{minipage}
\hfill
\begin{minipage}[h]{0.3\linewidth}
\center{\includegraphics[width=1\linewidth, height=0.8\linewidth]{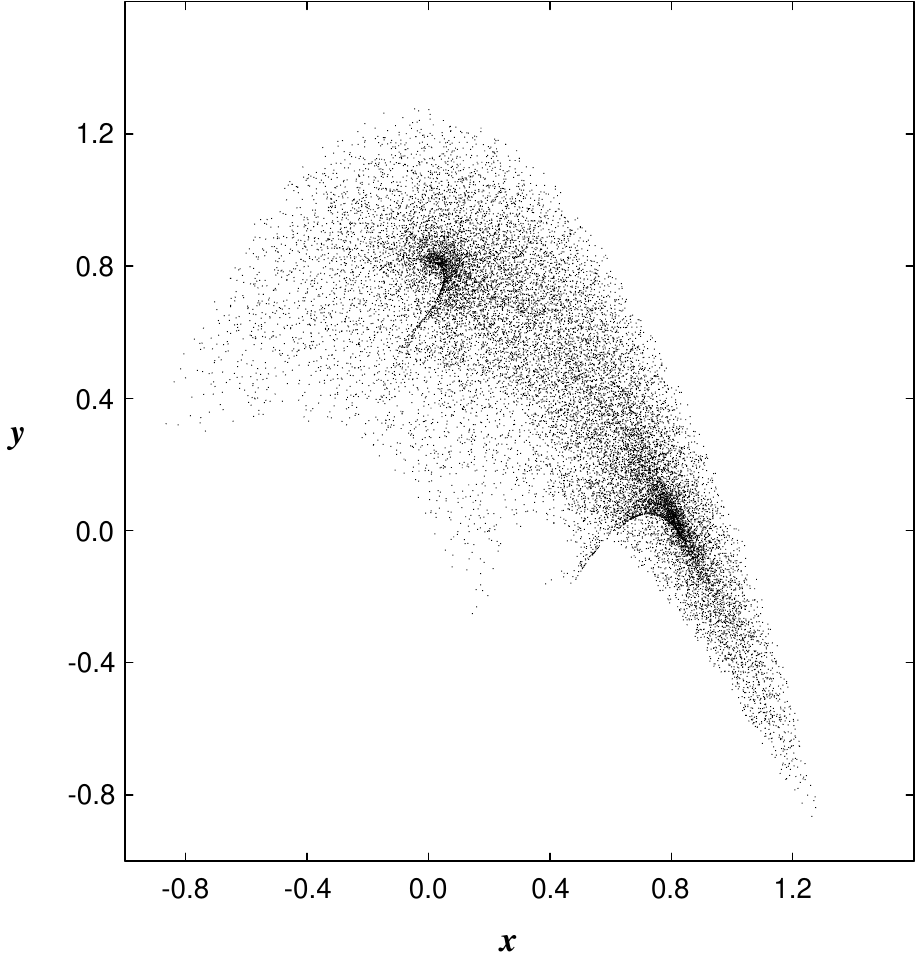} \\ (f) $M_1=0.1$}
\end{minipage}
\caption{
{\footnotesize Evolution of attractors in the 3D H\'enon map (\ref{HM1}) with $B=0.7, M_2=0.85$, as $M_1$ varies:
(a) period 2 point;
(b)  closed curve of period 2; (c)--(f)  strange attractors. All the figures are in the same scale.}}
\label{fig1lor}
\end{figure}

We now present numerics that illustrates the theory above. We consider, first, map (\ref{HM1}) with the Jacobian $B=0.7$
(i.e. the dissipation is weak enough). In Fig.~\ref{fig1lor} the corresponding phase portraits
(numerical iterations of a single initial condition) are shown for fixed $M_2=0.85$ and varying $M_1$.
The transition to a Lorenz-like attractor proceeds in the following steps:
the orbit of period 2 (Fig.~\ref{fig1lor}a) gives rise
to a stable two-component closed invariant curve (Fig.~\ref{fig1lor}b), which then gets destroyed by a ``collision''
with a saddle two-component invariant curve that was formed from a homoclinic butterfly to the saddle fixed point,
and a Lorenz-like attractor is formed
(Figs.~\ref{fig1lor}c,d). This scenario is similar to what one should observe in a periodically perturbed Morioka-Shimizu
system.

As $M_1$ grows, the attractor grows in size and evolves into a strange
attractor ``without holes'' (Fig.~\ref{fig1lor}f). It
reminds the attractor of Lorenz model after the absorbtion of the saddle-foci. This
effect is related to a creation of a heterodimensional cycle
where the one-dimensional unstable manifold of the saddle fixed point intersects with
the one-dimensional stable manifold of the saddle-focus orbit of period two; this cycle is
analogous to the ``Bykov contour'' of the Lorenz model \cite{Byk0,Byk1,Byk2}.
This bifurcation (e.g. in the Lorenz model \cite{ABS80,PY,BS89,BSS} and in the Shimizu-Morioka
model \cite{ShA86,ShA93,SST93}) leads to creation of stable periodic orbits. By analogy, we expect
that the same is true for the ``no hole'' attractor in the Henon map (\ref{HM1}), i.e. it loses the pseudohyperbolic structure
and becomes a quasiattractor.

\begin{figure}[h]
\begin{minipage}[h]{0.3\linewidth}
\center{\includegraphics[height=0.7\linewidth]{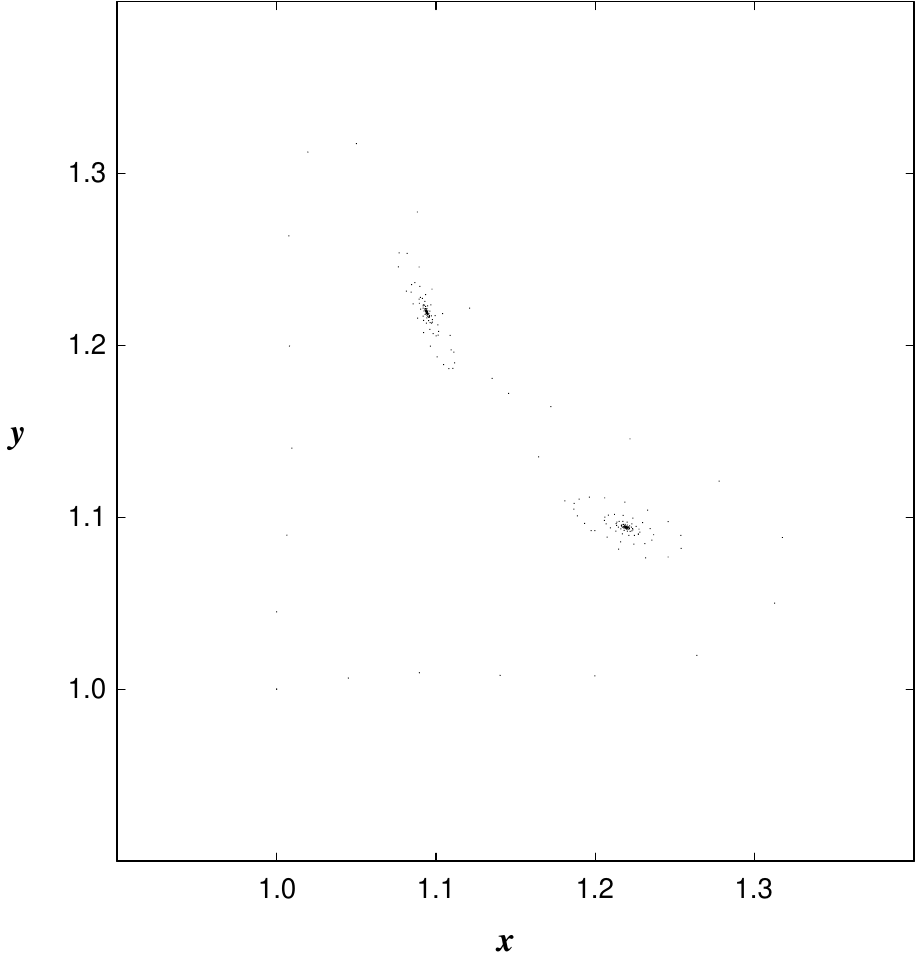} \\ (a)}
\end{minipage}
\hfill
\begin{minipage}[h]{0.3\linewidth}
\center{\includegraphics[height=0.7\linewidth]{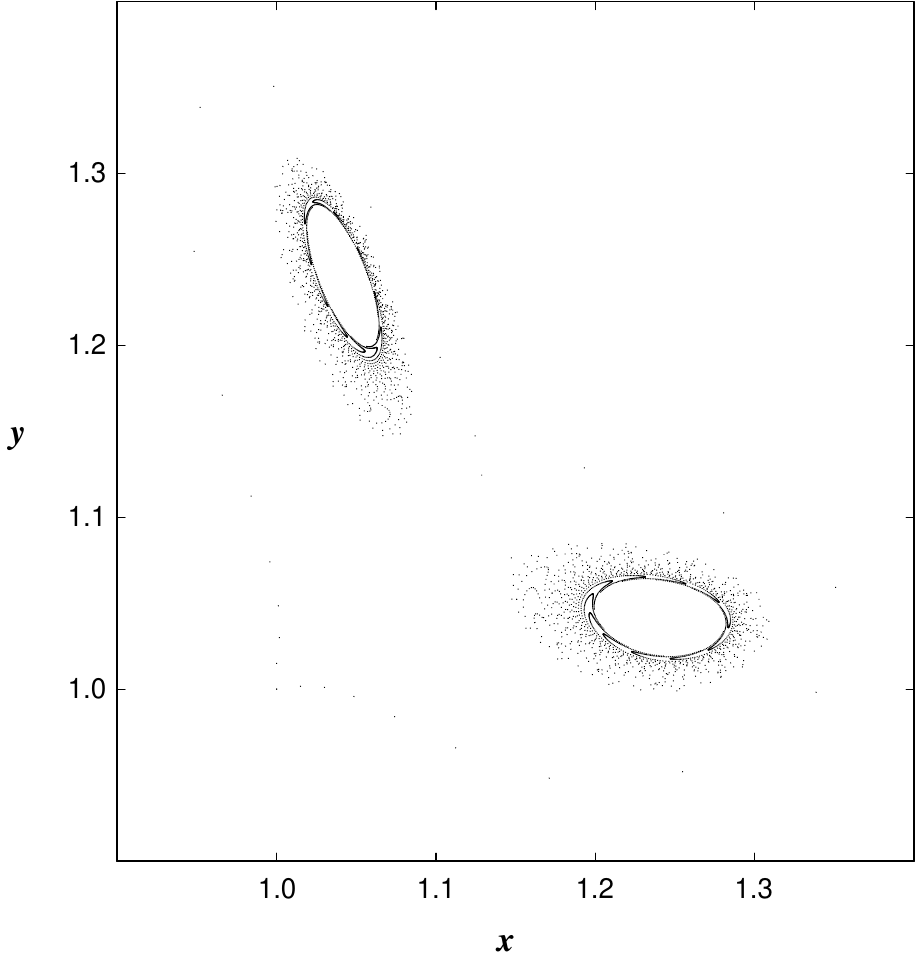} \\ (b)}
\end{minipage}
\hfill
\begin{minipage}[h]{0.3\linewidth}
\center{\includegraphics[height=0.7\linewidth]{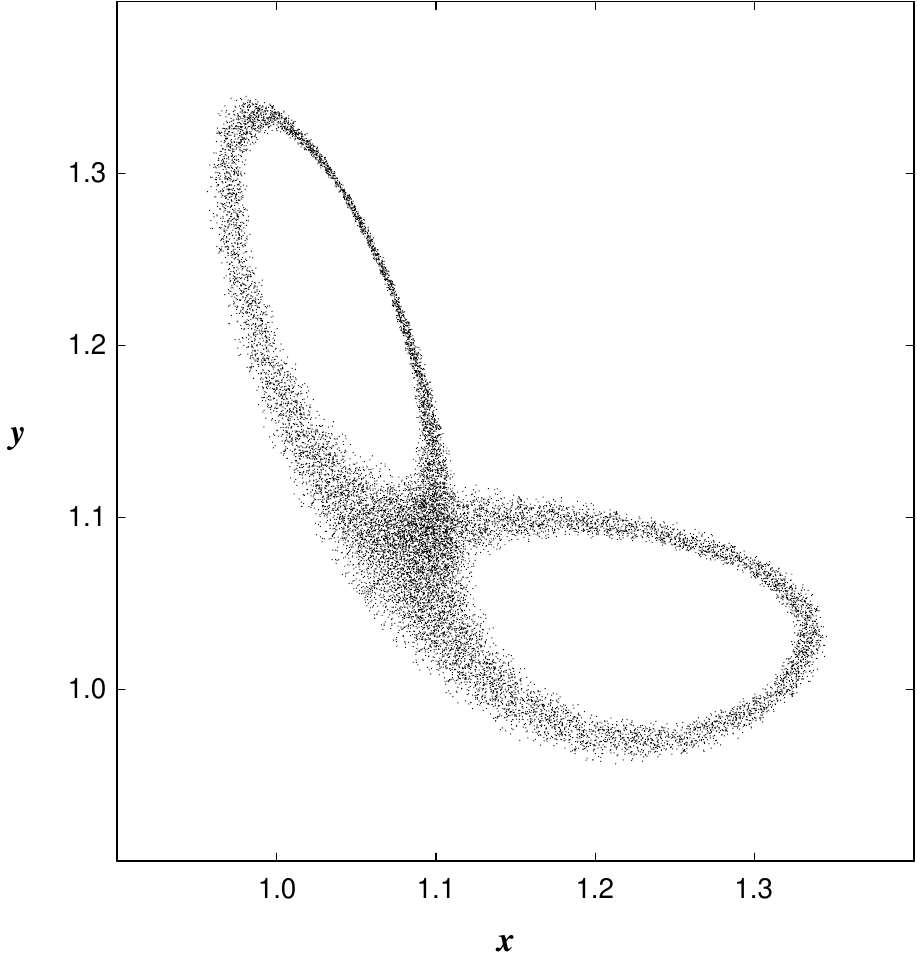} \\ (c)}
\end{minipage}
\vfill
\begin{minipage}[h]{0.3\linewidth}
\center{\includegraphics[height=0.7\linewidth]{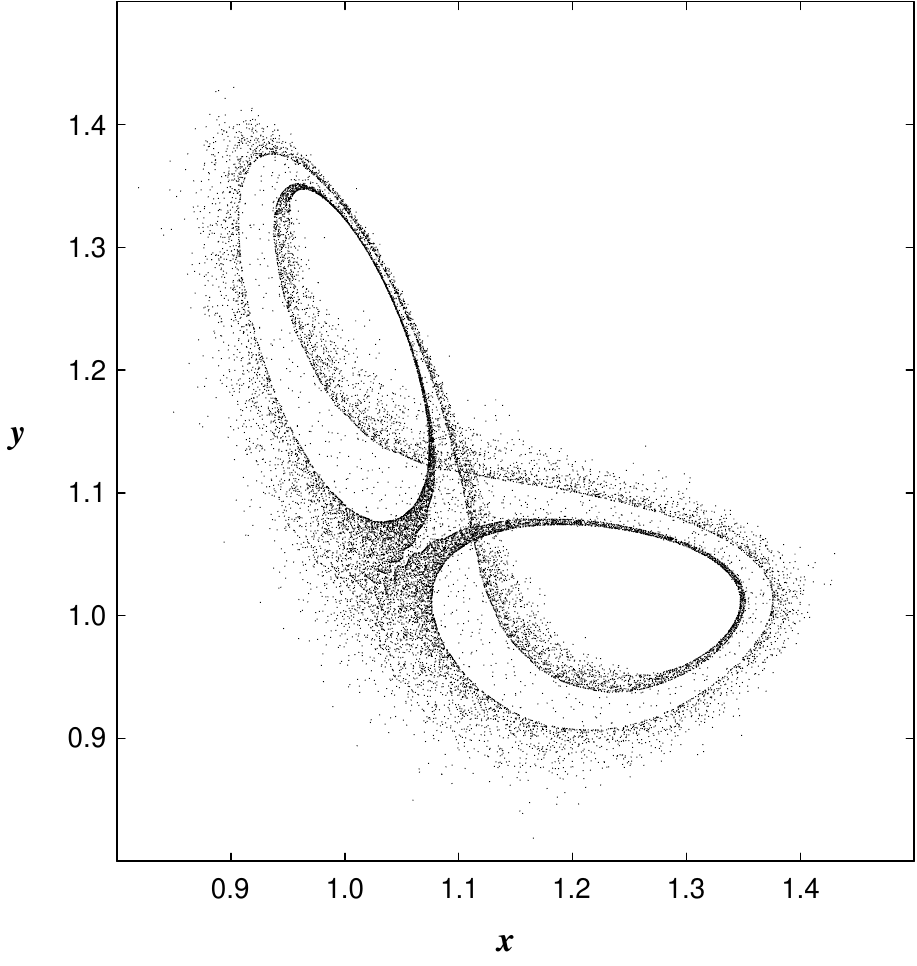} \\ (d)}
\end{minipage}
\hfill
\begin{minipage}[h]{0.3\linewidth}
\center{\includegraphics[height=0.7\linewidth]{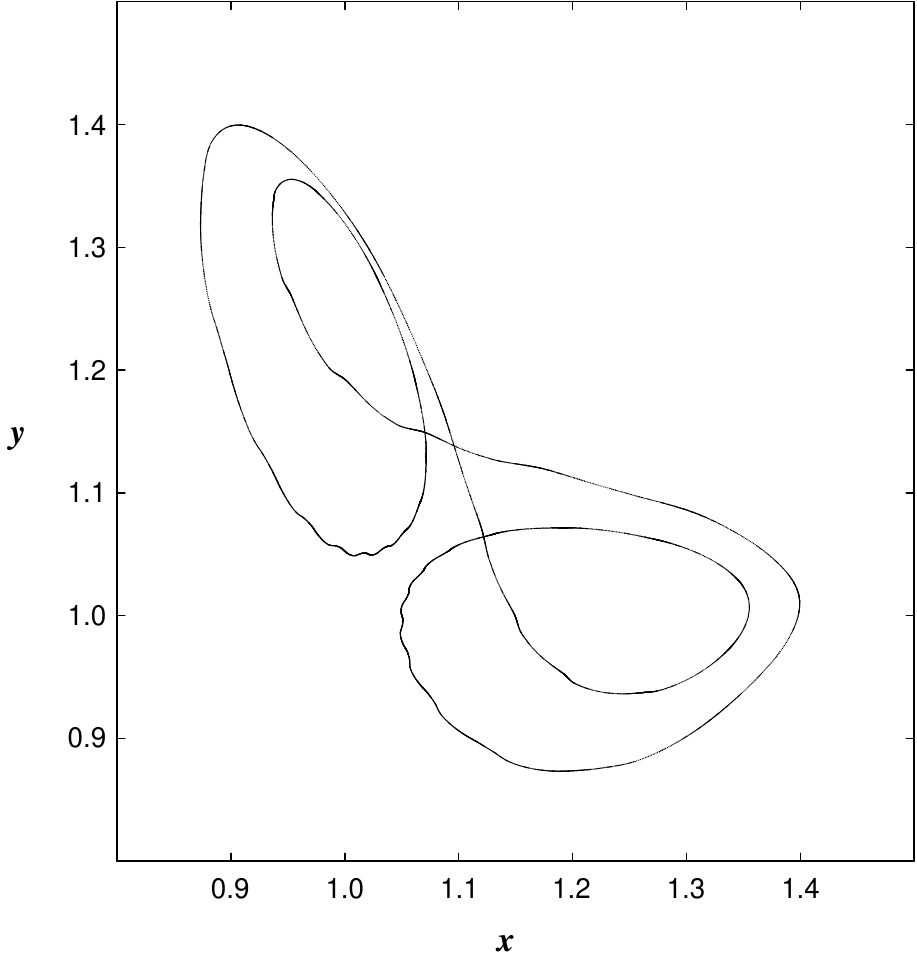} \\ (e)}
\end{minipage}
\hfill
\begin{minipage}[h]{0.3\linewidth}
\center{\includegraphics[height=0.7\linewidth]{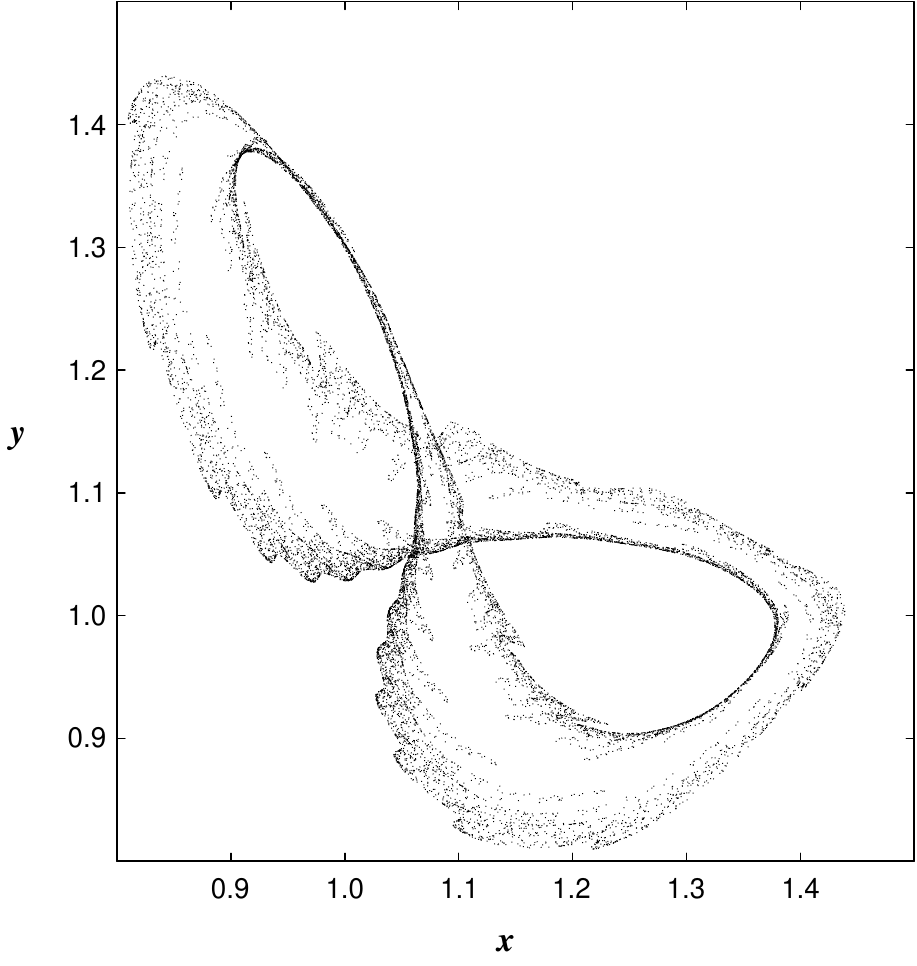} \\ (f)}
\end{minipage}
\caption{
{\footnotesize Creation and destruction of the Lorenz-like attractor in map (\ref{mapc})
at $\beta = -3, \; \delta = -3, \; \gamma = 1, \; M_0 = -2, \; B = 0.7, \; A = 0.1$ as
$M_1$ varies from $4.365$ down to $4.235$.
(a) cycle of period 2 (after period doubling); (b) closed curve of period 2
(after Andronov-Hopf bifurcation of the cycle), one can
also see the location of the saddle curve of period 2 (cf. Figs.~\ref{Fig-LSMcase}(d')--(e'));
(c) Lorenz-like attractor; (d) Lorenz-like attractor with a lacuna; (e)--(f) bifurcation
stages after the destruction of the attractor (stable invariant curve and torus-chaos).}
\vspace*{-5mm}} \label{fig2Lor}
\end{figure}

Next, we show results of numerical simulations of map (\ref{mapc}). We start with the case
$\beta = -3, \; \delta = -3, \; \gamma = 1, \; M_0 = -2$.
At $M_1 = 4, \; B = 1, \; A = 0$ this map has a fixed point at $x = y = z = 1$ with the
multipliers of $(-1, -1, +1)$. It is easy to check that condition (\ref{gabeta})
is satisfied, so we can expect the Lorenz-like attractor for parameter values close to these.
We choose $B = 0.7, \; A = 0.1$, and vary $M_1$ from $4.365$ down to $4.235$.
The results are shown in Fig.~\ref{fig2Lor}. The attractor forms in the way similar to the previous case.
The destruction of the attractor (Figs.~\ref{fig2Lor}d--f) proceeds via formation of a lacuna where
a stable closed invariant curve emerges (Fig.~\ref{fig2Lor}e) which next breaks-down and forms
a strange attractor (Figs.~\ref{fig2Lor}f) of the ``wriggled'' shape
typical for the  ``torus-chaos'' quasiattractor \cite{AfrShRing2,CY,ArCh,AfrSh}.
Note that the route of the destruction of the Lorenz-like attractor via formation of a lacuna, which we see in these figures,
reminds one of the scenarios of the disappearance of the Lorenz attractor that was described in \citet{ABS83}
and was also discovered in the Shimizu-Morioka model \cite{ShA86,ShA93,SST93}.

\begin{figure}[h]
\begin{minipage}[h]{0.3\linewidth}
\center{\includegraphics[height=0.6\linewidth]{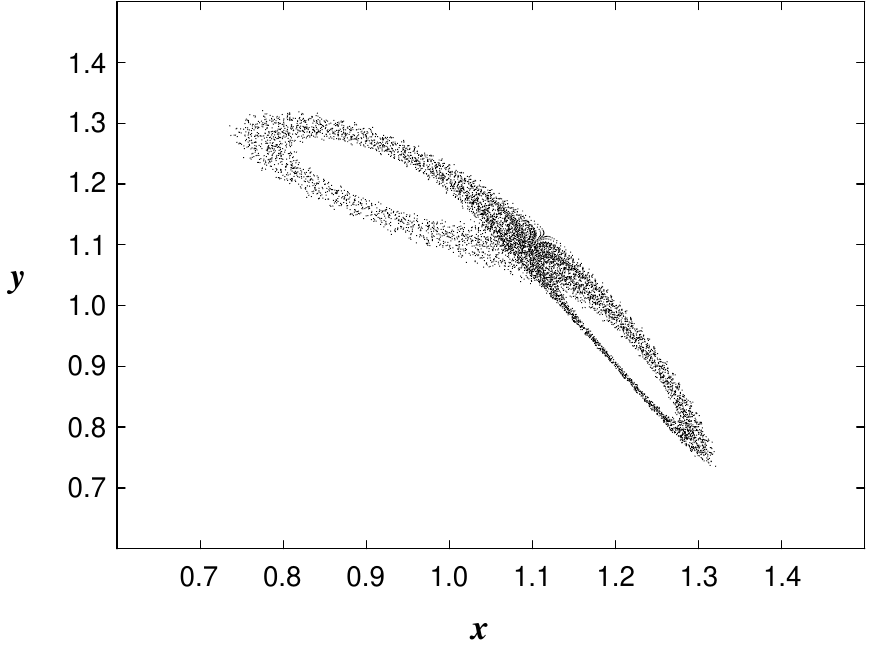} \\ (a)}
\end{minipage}
\hfill
\begin{minipage}[h]{0.3\linewidth}
\center{\includegraphics[height=0.6\linewidth]{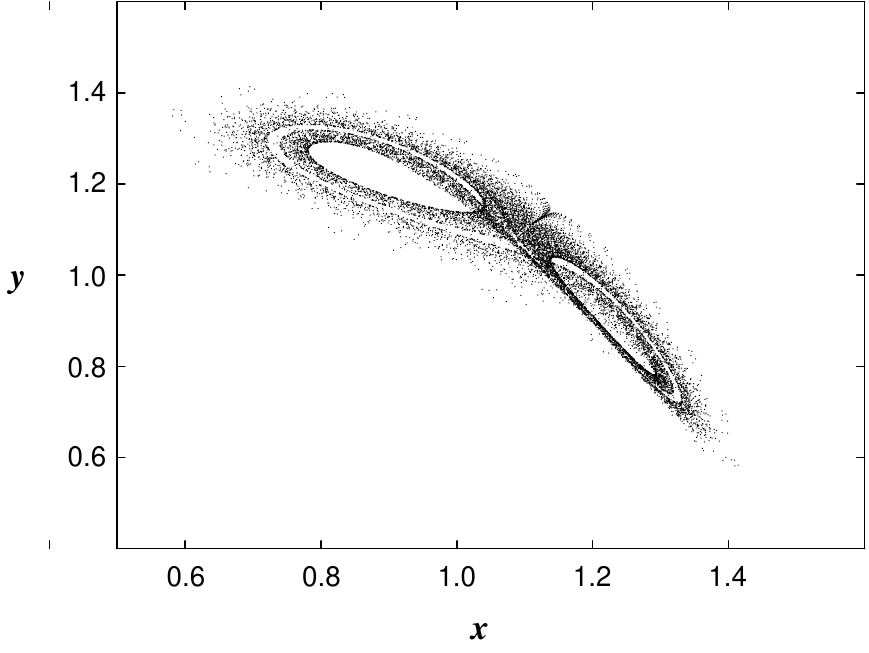} \\ (b)}
\end{minipage}
\hfill
\begin{minipage}[h]{0.3\linewidth}
\center{\includegraphics[height=0.6\linewidth]{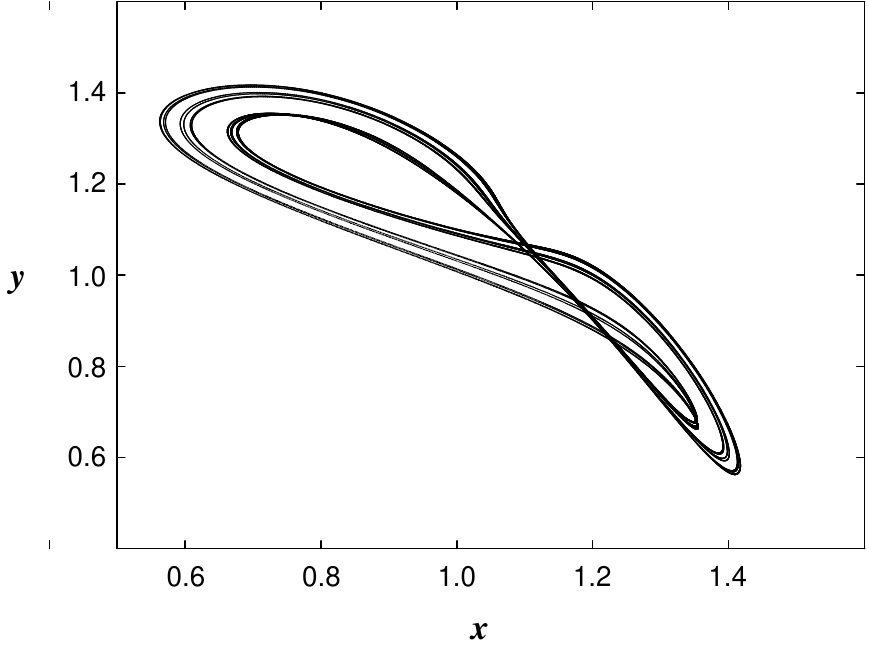} \\ (c)}
\end{minipage}
\vfill
\begin{minipage}[h]{0.3\linewidth}
\center{\includegraphics[height=0.6\linewidth]{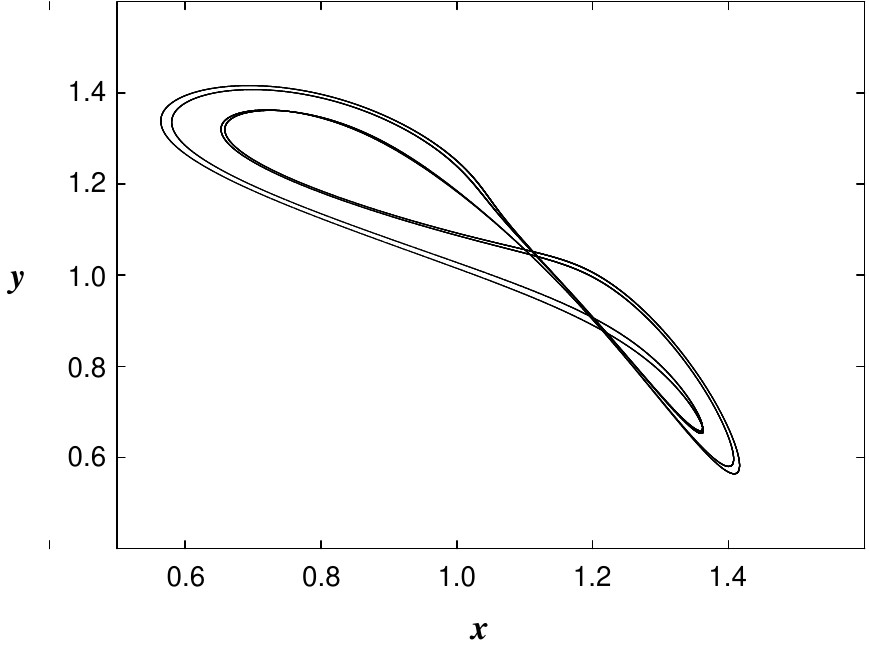} \\ (d)}
\end{minipage}
\hfill
\begin{minipage}[h]{0.3\linewidth}
\center{\includegraphics[height=0.6\linewidth]{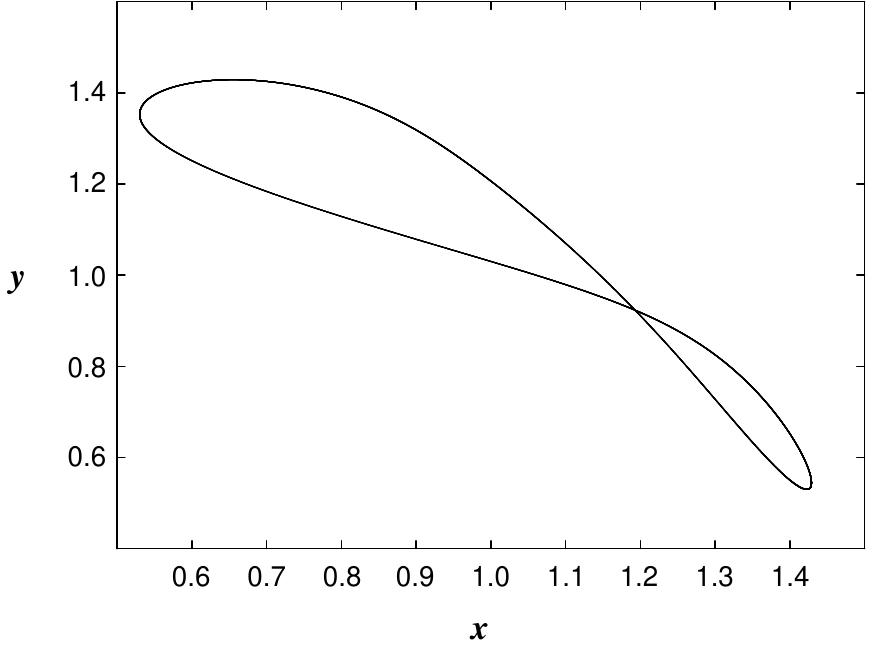} \\ (e)}
\end{minipage}
\hfill
\begin{minipage}[h]{0.3\linewidth}
\center{\includegraphics[height=0.6\linewidth]{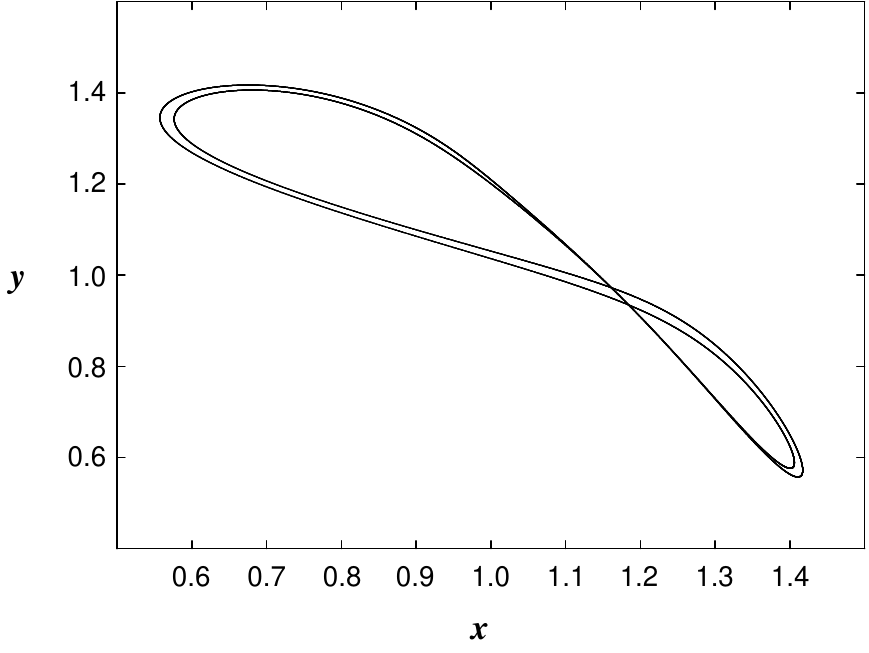} \\ (f)}
\end{minipage}
\caption{
{\footnotesize Plots of attractors of map (\ref{mapc}) for
$\beta = 2, \; \delta = 1/3, \; \gamma = 0, \; M_0 = 3.67, \; B = 0.7, \; A = -3.1$ as
$M_1$ varies from $-2.555$ to $-2.505$. (a)--(b) as in Fig.~\ref{fig2Lor}; (c) strange quasiattractor; (d)--(f)
stable closed invariant curves.}}
\label{fig3Lor}
\end{figure}

Another case corresponds to $\beta = 2, \; \delta = 1/3, \; \gamma = 0$. A fixed point at $x=y=z=1$ has multipliers
$(-1, -1, +1)$ at $M_0 = -2, \; M_1 = -2, \; B = 1, \; A = -3$. Again, condition (\ref{gabeta})
is satisfied. Numerics was performed at  $M_0 = 3.67, \; B = 0.7, \; A = -3.1$, with
$M_1$ varying from $-2.555$ to $-2.501$. The results are shown in Fig.~\ref{fig3Lor}.
The first stages (Figs.~\ref{fig3Lor}a--b) on the route to the Lorenz-like attractor (Fig.~\ref{fig3Lor}b)
are the same here as in the previous cases.
The destruction of the attractor proceeds via formation of a lacuna where a stable invariant curve emerges (Fig.~\ref{fig3Lor}d), which then gives place to
a strange quasiattractor (Fig.~\ref{fig3Lor}c). These stages are also similar to what is seen in Fig.~\ref{fig3Lor}. However, the quasiattractor
has now a different structure, and unravels via a backward cascade of torus-doubling bifurcations (Figs.~\ref{fig3Lor}d--f.
The last invariant curve
disappears by colliding with a saddle invariant curve at a saddle-node bifurcation at $M_1\sim -2.501$. See
\citet{Ch1,Los3,Bro1,Bro2,AnN} for the theory of the saddle-node and doubling bifurcations for invariant curves.
Numerous examples of such bifurcations in three-dimensional diffeomorphisms can be found in \citet{Vitolo}.

It is curious that, despite of our numerics being performed for parameter values sufficiently far from the bifurcation of a fixed point with the multipliers
(-1,-1,1), the bifurcations scenarios are quite similar to those one should have in its normal form, i.e. in the Shimizu-Morioka model with a small periodic
forcing. Namely, the Lorenz attractor in the Shimizu-Morioka model transforms into a strange quasiattractor in a variety of ways \cite{ShA86,ShA93,SST93},
depending on the choice of a path in the parameter plane, and this variety does
include an absorption of saddle-foci like in Fig.~\ref{fig1lor}, or formation of a lacuna with a consequent boundary crisis of the Lorenz attractor
and emergence of a quasiattractor which may be accompanied by period-doubling cascades or not, like in Figs.~\ref{fig2Lor},\ref{fig3Lor}. The differences
between the destruction of the Lorenz-like attractor in our maps and the destruction of the Lorenz attractor in the model flow are still visible
(mainly due to the effects of loss of smoothness and breakdown of invariant curves), but they do not seem to play a major role.

\section{Discrete Lorenz-like and figure-eight attractors in models of non\-ho\-lo\-no\-mic mechanics.}

\label{nhm}

In this Section we show how the strange attractors described in Section \ref{L8} emerge in the dynamics
of rigid bodies moving on a plane without slipping. This means that we consider a nonholonomic model of motion for which the
contact point of the body has zero velocity, i.e. $ \boldsymbol v
+ \boldsymbol{\omega} \times \boldsymbol r = 0 $, where $\boldsymbol r$ is the vector from the center
of mass $C$ to the contact point, $\boldsymbol v$ is the velocity of $C$
and $\boldsymbol{\omega}$ is the angular velocity. By introducing a coordinate frame rigidly rotating with the body
the equations of motion can be written in the form \cite{BM03}:
\begin{equation}
\begin{array}{l}\displaystyle
\dot{\boldsymbol M} =
\boldsymbol M\times\boldsymbol{\omega} +
m\dot{\boldsymbol r}\times (\boldsymbol{\omega}\times\boldsymbol r) + m{\mbox g}\boldsymbol r\times\boldsymbol{\gamma}, \qquad
\dot{\boldsymbol{\gamma}}  = \boldsymbol{\gamma} \times\boldsymbol{\omega},\\ \displaystyle
\boldsymbol M= [{\bf J} + m(\boldsymbol r, \boldsymbol r) {\mbox{{\bf  I}}} -m \boldsymbol r \cdot \boldsymbol r^T]\cdot \boldsymbol {\omega},\qquad \boldsymbol \gamma =- \nabla F(\boldsymbol r)\left/\left\|\nabla F(\boldsymbol r)\right\|\right.,
\end{array}
\label{eq:2}
\end{equation}
where $\boldsymbol M$ is the angular momentum with respect to the contact point,
$\boldsymbol{\gamma}$ is the unit vector normal to the surface of the body at this point (all the vectors are
taken in the rotating frame), $F$ is the function which defines the shape of the body such that $F(\boldsymbol r)=0$
is the equation of its surface, $m {\mbox g}$ is the value of the gravity force, ${\bf J}$ is the
inertia tensor, ${\mbox{{\bf I}}}$ is the 3x3 identity matrix and $(\cdot)$ means the matrix product. We choose
the axes of the rotating coordinate frame to coincide with the principal axes of inertia, i.e. ${\bf J} = diag(J_1, J_2, J_3)$.

Equation (\ref{eq:2}) admits two conserved quantities, the energy integral $\displaystyle
E = \frac{1}{2}({\boldsymbol M}, \boldsymbol{\omega}) - m{\mbox g}({\boldsymbol r},\boldsymbol{\gamma} )$ and
$(\boldsymbol{\gamma} , \boldsymbol{\gamma} ) = 1$. By restricting system (\ref{eq:2}) to a constant energy level,
we obtain a four-dimensional system of differential equations. By choosing an appropriate cross-section, we obtain
a three-dimensional Poincare map which depends on the value of energy $E$. Below we study two different examples
of how attractors of this map evolve as $E$ changes.

\subsection{Discrete Lorenz attractors in a Celtic stone dynamics.}
\vspace*{-.5cm}
\begin{figure}[htb]
\center{\includegraphics[width=0.7\textwidth]{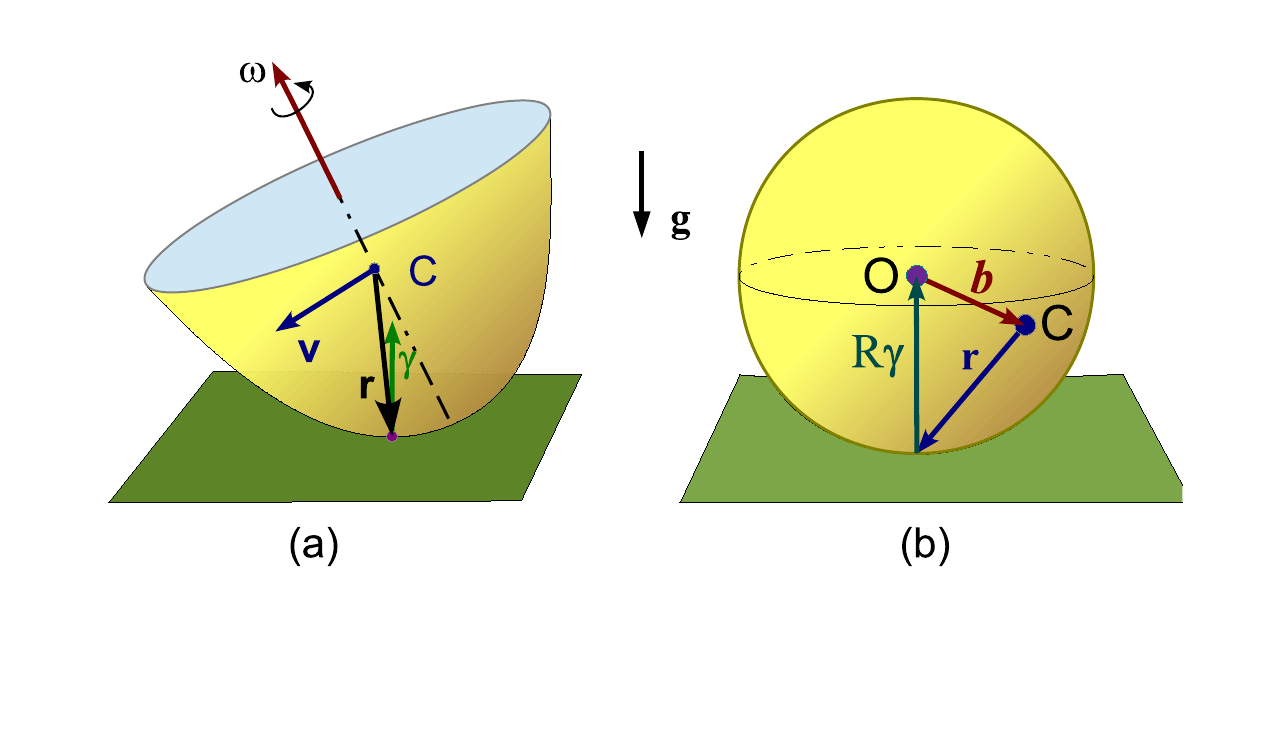}}
\vspace*{-1.5cm}
\caption{{\footnotesize (a) Celtic stone, (b) Unbalanced ball}}
\end{figure}

\begin{figure}[h]
\begin{minipage}[h]{0.3\linewidth}
\center{\includegraphics[height=0.8\linewidth]{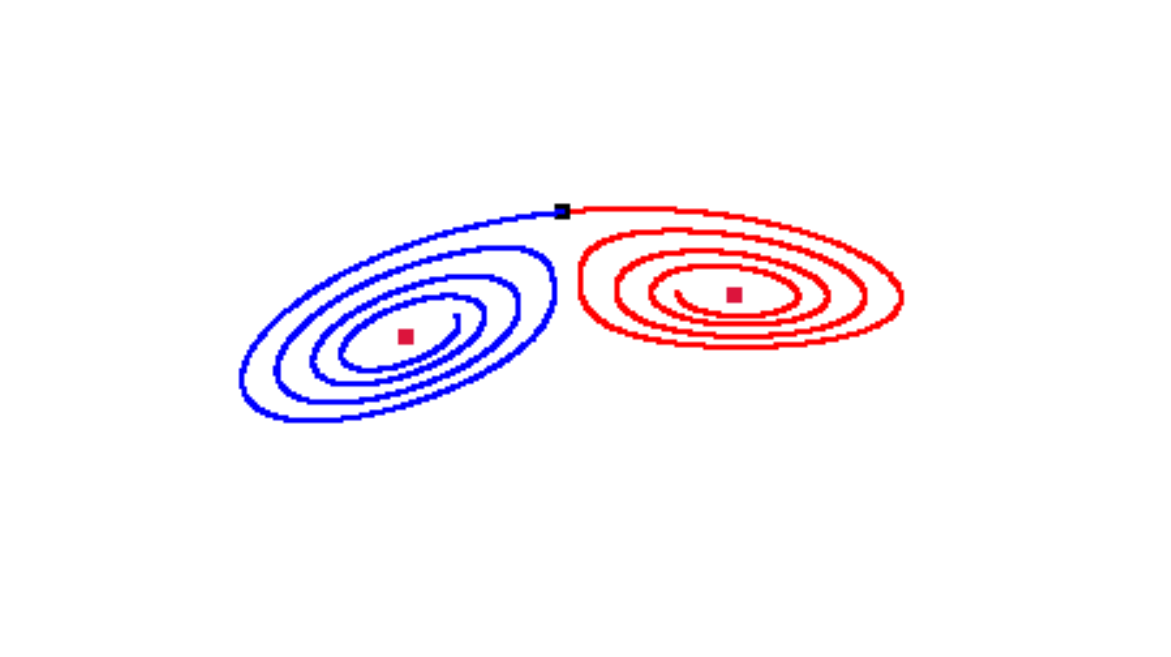} \\ \vspace*{-1.4cm} (a) E=748.4}
\end{minipage}
\hfill
\begin{minipage}[h]{0.3\linewidth}
\center{\includegraphics[height=0.8\linewidth]{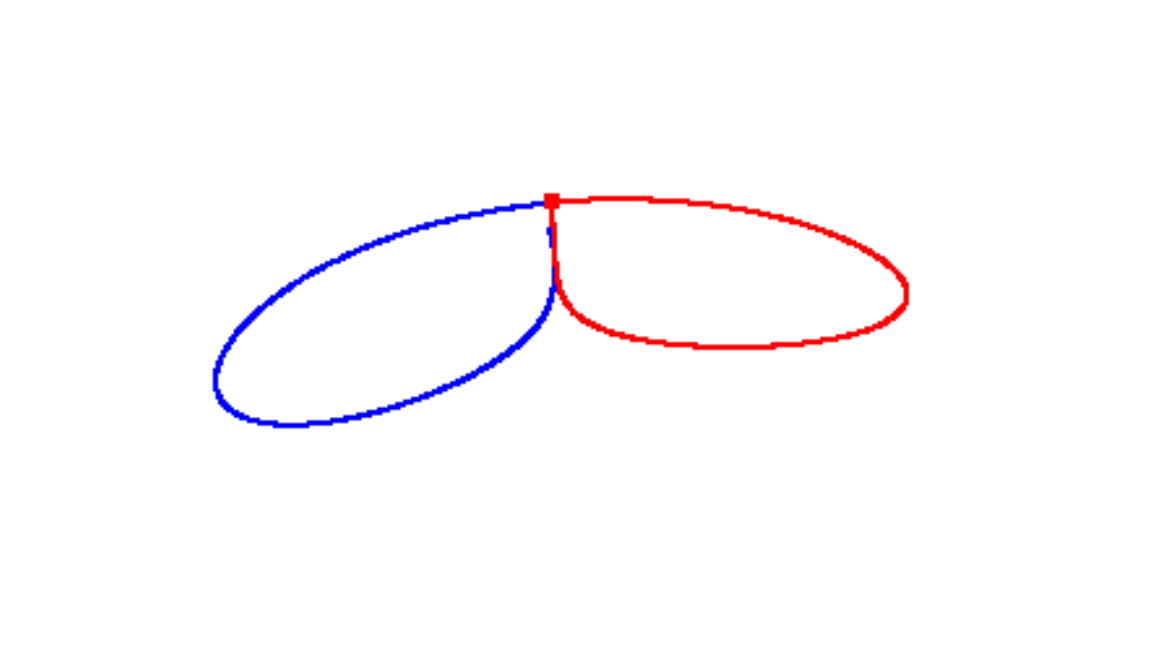} \\ \vspace*{-1.4cm} (b) E=748.4395}
\end{minipage}
\hfill
\begin{minipage}[h]{0.3\linewidth}
\center{\includegraphics[height=0.8\linewidth]{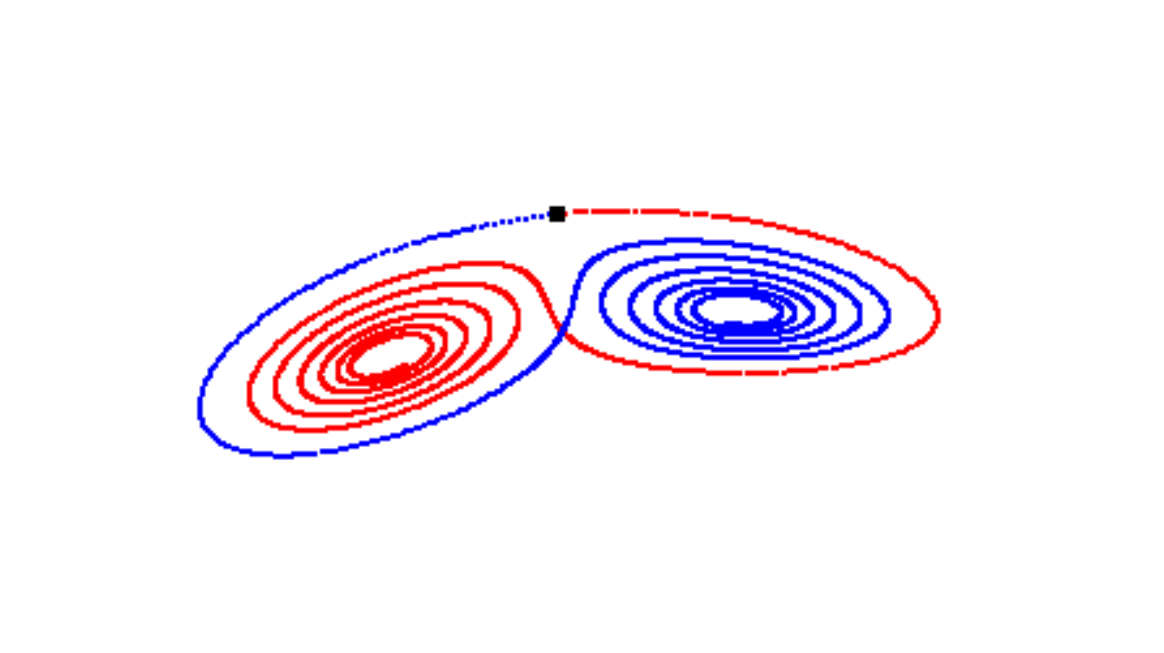} \\ \vspace*{-1.4cm} (c) E=748.5}
\end{minipage}
\vfill
\begin{minipage}[h]{0.3\linewidth}
\center{\includegraphics[height=0.8\linewidth]{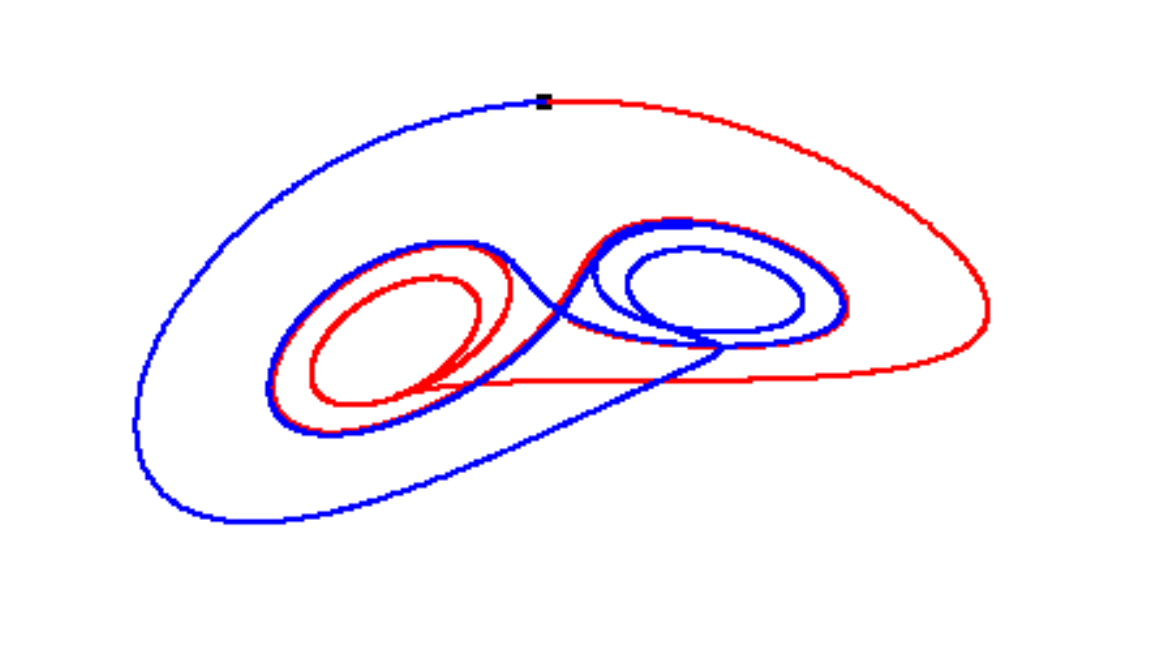} \\ \vspace*{-1cm} (d) E = 750.0}
\end{minipage}
\hfill
\begin{minipage}[h]{0.3\linewidth}
\center{\includegraphics[height=0.8\linewidth]{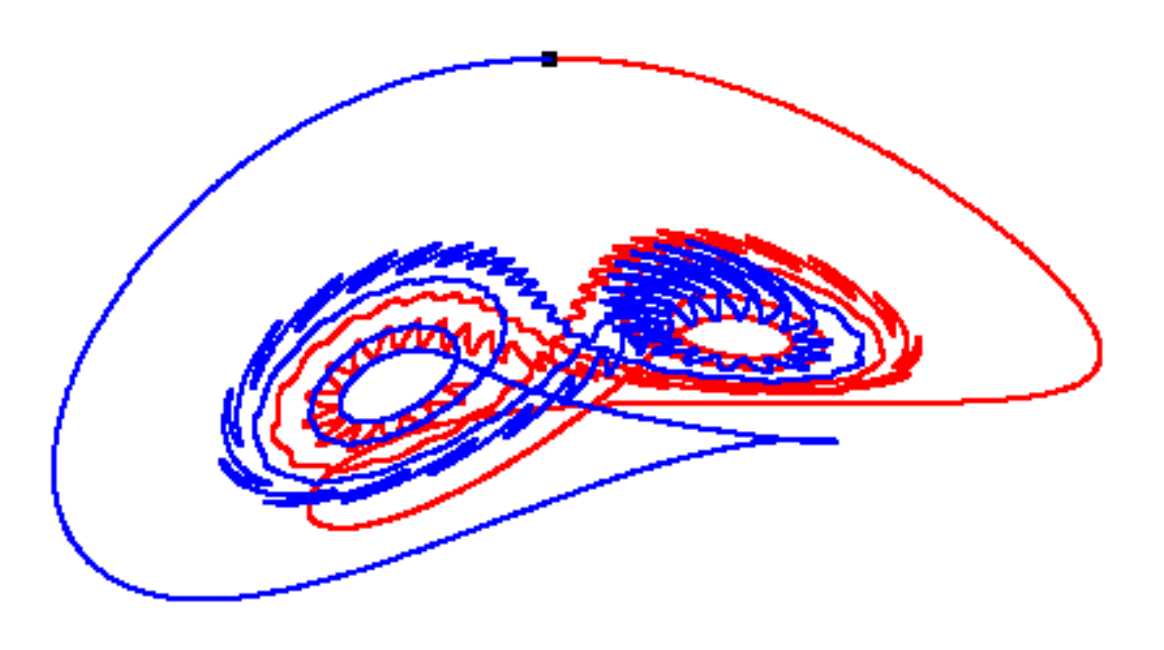} \\ \vspace*{-1cm} (e) E=752.0}
\end{minipage}
\hfill
\begin{minipage}[h]{0.3\linewidth}
\center{\includegraphics[height=0.8\linewidth]{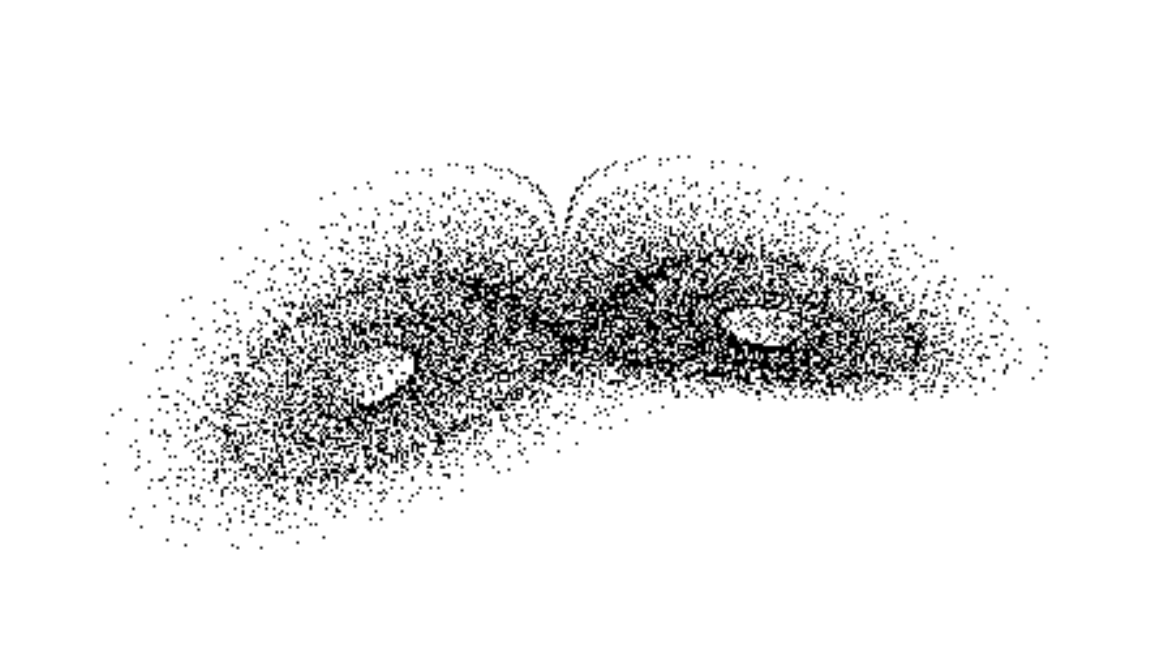} \\ \vspace*{-1cm} (f) E=752.0}
\end{minipage}
\vfill
\begin{minipage}[h]{0.3\linewidth}
\center{\includegraphics[height=0.8\linewidth]{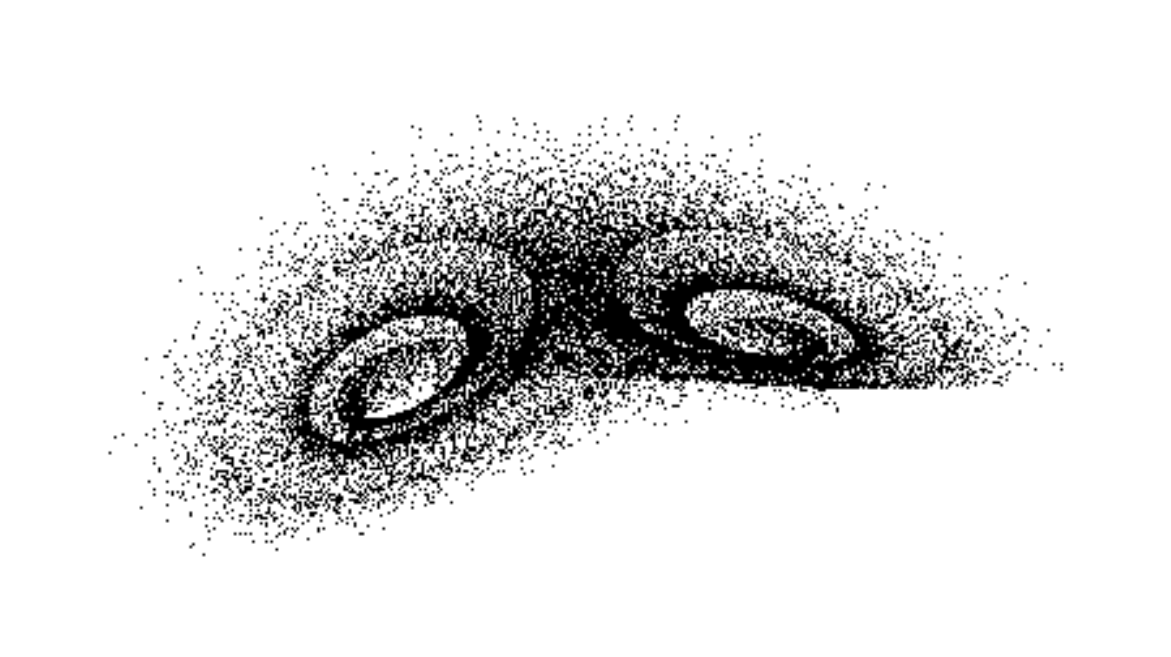} \\ \vspace*{-1cm} (g) E = 754.0}
\end{minipage}
\hfill
\begin{minipage}[h]{0.3\linewidth}
\center{\includegraphics[height=0.8\linewidth]{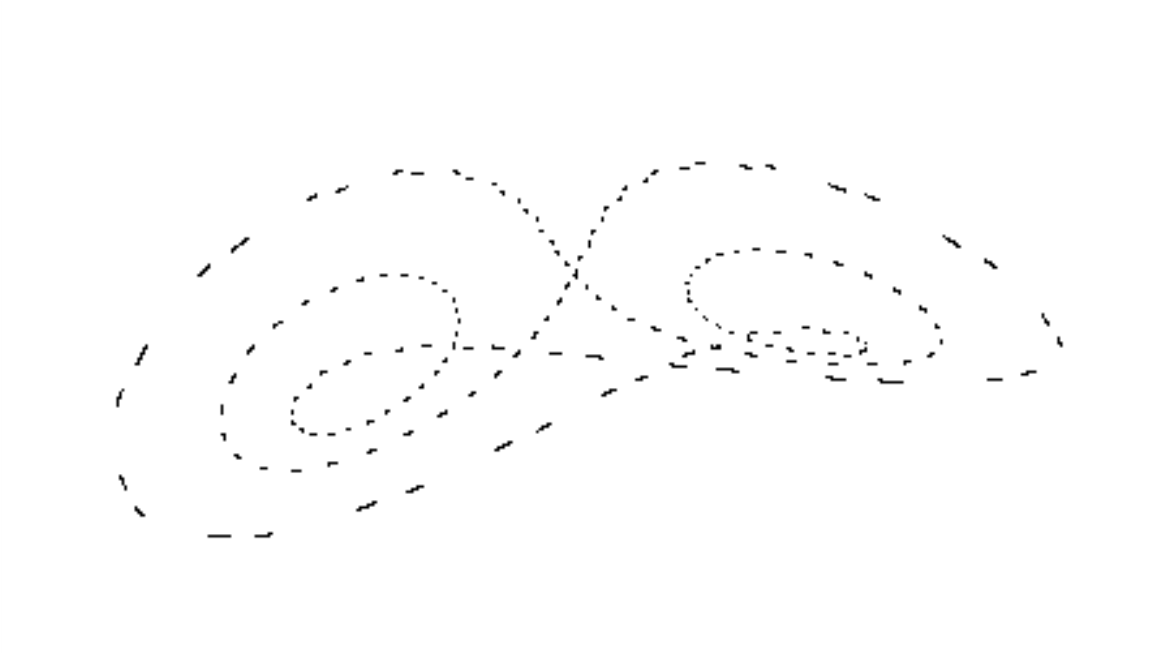} \\ \vspace*{-1cm} (h) E=755.0}
\end{minipage}
\hfill
\begin{minipage}[h]{0.3\linewidth}
\center{\includegraphics[height=0.8\linewidth]{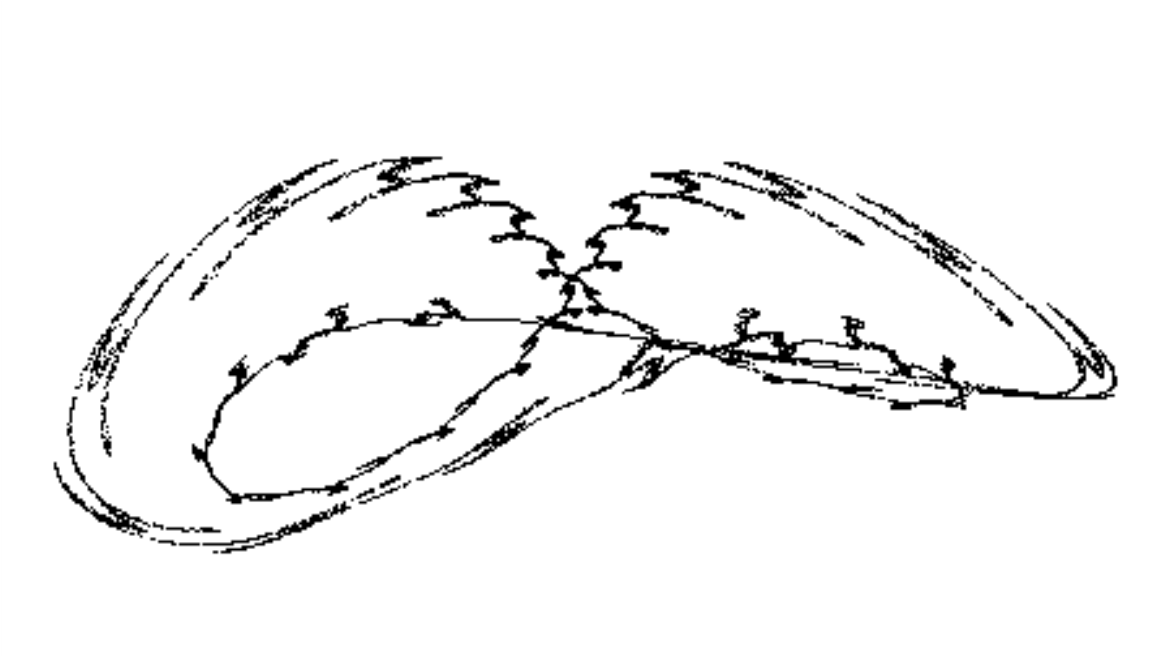} \\ \vspace*{-1cm} (i) E=765.0}
\end{minipage}
\caption{{\footnotesize The main stages of evolution of the Lorenz-like attractor
in the Poincare map for the Celtic stone. Figs. e--i show iterations of a single point,
and Figs. a--e show the unstable manifold of the saddle fixed point $O$.}}
\label{fig3-2a}
\end{figure}

A Celtic stone is a rigid body such that one of its
inertia axes is vertical and the two others are rotated by an angle $\delta$ with respect
to the horisontal geometrical axes. Namely, we consider a Celtic stone in the shape of elliptic paraboloid, i.e.
$\displaystyle F({\boldsymbol r^*}) = \frac{1}{2}\left(\frac{r_1^{*2}}{a_1} + \frac{r_2^{*2}}{a_2}\right) - (r^*_3 + h) = 0$,
where $a_1$ and $a_2$ are the principal radii of curvature at the paraboloid vertex $(0,0,-h)$ and
$\boldsymbol r^*=\left(\begin{array}{ccc}
\cos\delta & \sin\delta & 0 \\
-\sin\delta & \cos\delta &  0 \\
0 & 0 & 1\end{array}\right) \boldsymbol r$. We take
$J_1 =2, J_2 = 6, J_3 =7, m=1, {\rm g} = 100, a_1 = 9, a_2=4,  h=1, \delta = 0.485$.
Figure~\ref{fig3-2a} illustrates the evolution of the attractor of the Poincare map as the energy $E$ grows from $E = 748$ to $E= 765$.\\
\vspace*{-1cm}

Initially the attractor is a stable fixed point $O$. At $E \sim 747.61$ this point
undergoes a period-doubling bifurcation and becomes a saddle; the stable orbit $P=(p_1,p_2)$
of period two becomes an attractor, Fig.~\ref{fig3-2a}a.
At $E=E_2 =748.4395$ a homoclinic butterfly of the unstable
manifold of the saddle $O$ has been formed, Fig.~\ref{fig3-2a}b; as $E$ grows, this homoclinic structure
gives rise to a saddle periodic curve $L = (L_1,L_2)$ of two components,
$L_1$ that surrounds the point $p_1$, and and $L_2$ that surrounds $p_2$.
At the same time, the unstable manifod of $O$ tends to the stable periodic orbit $P$, Fig.~\ref{fig3-2a}c.
At $E\sim 748.97$ (not shown in the Figure) the separatrices touch the stable
manifold of the curve $L$ and then leave it, after which the discrete Lorenz-like attractor is formed.
Almost immediately after that, at $E \sim 748.98$, the
period-$2$ orbit $P$ loses stability at a subcritical torus bifurcation: the saddle periodic closed curve $L$ merges with $P$.
the cycle becomes a saddle and the curve disappears. The discrete Lorenz-like attractor is shown in Figs.~\ref{fig3-2a}d-g.
Note that the scenario of the attractor formation is very similar to that in the Lorenz system subject to a small periodic
perturbation (see Figs.~\ref{Fig-LSMcase}a-e.). Note also that close to the moment of the attractor creation (Fig.~\ref{fig3-2a}d)
the behavior of the unstable manifold of the fixed point $O$ is quite similar to the behavior of the separatrices of the
saddle equilibrium state in the Lorenz model (here we have a difference with the Henon-like maps described in Section \ref{numerical}
where the transition to the Lorenz-like attractor were similar to that in the Shimizu-Morioka model).

As $E$ grows the unstable manifold starts forming visible wriggles (see Fig.~\ref{fig3-2a}e),
so the dynamics of the discrete Lorenz-like attractor is no longer ``flow-like'', eventhough
it still looks quite similar to the classical Lorenz attractor (see Figs.~\ref{fig3-2a}f,g).
In order to check the pseudohyperbolicity of the attractor, we computed the multipliers
of the saddle fixed point $O$ at $E=752.0$: $\lambda_1 = -1.312$, $\lambda_2=0.996$, $\lambda_3 = -0.664$;
the spectrum of Lyapunov exponents for a randomly chosen trajectory is
$\Lambda_1 = 0.0248;\;\; \Lambda_3 = - 0.2445,\;\; 0.00007<\Lambda_2<0.00015$. Evidently, the necessary conditions
for area expansion, $|\lambda_1\lambda_2|>1$ and $\Lambda_1 + \Lambda_2 >0$, are fulfilled, so we, probably, have
a true strange attractor here.

Figures~\ref{fig3-2a}h,i show the destruction of the discrete Lorenz attractor. At $E$ grows, a stable invariant curve
is formed in a lacuna (Fig.~\ref{fig3-2a}h); later, the invariant curve gets destroyed and we see a characteristic shape of the
torus-chaos quasiattractor (Fig.~\ref{fig3-2a}i). The latter disappears at $E > 790$ and the orbits tend to a new stable regime,
a spiral attractor, observed in \citet{GGK12,KJSS12}.

\subsection{Figure--eight attractor in the dynamics of the unbalanced ball}

\begin{figure}[h]
\begin{minipage}[h]{0.3\linewidth}
\center{\includegraphics[height=.8\linewidth]{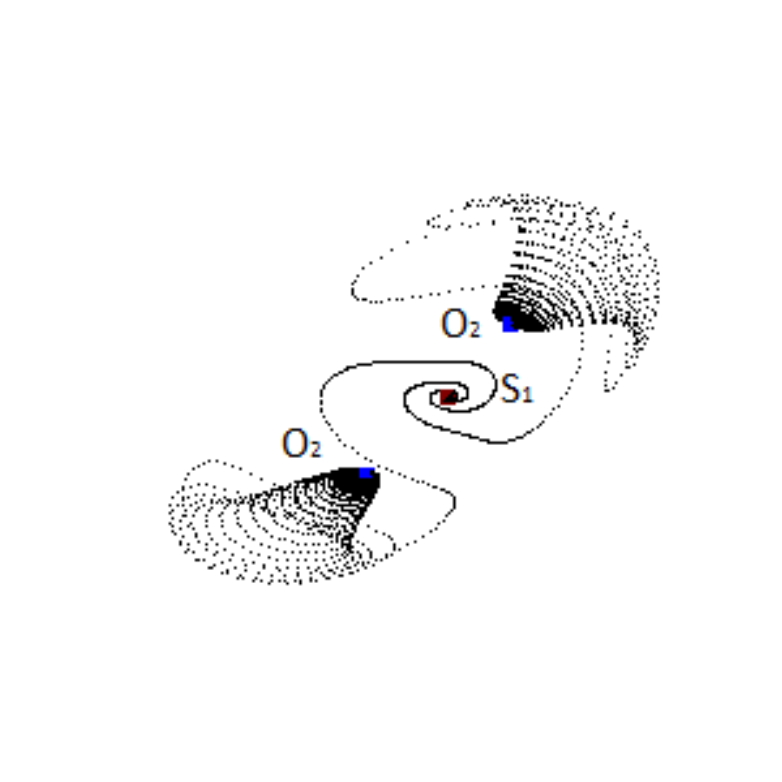} \\ \vspace*{-.5cm} (a) $E = 455$}
\end{minipage}
\hfill
\begin{minipage}[h]{0.3\linewidth}
\center{\includegraphics[height=.8\linewidth]{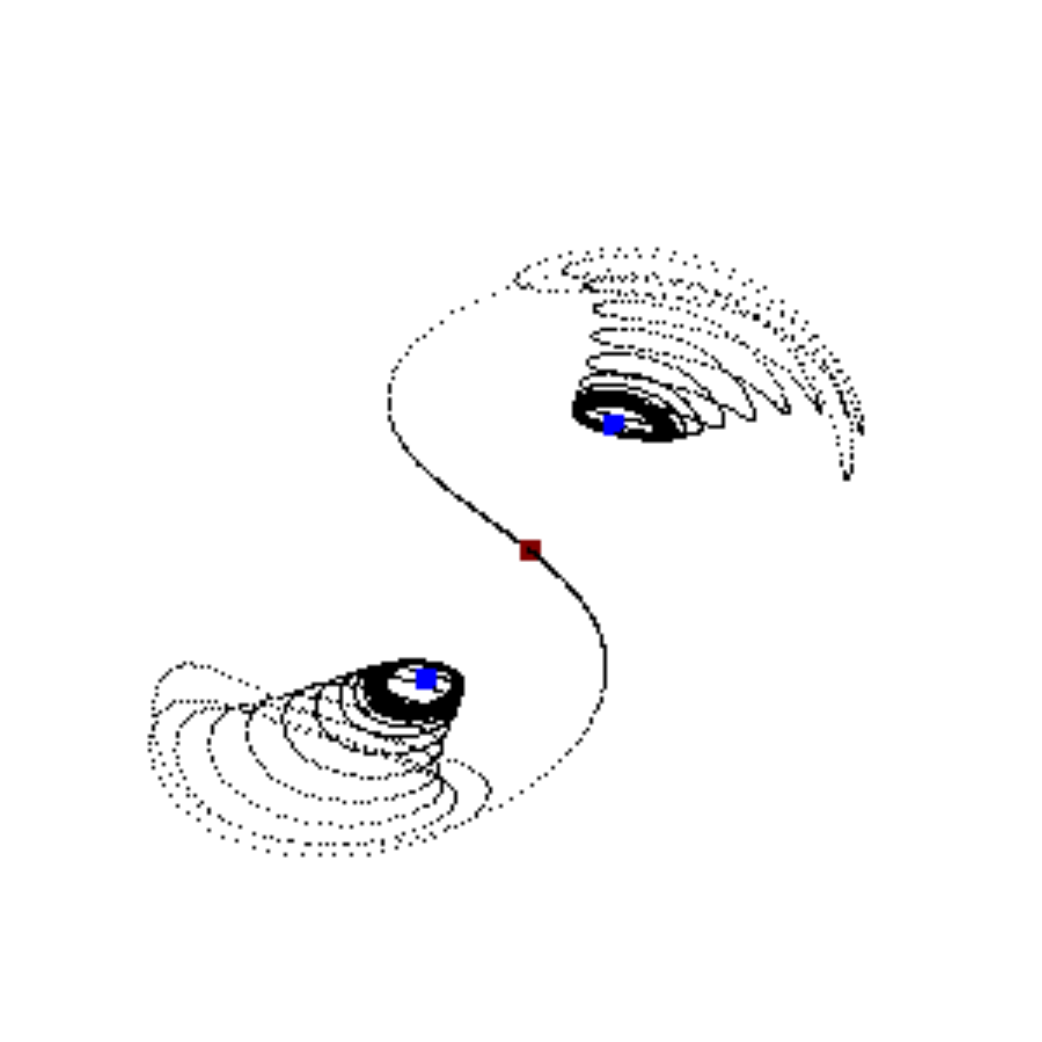} \\ \vspace*{-.5cm} (b)  $E = 457$}
\end{minipage}
\hfill
\begin{minipage}[h]{0.3\linewidth}
\center{\includegraphics[height=.8\linewidth]{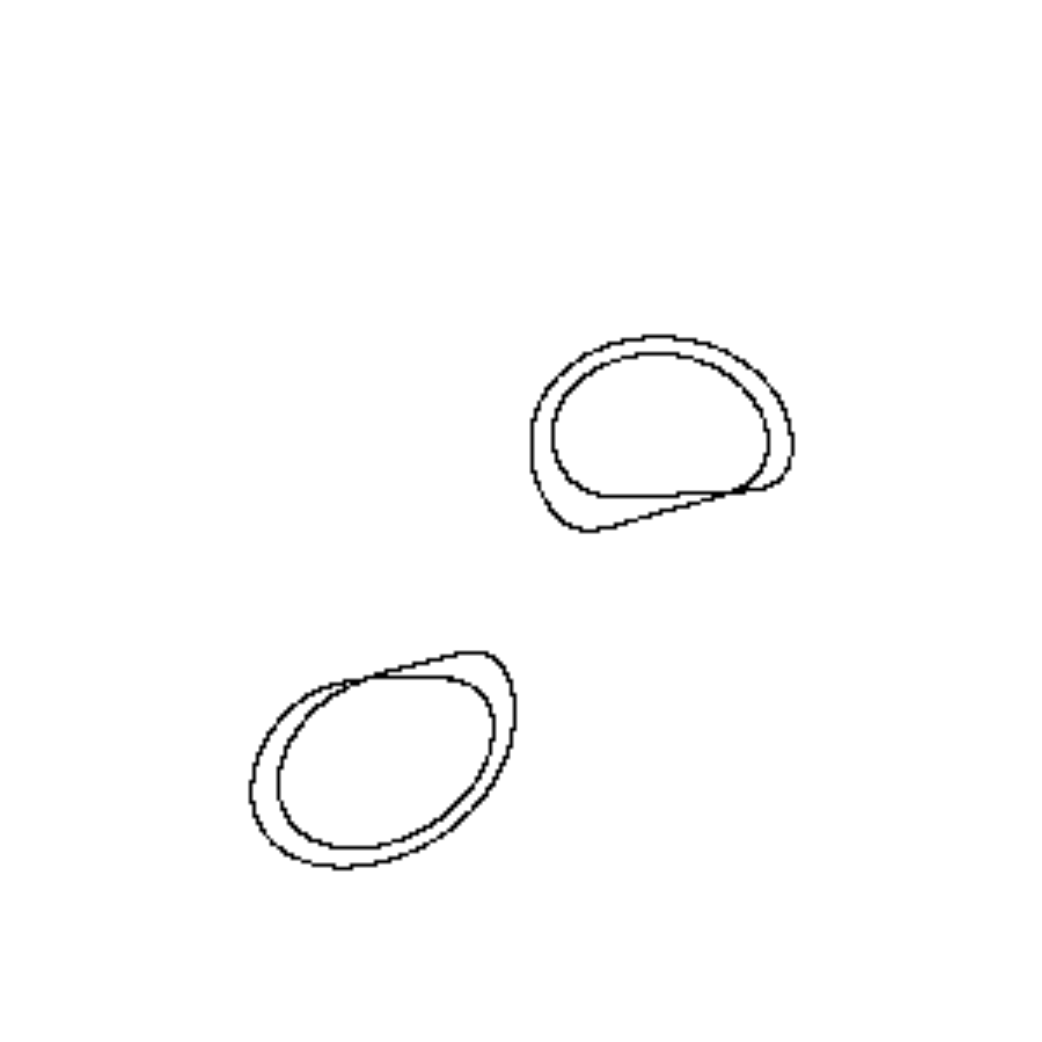} \\ \vspace*{-.5cm} (c) $E = 457.904$}
\end{minipage}
\vfill
\begin{minipage}[h]{0.3\linewidth}
\center{\includegraphics[height=.8\linewidth]{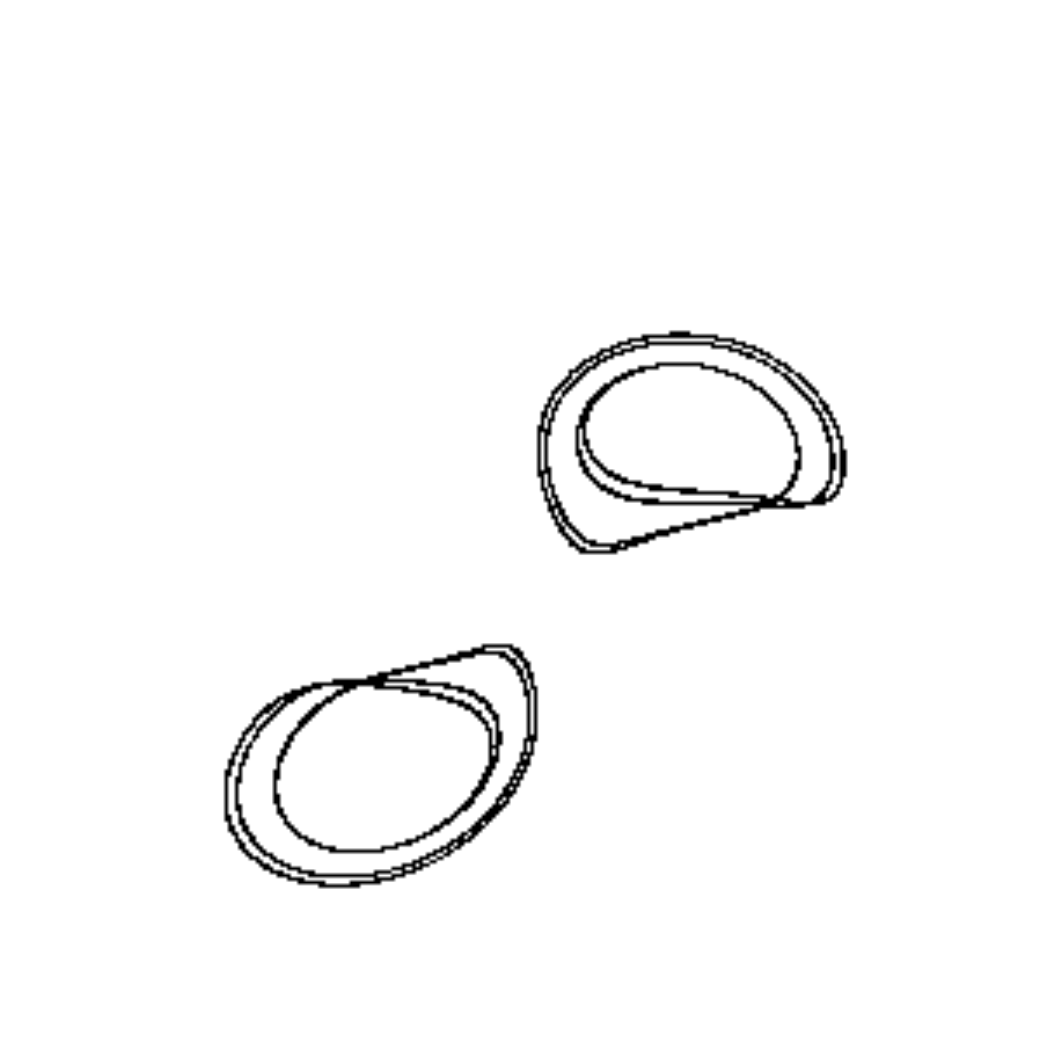} \\ (d) $E = 457.910$}
\end{minipage}
\hfill
\begin{minipage}[h]{0.3\linewidth}
\center{\includegraphics[height=.8\linewidth]{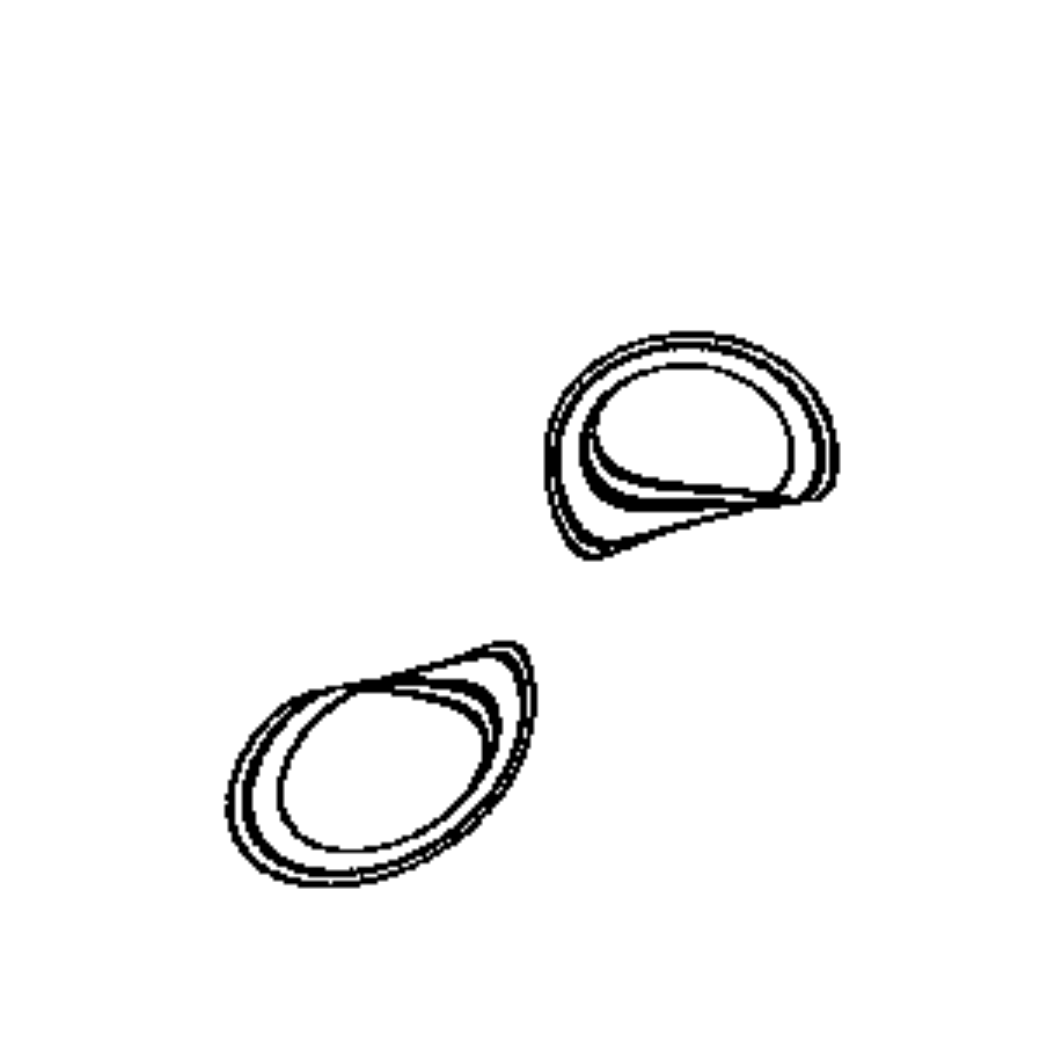} \\ (e) $E = 457.911$}
\end{minipage}
\hfill
\begin{minipage}[h]{0.3\linewidth}
\center{\includegraphics[height=.8\linewidth]{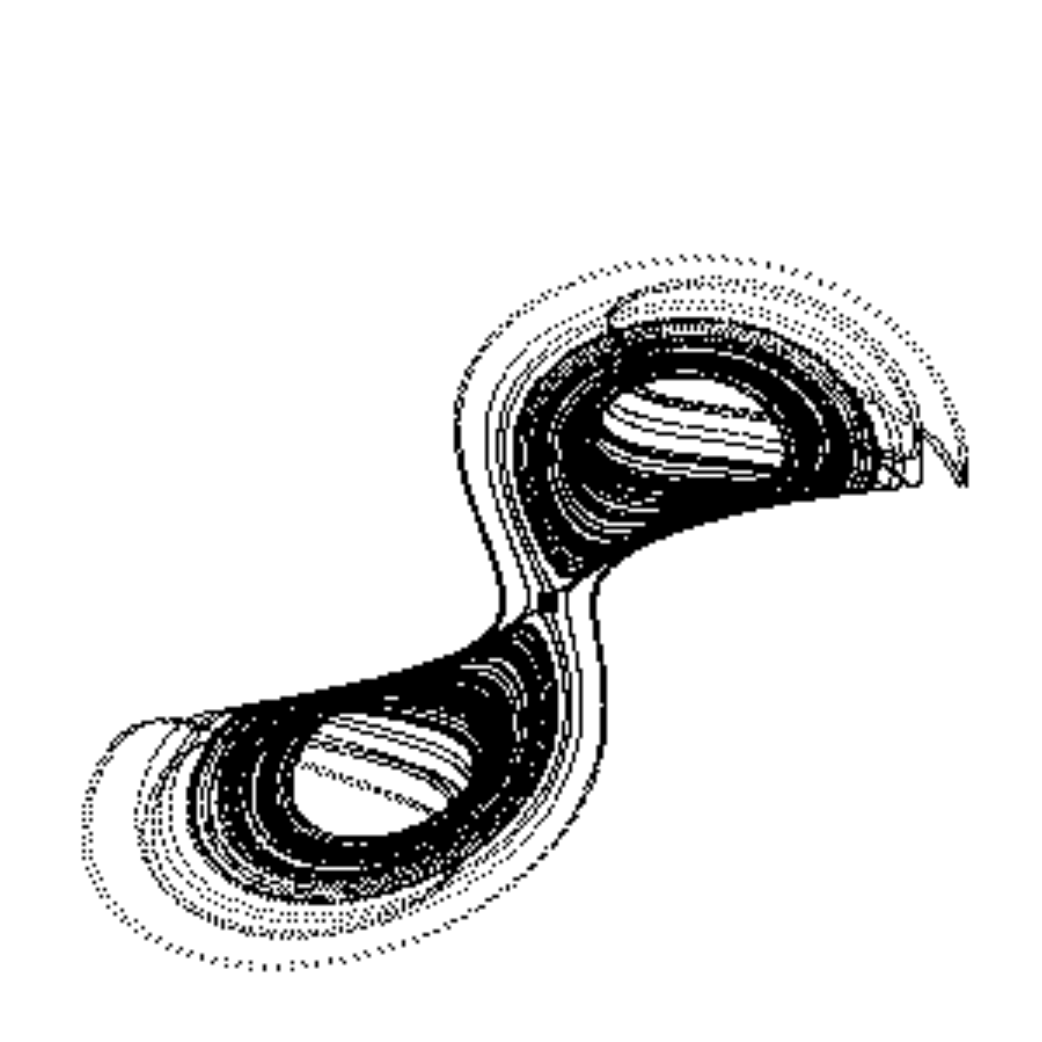} \\ (f) $E = 457.913$}
\end{minipage}
\caption{The main stages of the evolution to the figure-eight attractor.}
\label{Fig:8_Biffurcations_3}
\end{figure}

A model of an unbalanced ball (a ball with displaced center of gravity) rolling on the plane
is given by Eqs. (\ref{eq:2}) with $\displaystyle F(\boldsymbol r) = (\boldsymbol r - \boldsymbol b)^2 - R^2$,
where $\boldsymbol b$ is the vector of the displacement of the center of mass from the geometric center of the ball; $R$
 is the ball's radius. We choose the following parameters: $J_1 =2, J_2 = 6, J_3 =7, m=1, {\rm g} = 100, R = 3, b_1 = 1, b_2 = 1.5, b_3 = 1.9$.
A figure--eight, seemingly pseudo-hyperbolic attractor was numerically found in this model in \citet{BKS14}
\footnote{A model of an unbalanced rubber ball (i.e. the unbalanced ball that moves without spinning) was considered in \citet{Kazakov13}.
The additional nonholonomic constraint (no spinning) reduces dimension of the problem, i.e. the Poincar\'e map becomes two-dimensional.
Still the dynamics of the system remains very complex. In particular, coexisting strange attractors and
repellers, as well as mixed dynamics \cite{GST97,DGGTS} were found.}. Figure~\ref{Fig:8_Biffurcations_3} shows
the development of the attractor of the Poincare map in the model as the energy $E$ varies from
$E=455.0$ to $E = 457.913$.\\
\vspace*{-1cm}

At first, for $E_1\simeq 417.5 < E < E_2 \simeq 455.95$ (Fig.~\ref{Fig:8_Biffurcations_3}a)
the attractor is a period-2 orbit  $(O_1,O_2)$ that emerges at $E=E_1$ along
with a saddle orbit $S=(s_1,s_2)$ as a result of a saddle-node bifurcation. Simultaneously, the system has a saddle
fixed point $S_1$: this point, a saddle-focus then a saddle, has a two-dimensional unstable manifold; then
at $E = E_3\simeq 456.15$, the fixed point becomes a saddle with one-dimensional unstable manifold
as a result of a subcritical period-doubling bifurcation when the saddle orbit $(s_1,s_2)$ merges to $S_1$.
At $E=E_2\simeq 455.95$ the orbit $(O_1,O_2)$  loses the stability at a supercritical Andronov-Hopf bifurcation
and a stable period-2 closed  curve appears. Thus, at $E>E_3$ the one-dimensional unstable separatrices of the saddle
fixed point $S_1$ (with multipliers $\lambda_1<-1, |\lambda_{2,3}|<1$
and $\lambda_{2}\lambda_{3}<0$) wind up  onto a stable closed curve of  period-2, Fig.~\ref{Fig:8_Biffurcations_3}b.
Next, several doublings of the invariant curve take place, see Figs.~\ref{Fig:8_Biffurcations_3}c--e.
The further growth of $E$ leads to a figure-eight attractor, Fig.~\ref{Fig:8_Biffurcations_3}f.

Note that at $E = 457.913$, the fixed point $S_1$ has the multipliers $\lambda_1 \simeq -1.00907, \lambda_{2} \simeq  -0.99732,
\lambda_3 \simeq 0.98885$. Thus, the area-expansion conditions $|\lambda_1\lambda_2|>1$ is fulfilled.
Moreover, the Lyapunov exponents for a random trajectory in the attractor
are as follows: $\Lambda_1 \simeq 0.00063, \Lambda_2 \simeq -0.00003, \Lambda_3 \simeq -0.00492$,
which gives $\Lambda_1 +\Lambda_2>0$ and hints the pseudohyperbolicity.

\section*{Acknowledgments} This work was supported by grants of RFBR 13-01-00589, 13-01-97028-r-povoljie, and 14-01-344, by
basic funding of Russian Ministry of Education and Science, and by the grant of RSF 14-41-00044.


\begin{thebibliography}{40}
\bibitem[Afraimovich {\it et al}.(1977)]{ABS77}
Afraimovich, V.S.,  Bykov, V.V. \&  Shilnikov, L.P. [1977] ``The origin and
structure of the Lorenz attractor,'' {\it Sov. Phys. Dokl.} {\bf 22}, 253-255.
%
\bibitem[Afraimovich {\it et al}.(1980)]{ABS80}
Afraimovich, V.S.,  Bykov, V.V. \&  Shilnikov, L.P. [1980] ``On the existence of stable periodic orbits in the Lorenz model,''
{\it Russ. Math. Survey} {\bf 35}, 164--165.
%
\bibitem[Afraimovich {\it et al}.(1982)]{ABS83}
Afraimovich, V.S.,  Bykov, V.V. \&  Shilnikov, L.P. [1982] ``On attracting
structurally unstable limit sets of lorenz attractor type,'' {\it Trans. Mosc. Math. Soc.} {\bf 44}, 153-216.
%
\bibitem[Afraimovich \& Shilnikov(1974a)]{AfrShRing1} Afraimovich, V.S. \& Shilnikov, L.P. [1974a]
``On small periodic perturbations of autonomous systems,'' {\it Soviet Math., Dokl.} {\bf 15}, 206-211.
%
\bibitem[Afraimovich \& Shilnikov(1974b)]{AfrShRing2}
Afraimovich, V.S. \& Shilnikov, L.P. [1974b] ``On some global bifurcations connected with the disappearance of saddle-node fixed point,''
{\it Soviet Math., Dokl.} {\bf 15}, 1761-1765.
%
\bibitem[Afraimovich \& Shilnikov(1977)]{AfrShRing3} Afraimovich, V.S. \& Shilnikov, L.P. [1977] ``The
ring principle in problems of interaction between two self-oscillating systems,''
{\it J. Appl. Math. Mech.} {\bf 41}, 632-641.
%
\bibitem[Aframovich \& Shilnikov(1983))]{ASh83a}
Aframovich, V.S. \&  Shilnikov, L.P. [1983] ``Strange attractors and quasiattractors''  {\it
Nonlinear Dynamics and Turbulence}, eds. Barenblatt, G.I., Iooss, G. \& Joseph, D. (Boston, Pitmen).
%
\bibitem[Afraimovich \& Shilnikov(1983a)]{AfrSh}  Afraimovich, V.S. \& Shilnikov, L.P. [1983a]
``On invariant two-dimensional tori, their breakdown and stochasticity,'' {\it
Methods of the Qualitative Theory of Differential Equations, Gorky}, 3--26.
[English translation in: Amer. Math. Soc. Transl., 149 (1991), 201--212].
%
\bibitem[Afraimovich \& Vozovoi(1988)]{AfrVoz1}  Afraimovich, V.S. \& Vozovoi, L.P. [1988] ``Mechanism
of the generation of a two-dimensional torus
upon loss of stability of an equilibrium state,'' {\it Soviet Phys. Doklady} {\bf 33}, 720--723.
%
\bibitem[Afraimovich \& Vozovoi(1989)]{AfrVoz2}  Afraimovich, V.S. \&  Vozovoi, L.P. [1989]
``The mechanism of the hard appearance of a two-frequency oscillation mode in the case
of Andronov-Hopf reverse bifurcation,'' {\it J. Appl. Math. Mech.} {\bf 53}, 24--28.
%
\bibitem[Anishenko \&  Nikolaev(2005))]{AnN}
Anishenko, V.S. \&  Nikolaev, S.M. [2005] ``Bifuraction doubling of the two-dimension torus,''  {\it
Letters to JTPH}, {\bf 31}, 88--94.
%
\bibitem[Arneodo {\em et al}.(1981)]{Arn1} Arneodo, A., Coullet, P. \& Tresser, C. [1981]
``Possible new strange attractors with spiral structure,'' {\it Communications in Mathematical Physics} {\bf 79}, 573--579.
%
\bibitem[Arneodo {\em et al}.(1985)]{Arn2}  Arneodo, A., Coullet, P.H. \& Spiegel, E.A. [1985]
``The dynamics of triple convection,'' {\it Geophys. Astrophys. Fluid Dynamics} {\bf 31}, 1--48.
%
\bibitem[Arnold(1977)]{Arn77}  Arnold, V.I. [1977]
``Loss of stability of self-oscillations close to resonance and
versal deformations of equivariant vector fields,'' {\it Funct. Anal. Appl.} {\bf 11}, 85--92.
%
\bibitem[Aronson {\em et al}.(1982)]{ArCh}  Aronson, D.G., Chory, M.A., Mcgehee, R.P. \& Hall, G.R. [1982]
``Bifurcation from an invariant circle for two-parameter families of maps of the plane,''
{\it  Commun. Math. Phys.} {\bf 83}, 303--354.
%
\bibitem[Bamon {\em et al}.(2005)]{Bamon}  Bamon, R.,  Kiwi, J. \&  Rivera, J. [2005]
``Wild Lorenz like attractors,'' Preprint , {\bf arXiv:math/0508045}.
%
\bibitem[Barrio {\em et al}.(2012)]{BSS}
Barrio, R., Shilnikov, A. \& Shilnikov, L. [2012] ``Kneadings, symbolic dynamics, and painting Lorenz chaos,'' {\em
Int. J. Bifurcation and Chaos} {\bf 22}, 1230016.
%
\bibitem[Belyakov(1980)]{Bel70}
Belyakov, L.A. [1980] ``A case of the generation of a periodic orbit motion with homoclinic curves,'' {\it Math. Notes} {\bf 15}, 336-341.
%
\bibitem[Belykh {\em et al}.(2005)]{Belykh}  Belykh, V.,  Belykh, I. \&   Mosekilde, E.
[2005] ``The hyperbolic Plykin attractor can exist in neuron models,'' {\it Int. Journal of Bifurcation and Chaos} {\bf 15, No 11}, 3567-3578.
%
\bibitem[Benedicks \& Carleson(1991)]{BC}
Benedicks, M. \& Carleson, L. [1991] ``The dynamics of the Henon map,'' {\it  Ann. Math.} {\bf 133}, 73-169.
%
\bibitem[Bonatti {\em et al}.(2005)]{Bonatti} Bonatti, C., Diaz, L. \& Viana, M. [2005] Dynamics Beyond Uniform
Hyperbolicity: A Global Geometric and Probabilistic
Perspective, {\it Encyclopaedia of Mathematical Sciences} {\bf 102} (Springer).
%
\bibitem[Borisov{\em et al}.(2012)]{KJSS12}
Borisov, A.V., Jalnine, A.Yu., Kuznetsov, S.P., Sataev, I.R. \& Sedova J.V. [2013]
``Dynamical phenomena occuring due to phase volume compression in nonholonomic model of the rattleback,''
{\it Regular and Chaotic Dynamics} {\bf 17}, 512--532.
%
\bibitem[Borisov {\em et al}.(2014)]{BKS14}
Borisov, A.V., Kazakov, A.O. \& Sataev, I.R. [2014]
``Regular and chaotic phenomena in the nonholonomic model of the unbalanced ball roling on a plane,''
{\it Regular and Chaotic Dynamics} {\bf 18}, 512--532.
%
\bibitem[Borisov \& Mamaev(2003)]{BM03}
Borisov, A.V. \& Mamaev, I.S. [2003]
``Strange attractors in rattleback dynamics,''
{\it Physics-Uspekhi} {\bf 46}, 393--403.
%
\bibitem[Bosh \& Simo(1993)]{whoelse}
Bosh, M. \& Simo, C. [1993] ``Attractors in a Shilnikov-Hopf scenario and a related one-dimensinal map,'' {\it Physica D} {\bf 62}, 217-229.
%
\bibitem[Braaksma {\em et al}.(1990)]{Bro2} Braaksma, B.L.J., Broer, H.W. \&  Huitema, G.B. [1990]
``Toward a quasi-periodic bifurcation theory,''
{\it Memoirs AMS} {\bf 83}, 83-175.
%
\bibitem[Broer {\em et al}.(1990)]{Bro1}  Broer, H.W., Huitema, G.B. \& Takens, F. [1990] ``Unfoldings of quasi-periodic tori,''
{\it Memoirs AMS} {\bf 83}, 1--82.
%
\bibitem[Bykov(1978)]{Byk0} Bykov, V.V. [1978] ``On the structure of a neighborhood of a separatrix contour with a saddle-focus,''
{\it Methods of the Qualitative Theory of Differencial Equations, Gorky}, 3--32.
%
\bibitem[Bykov(1980)]{Byk1} Bykov, V.V. [1980] ``On bifurcations of dynamical systems close to systems
with a separatrix contour containing a saddle-focus,'' {\it Methods of the Qualitative Theory of Differencial Equations, Gorky},
44--72.
%
\bibitem[Bykov(1993)]{Byk2} Bykov, V.V. [1993] ``The bifurcations of separatrix contours and
chaos'' {\it Physica D} {\bf 62}, 290--299.
%
\bibitem[Bykov \& A.Shilnikov(1989)]{BS89}  Bykov, V.V. \& Shilnikov, A.L. [1989] ``On the boundaries of the domain of existence
of the Lorenz attractor,'' {\it Methods of the Qualitative Theory of Differencial Equations, Gorky},
151--159 [English translation in Selecta Math. Soviet., 11 (1992), 375--382].
%
\bibitem[Chenciner(1985)]{Ch1}  Chenciner, A. [1985] ``Bifurcations de points fixes elliptiques. I. Courbes invariantes,''
{\it IHES Publ. Math.} {\bf 61} , 67--127; ``II. Orbites periodiques et ensembles de Cantor
invariants,'' {\it Invent. Math.} {\bf 80} (1985), 81--106; ``III. Orbites periodiques de ``petites''
periodes et elimination resonante des couples de courbes invariantes,'' {\it IHES Publ. Math.} {\bf 66} (1987),
5--91.
%
\bibitem[Curry \& Yorke(1978)]{CY}  Curry, J.H., Yorke, J.A. [1978] ``A transition from Hopf
bifurcation to chaos: computer experiments with maps in $R^2$,''
{\it The Structure of Attractors in Dynamical Systems}, eds. Martin, J.C., Markley, N.G. \& Perrizo, W.
{\it Lecture Notes Math.} {\bf 668} (Springer, Berlin), 48--66.
%
\bibitem[Delshams {\it et al}.(2013)]{DGGTS}
Delshams, A., Gonchenko, S.V., Gonchenko, V.S., Lazaro, J.T. \& Sten'kin, O. [2013] ``Abundance of attracting, repelling and
elliptic periodic orbits in two-dimensional reversible maps,'' {\it Nonlinearity} {\bf 26}, 1--33.
%
\bibitem[Gavrilov(1977)]{Gav77} Gavrilov, N.K. [1977] ``On bifurcations of periodic orbits near inner resonance 1:3,''
{\it Investigations on stability and the theory of oscillations. Yaroslavl}, 192--199.
%
\bibitem[Gonchenko \& Gonchenko(2013)]{GG13}
Gonchenko, A.S. \& Gonchenko, S.V. [2013] ``On existence
of Lorenz-like attractors in a nonholonomic model of a Celtic stone''{\it  Rus. Nonlin. Dyn.} {\bf 9}, 77--89.
%
\bibitem[Gonchenko {\em et al}.(2012)]{GGK12}
Gonchenko, A.S., Gonchenko, S.V. \& Kazakov, A.O. [2012]
``On new aspects of chaotic dynamics of Celtic stone'','' {\it Rus. Nonlinear Dyn.} {\bf 8}, 507--518.
%
\bibitem[Gonchenko {\em et al}.(2013)]{GGK13}
Gonchenko, S.V., Gonchenko, A.S. \& Kazakov, A.O. [2013] ``Richness of chaotic dynamics in nonholonomic
models of a Celtic stone,'' {\it Regular and Chaotic Dynamics} {\bf 15}, 521--538.
%
\bibitem[Gonchenko {\em et al}.(2012a)]{GGS13}
Gonchenko, A.S., Gonchenko, S.V. \& Shilnikov, L.P. [2012a] ``Towards scenarios of chaos appearance in three-dimensional maps,''
{\it Rus. Nonlinear Dyn.} {\bf 8}, 3--28.
%
\bibitem[Gonchenko {\em et al}.(2005)]{GOST05}  Gonchenko, S.V., Ovsyannikov, I.I., Simo, C. \& Turaev, D. [2005]
``Three-dimensional Henon-like maps and wild Lorenz-like
attractors,'' {\it Bifurcation and Chaos} {\bf 15}, 3493--3508.
%
\bibitem[Gonchenko {\em et al}.(2013)a]{GGOT13}
Gonchenko, S.V., Gonchenko, A.S., Ovsyannikov, I.I. \& Turaev, D.V. [2013a] ``Examples  of  Lorenz-like Attractors
in  Henon-like Maps,'' {\it Math. Model. Nat. Phenom.} {\bf 8}, 32--54.
%
\bibitem[Gonchenko {\em et al}.(2012a)] {GOT12} Gonchenko, S.V., Ovsyannikov, I.I. \& Turaev D. [2012a]
``On the effect of invisibility of stable periodic orbits at homoclinic bifurcations,''
{\it Physica D} {\bf 241}, 1115--1122.
%
\bibitem[Gonchenko {\em et al}.(1993)]{GST93c}
Gonchenko, S.V., Turaev, D.V. \& Shilnikov, L.P. [1993] ``Dynamical phenomena
in multi-dimensional systems with a non-rough Poincare homoclinic
curve,'' {\it Russ. Acad. Sci. Dokl. Math.} {\bf 47}, 410--415.
%
\bibitem[Gonchenko {\em et al}.(1996)]{GST96}
Gonchenko, S.V., Shilnikov, L.P. \& Turaev, D.V. [1996] ``Dynamical phenomena
in systems with structurally unstable Poincare homoclinic orbits,'' {\it Chaos} {\bf 6}, 15--31.
%
\bibitem[Gonchenko {\em et al}.(1997)]{GST97}
Gonchenko, S.V., Shilnikov, L.P. \& Turaev, D.V. [1997] ``On Newhouse regions of two-dimensional diffeomorphisms close to a
diffeomorphism with a nontransversal heteroclinic cycle,'' {\it Proc. Steklov Inst. Math.} {\bf 216}, 70-–118.
%
\bibitem[Gonchenko {\em et al}.(2008)]{GST08}
Gonchenko, S.V., Shilnikov, L.P. \& Turaev  D.V. [2008] ``On dynamical
properties of multidimensional diffeomorphisms from Newhouse
regions.~I,'' {\it Nonlinearity} {\bf 21}, 923-972.
%
\bibitem[Guckenheimer(1976)]{G76}
Guckenheimer, J. [1976] ``A strange, strange attractor,'' {\it The Hopf Bifurcation Theorem and its Applications},
eds. Marsden, J. \& McCracken, M. (Springer-Verlag), 368-381.
%
\bibitem[Guckenheimer \&  Williams(1979)]{GW79}
Guckenheimer, J. \&  Williams, R.F. [1979] ``Structural stability of Lorenz attractors,'' {\it
IHES Publ. Math.} {\bf 50}, 59--72.
%
\bibitem[Kazakov(2013)]{Kazakov13}
Kazakov, A.O. [2013] ``Strange attractors and mixed dynamics
in the problem of an unbalanced rubber ball rolling on a plane,'' {\it Regular and Chaotic Dynamics} {\bf 15}, 508--520.
%
\bibitem[Khibnik {\em et al}.(1993)]{Chua}  Khibnik, A.I., Roose, D. \& Chua, L.O. [1993] ``On periodic orbits
and homoclinic bifurcations in Chua's circuit with smooth nonlinearity,'' {\it Int. J. Bifurcation and Chaos} {\bf 3}, 363--384.
%
\bibitem[Kuznetsov(1998)]{Kuznetsov}  Kuznetsov Yu.A. [1998] ``Elements of applied bifurcation theory'' (Springer-Verlag,
NY, Berlin, Heidelberg).
%
\bibitem[Los(1989)]{Los3}  Los, J.E.[1989] ``Non-normally hyperbolic invariant curves for maps in $R^3$ and doubling bifurcation,''
{\it Nonlinearity} {\bf 2}, 149--174.
%
\bibitem[Mora \&  Viana(1993)]{MV93}
Mora, L. \&  Viana, M. [1993] ``Abundance of strange attractors,'' {\it Acta Math.} {\bf 171}, 1--71.
%
\bibitem[Newhouse(1974)]{N74}
Newhouse, S.E. [1974] ``Diffeomorphisms  with  infinitely  many  sinks,'' {\it Topology} {\bf 13}, 9--18.
%
\bibitem[Ovsyannikov \& Shilnikov(1987)]{OSh}
Ovsyannikov, I.M. \& Shilnikov, L.P. [1987] ``On systems with a saddle-focus
homoclinic curve,'' {\it Math. USSR Sbornik} {\bf 58}, 91-102.
%
\bibitem[Petrovskaya \& Yudovich(1980)]{PY}
Petrovskaya, N.V. \& Yudovich, V.I.,[1980] ``Homoclinic loops of
the Saltzman-Lorenz system,'' {\it Methods of the Qualitative Theory of Differential Equations, Gorky}, 73--83.
%
\bibitem[A.Shilnikov(1986)]{ShA86}
Shilnikov, A.L.[1986] ``Bifurcation and chaos in the Morioka-Shimizu system,'' {\it Methods of the
Qualitative Theory of Differential Equations, Gorky}, 180--193 [English translation in Selecta
Math. Soviet., 10 (1991) 105--117]; II. {\it Methods of Qualitative Theory and Theory of
Bifurcations, Gorky} (1989), 130--138.
%
\bibitem[A.Shilnikov(1993)]{ShA93}
Shilnikov, A.L. [1993] ``On bifurcations of the Lorenz attractor in the
Shimuizu-Morioka model,'' {\it Physica D} {\bf 62}, 338--346.
%
\bibitem[Shilnikov \& Shilnikov(1991)]{AshSh}
Shilnikov, A.L. \& Shilnikov, L.P. [1991] ``On the nonsymmetric Lorenz model,'' {\it Int. J. Bifurcation and Chaos} {\bf 1}, 773--776.
%
\bibitem[Shilnikov {\em et al}.(1993)]{SST93}
Shilnikov, A.L., Shilnikov, L.P. \& Turaev, D.V. [1993] ``Normal forms and
Lorenz attractors,'' {\it Bifurcation and Chaos} {\bf 3}, 1123--1139.
%
\bibitem[Shilnikov(1965)]{Sh65}
Shilnikov, L.P. [1965] ``A case of the existence of a denumerate set of periodic motions,''  {\it Sov. Math. Docl.} {\bf 6}, 163--166.
%
\bibitem[Shilnikov(1967)]{Sh67}
Shilnikov, L.P. [1967] ``On a Poincare-Birkhoff problem,'' {\it Math. USSR Sb.} {\bf 3}, 91--102.
%
\bibitem[Shilnikov(1970)]{Sh70}
Shilnikov, L.P. [1970] ``A contribution to the problem of the structure of an extended
neighbourhood of a rough equilibrium state of saddle-focus type,'' {\it Math. USSR Sbornik} {\bf 10}, 91--102.
%
\bibitem[Shilnikov(1980)]{Sh80}
Shilnikov, L.P. [1980] ``Bifurcation theory and the Lorenz model,'' Addition 1 to the book Marsden, J.
\& Mac-Cracken, M., Bifurcation of cycle birth and its applications (Moscow, Mir), 317--335.
%
\bibitem[Shilnikov(1986)]{Sh86}
Shilnikov, L.P. [1986] ``Bifurcation theory and turbulence,'' {\it Methods of the Qualitative Theory of Differential Equations,
Gorky}, 150--163 [English translation in {\it Selecta Math. Sovietica} {\bf 10} [1991], 43--53].
%
\bibitem[Shilnikov {\em et al}.(1998,2001)]{book}
Shilnikov, L.P.,  Shilnikov, A.L., Turaev, D.V., Chua, L.O.[1998] Methods of qualitative theory in nonlinear dynamics. Part~I. [2001]
Part~II (World Scientific, Singapore).
%
\bibitem[Turaev(1996)]{T96} Turaev, D.V.[1996] ``On dimension of non-local
bifurcational problems,'' {\it Bifurcation and Chaos} {\bf 6}, 919--948.
%
\bibitem[Turaev \&  Shilnikov(1998)]{TS98}
Turaev, D.V. \& Shilnikov, L.P. [1998] ``An example of a wild strange attractor,'' {\it Sb. Math.} {\bf 189}, 291--314.
%
\bibitem[Turaev \& Shilnikov(2008)]{TS08}
Turaev, D.V. \& Shilnikov L.P.[2008] ``Pseudo-hyperbolisity and the problem
on periodic perturbations of Lorenz-like attractors,'' {\it Russian Dokl. Math.} {\bf 467}, 23--27.
%
\bibitem[Ures(1995)]{Ures95} Ures, R. [1995] ``On the approximation of H\'enon-like attractors by homoclinic tangencies,''
{\it Ergod. Th. Dyn. Sys.} {\bf 15}, 1223--1229.
%
\bibitem[Vitolo(2003)]{Vitolo} Vitolo, R.[2003] Bifurcations of attractors in 3D diffeomorphisms: a study in experimental mathematics,
{\it Doctoral thesis} (University of Groningen Press).
%
\bibitem[Williams(1977)]{W77}
Williams R.F. [1977] ``The structure of Lorenz attractors,'' {\it Lect. Notes Math.} {\bf 615}, 94--112.
\end{thebibliography}
\end{document}